\documentclass[prb,twocolumn,showpacs]{revtex4}

\usepackage{times}
\usepackage{float}
\usepackage{amsmath}
\usepackage{amsthm}
\usepackage{amssymb}
\usepackage{amsbsy}
\usepackage{graphicx}
\usepackage{pstricks}
\usepackage{color}
\usepackage{enumerate}
\usepackage{multirow} 
\usepackage{ulem}
\usepackage{wasysym}

\begin{document}


\title{Seeing the light : experimental signatures of emergent electromagnetism in a quantum spin ice}


\author{Owen Benton}
\address{H. H. Wills Physics Laboratory, University of Bristol, Tyndall Avenue, Bristol BS8 1TL, UK}


\author{Olga Sikora}
\address{H. H. Wills Physics Laboratory, University of Bristol, Tyndall Avenue, Bristol BS8 1TL, UK}
\address{Okinawa Institute of Science and Technology, 12-22 Suzaki, Uruma, Okinawa, 904-2234, Japan}


\author{Nic Shannon}
\address{H. H. Wills Physics Laboratory, University of Bristol, Tyndall Avenue, Bristol BS8 1TL, UK}
\address{Clarendon Laboratory, University of Oxford, Parks Road, Oxford OX1 3PU, UK}
\address{Okinawa Institute of Science and Technology, 12-22 Suzaki, Uruma, Okinawa, 904-2234, Japan}


\begin{abstract}
The ``spin ice'' state found in the rare earth pyrochlore magnets
Ho$_2$Ti$_2$O$_7$ and Dy$_2$Ti$_2$O$_7$ offers a beautiful realisation of classical 
magnetostatics, complete with magnetic monopole excitations.  
It has been suggested that in ``quantum spin ice"  materials, quantum-mechanical 
tunnelling between different ice configurations could convert the magnetostatics 
of spin ice into a quantum spin liquid which realises a fully dynamical, 
lattice-analogue of quantum electromagnetism. 
Here we explore how such a state might manifest itself in experiment, within the 
minimal microscopic model of a such a quantum spin ice.
We develop a lattice field theory for this model, and use this to make explicit
predictions for the dynamical structure factor which would be observed in neutron
scattering experiments on a quantum spin ice.
We find that ``pinch points'', 
which are the signal feature of a classical spin ice, fade away as a quantum ice is 
cooled to its zero-temperature ground state.
We also make explicit predictions for the ghostly, linearly dispersing magnetic excitations 
which are the ``photons'' of this emergent electromagnetism.
The predictions of this field theory are shown to be in quantitative agreement with
Quantum Monte Carlo simulations at zero temperature.
\end{abstract}


\pacs{
75.10.Jm 
75.10.Kt, 
11.15.Ha, 
}


\maketitle

\section{Introduction}
\label{intro}


The idea that a strongly interacting quantum magnet might support a spin liquid phase which remains 
disordered even at zero-temperature has fascinated --- and frustrated --- 
physicists ever since the seminal ``resonating valence bond'' (RVB) paper of Anderson in 1973~\cite{anderson73}.
Such a phase, 
it was argued, need not support the spin waves found in conventional 
magnets, but could instead exhibit ``spinons'' with fractional quantum numbers.
Forty years later, the search for quantum spin liquids goes on, but with strong grounds for \mbox{encouragement :}
a growing number of quantum magnets have been identified which {\it do not} order down to the lowest
temperatures measured, many of which have low-temperature properties which hint at spinons~\cite{lee08,balents10}.  
At the same time, the ``spin ice'' materials Ho$_2$Ti$_2$O$_7$ and Dy$_2$Ti$_2$O$_7$ have emerged
as text-book examples of classical (i.e.~entropy-driven) spin liquids~\cite{harris97,bramwell01,gardner10}
These highly-frustrated magnetic insulators 
show algebraic correlations of spins over macroscopic 
distances~\cite{huse03,henley05,fennell09,henley10} 
and support magnetic monopole excitations which provide classical analogues to 
the spinons envisaged by 
Anderson~\cite{ryzhkin05, castelnovo08, bramwell09, morris09, kadowaki09, jaubert09, giblin11}.


Recently, the idea of a ``quantum spin ice'' has also attracted considerable interest.
The family of rare earth pyrochlores to which Ho$_2$Ti$_2$O$_7$ and 
Dy$_2$Ti$_2$O$_7$ belong includes other systems 
in which quantum effects play a much more important role~\cite{gardner10}.
Perhaps the most widely studied system of this type is Tb$_2$Ti$_2$O$_7$.
Like the classical spin ices, 
the magnetism of Tb$_2$Ti$_2$O$_7$ is  controlled by the competition 
between strong Ising anisotropy, and dipolar interactions which are ferromagnetic 
on nearest-neighbour bonds, so it is expected to be an ``ice''.  
However, in Tb$_2$Ti$_2$O$_7$, anisotropic exchange interactions also 
play an important role, and endow the spins with 
dynamics~\cite{gingras00,enjalran04,molavian07,molavian09}.
A diffuse, liquid-like structure is observed in neutron scattering for a wide range of 
temperatures, with no conventional magnetic order observed down to $50$mK, 
despite the fact that the typical scale of interactions between spins 
is closer to $11$K~\cite{gardner03,gardner01}.
Muon spin rotation experiments, meanwhile,  suggest that spins continue to fluctuate 
down to the lowest temperatures~\cite{gardner99}, and the most recent quasi-elastic 
neutron scattering experiments find evidence of power-law spin correlations at 
$50$mK~\cite{fennel12}.
Taken together, these facts make Tb$_2$Ti$_2$O$_7$ a prime example of a 
three-dimensional, quantum spin liquid.


The magnetism of Yb$_2$Ti$_2$O$_7$ has also proved very interesting, with 
neutron scattering finding no evidence of order at temperatures above $210$mK, 
and evidence for frustrated, anisotropic exchange interactions favouring significant 
dynamics within an ``ice-like'' manifold of states
~\cite{thompson11, ross11-PRX,ross11-PRB,chang-arXiv,applegate-arXiv}.  
Comparable studies of Pr$_2$Sn$_2$O$_7$ suggest that it does not order 
down to $500$mK, but with spins continuing to fluctuate~\cite{zhou08,onoda10,onoda11-PRB}.
There is also reason to believe that other Pr metal oxides, including Pr$_2$Zr$_2$O$_7$, 
may prove a worthwhile hunting ground for quantum spin 
liquids~\cite{onoda10, onoda11-PRB, matsuhira09,lee-arXiv}.
And, while the dynamics of the ``classical'' spin ices Ho$_2$Ti$_2$O$_7$ and 
Dy$_2$Ti$_2$O$_7$ become very slow at low temperatures, neither system has ever 
been observed to order, despite the fact that the 
dipolar interactions present in these systems are expected to favour an ordered 
state~\cite{gingras01,melko04}.
%
%
All of this begs the question of how the classical spin liquid found in spin ice 
might evolve into a quantum spin liquid as quantum effects become more important ?


\begin{figure}
\centering
\includegraphics[width=0.5\textwidth]{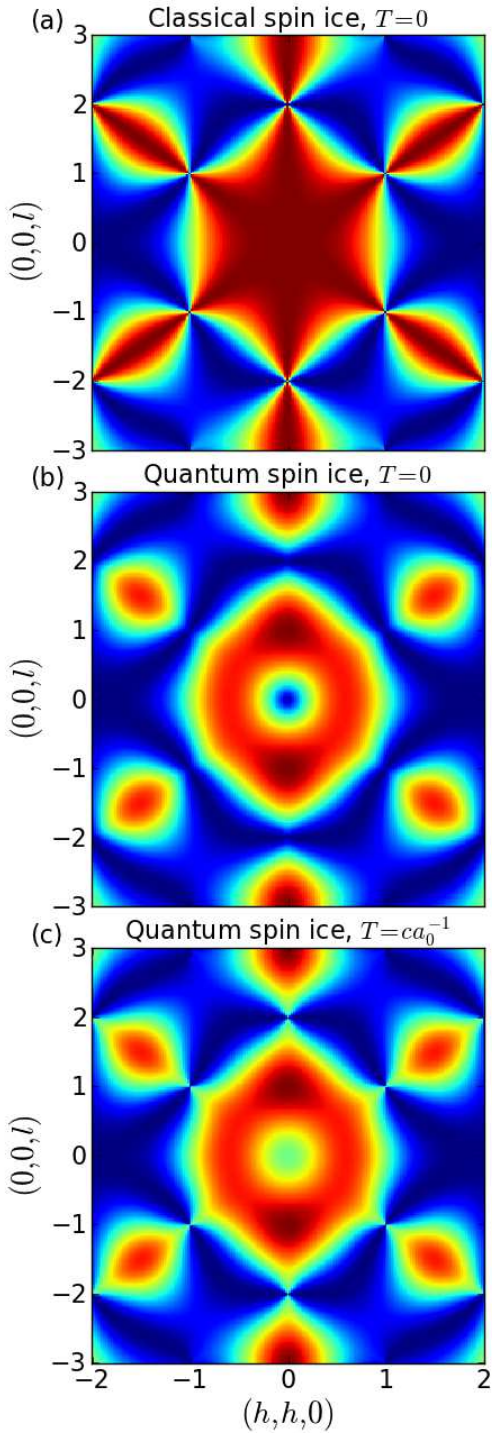}
\caption{
(Color online). 
Spin correlations in a spin ice, as measured by quasi-elastic neutron scattering : 
(a) Correlations within the classical spin ice configurations,  
showing the characteristic ``pinch point'' singularities.
(b) Correlations in a quantum ice at $T=0$, showing the suppression of pinch points
by quantum fluctuations.
(c) Correlations in a quantum ice at an intermediate temperature $T=ca_0^{-1}$, 
showing how pinch points are progressively restored by the thermal excitation of 
magnetic photons.
In all cases, results are shown for equal-time structure factors in the 
$(h,h,0)$ plane, for a polarised neutron scattering experiment in the spin-flip channel 
considered by Fennell~{\it et al}~\protect\mbox{\cite{fennell09}}.
Temperature is measured in units where $c$ is the speed of light associated 
with magnetic ``photon'' excitations, $a_0$ the lattice constant, and $\hbar = k_B =1$.
}
\label{fig:pinchpoint}
\end{figure}


In fact spin ice is just one example of a much broader class of systems which obey the 
``ice rules''.
First introduced by Bernal and Fowler in 1933 to describe the 
correlations of protons in water 
ice~\cite{bernal33}, the ice rules have since found application in models of frustrated charge 
order~\cite{anderson58,fulde02}, proton bonded ferroelectrics~\cite{youngblood80} and 
dense polymer melts~\cite{kondev98}.
All of these systems possess a local ``two-in, two-out'' 
constraint, which can most conveniently be written 
in terms of a zero-divergence condition on a notional magnetic field 
\begin{eqnarray}
\label{eq:ice-rules}
\nabla \cdot {\bf B} = 0 \, .
\end{eqnarray}


In the case of spin ice, ${\bf B}$ has the physical meaning of the local magnetisation 
of the system, and we can associate a field ${\bf B}_i$ with each spin
on the lattice.
For this reason, spin ice offers a beautiful realisation of classical magnetostatics, 
with local violations of the ice rules entering as point magnetic charges 
(magnetic monopoles~\cite{ryzhkin05, castelnovo08, bramwell09, morris09,
 kadowaki09, jaubert09, giblin11}) 
and spin correlations which exhibit ``pinch point'' singularities  in k-space
\begin{eqnarray}
 \langle S_{\mu} (-\mathbf{k})   S_{\nu} (\mathbf{k}) \rangle_{\sf classical}  
    \propto 
    \left( \delta_{\mu \nu} - \frac{k_{\mu} k_{\nu}}{k^2} \right)  \, ,
\end{eqnarray}
[Fig.~\ref{fig:pinchpoint}(a)] corresponding to algebraic (dipolar) correlations 
in real space~\cite{huse03,henley05,fennell09,henley10,youngblood80}.
Since the ice rules can be satisfied by an exponentially large number of proton 
(spin, charge, polymer\ldots) configurations~\cite{pauling35}, they explain 
the residual entropy observed in both water ice~\cite{giauque36} and 
spin ice~\cite{ramirez99} at low temperatures.
Given this enormous reservoir of entropy, both spin ice and water ice 
are natural places to look for a quantum liquid ground state.

\begin{figure}
\centering
\includegraphics[width=0.5\textwidth]{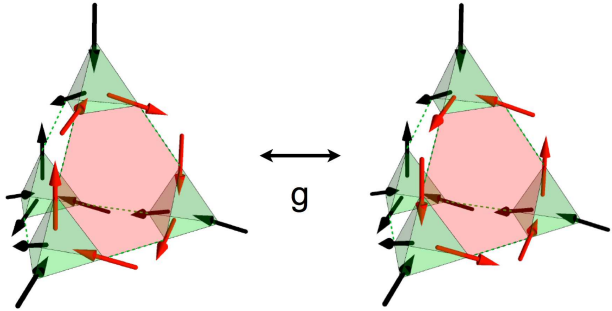}
\caption{
(Color online). 
An illustration of the simplest tunnelling process between different 
spin-ice configurations.
The ice rules dictate that each tetrahedron within the lattice has two
spins which point ``in'', and two which point ``out''.  
Where these spins form a closed loop on a hexagonal plaquette
--- here shaded red --- the sense of each spin within the loop can be reversed 
to give a new configuration which also obeys the ice rules.}
\label{fig:tunnelling}
\end{figure}

The key ingredient needed to convert a classical ice into a quantum 
liquid is tunnelling between different ice configurations [Fig.~\ref{fig:tunnelling}].
This opens the door to a ``quantum ice''~: a unique, quantum mechanical ground state, 
formed through the coherent superposition of an exponentially large number of classical ice 
configurations.   
Such a state could have a vanishing entropy at zero temperature, and so satisfy the 
third law of thermodynamics, without sacrificing the algebraic correlations and fractional 
excitations (magnetic monopoles) associated with the degeneracy of the ice states.
If realised in a spin ice, it would provide a concrete, three--dimensional example of the
long-sought quantum spin liquid.   


Precisely this scenario was proposed by Moessner and Sondhi in the context of 
three-dimensional quantum dimer models~\cite{moessner03}, by Hermele, Balents 
and Fisher in a quantum, ice-type model derived from an easy-axis antiferromagnet on 
a pyrochlore lattice~\cite{hermele04}, and by Castro-Neto, Pujol and 
Fradkin in the context of a simplified model of water ice~\cite{castro-neto06}.
All of these models included tunnelling between ice (or dimer) configurations of the type 
illustrated in Fig.~\ref{fig:tunnelling}.
In a spin ice, the dominant tunnelling process involves flipping loops 
of spins which point nose-to-tail on an hexagonal plaquette, and 
the resulting dynamics are described symbollically by
\begin{eqnarray}
\label{eq:Htunnelling-symbolic}
\mathcal{H}_{\sf tunnelling} =  
   -g \sum_{\hexagon} 
   \big[
   |\! \circlearrowright \rangle\langle \circlearrowleft\! | + 
   |\! \circlearrowleft \rangle\langle \circlearrowright\! | 
   \big]
\end{eqnarray}
where $g$ is the strength of the tunnelling matrix element, 
and $\mathcal{H}_{\sf tunnelling}$ acts on the space of all possible 
ice (or dimer) configurations.


Both Moessner and Sondhi~\cite{moessner03} and Hermele {\it et al.}~\cite{hermele04}
also introduced an additional control parameter $\mu$ to the Hamiltonian 
\begin{eqnarray}
\label{eq:Hmu-symbolic}
\mathcal{H}_{\mu} &=& \mathcal{H}_{\sf tunneling} + \delta\mathcal{H}_{\mu} \, ,
\end{eqnarray}
where
\begin{eqnarray}
\label{eq:deltaHmu}
\delta\mathcal{H}_{\mu} &=&  
   \mu  \sum_{\hexagon} 
   \big[
   |\! \circlearrowleft \rangle\langle \circlearrowleft\! | + 
   |\! \circlearrowright \rangle\langle \circlearrowright\! | 
   \big] \, .
\end{eqnarray}
This makes it possible to fine-tune the model to an exactly soluble 
Rokhsar-Kivelson (RK) point $g=\mu$, where the ground state wave 
is an equally-weighted sum of {\it all} 
possible ice (dimer) configurations~\cite{rokhsar88}. 
The authors then argued, by continuity, that a quantum liquid phase would 
occur for a finite range of parameters $\mu \lesssim 1$ bordering on the RK 
point~\cite{moessner03,hermele04}.


The most striking feature of this quantum liquid is ``light''.
Attempts to construct models with ``artificial light'' --- gapless photon excitations 
of an effective, low-energy $U(1)$ gauge field --- have a long history~\cite{foerster80}.
In recent years, it has been realised that large families of lattice models could, in principle, 
be described by such theories.
These include abstract models of ``quantum order''~\cite{wen02-PRL,wen02-PRB},
Bose-Hubbard models bordering on superfluidity~\cite{motrunich02}, 
systems of screened dipoles~\cite{wen03}, and suitably adapted sigma models~\cite{motrunich04}.
Reviews of these ideas can be found in papers by Montrunich and Senthil~\cite{motrunich05}
and Wen and Levin~\cite{levin05}.


\begin{figure}[h!]
\includegraphics[width=0.35\textwidth]{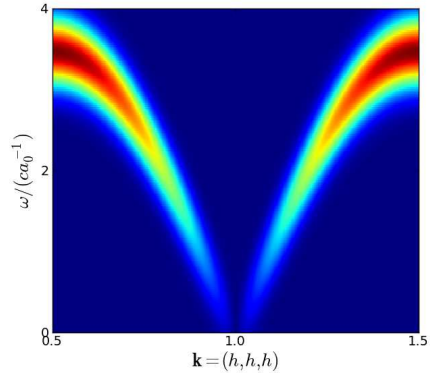}
\caption{
(Color online).
Ghostly magnetic ``photon'' excitation as it might appear in an inelastic neutron scattering
experiment on a quantum spin ice realising a quantum ice ground state.
The photon dispersion $\omega(\mathbf{k})$ is taken from lattice gauge theory 
developed in Section~\ref{subsection:photons} of this paper, convoluted 
with a Gaussian representing the finite energy resolution of the instrument. 
The intensity of scattering vanishes as $I \propto \omega(\mathbf{k})$  
at low energies.}
\label{fig:ghostlyphoton}
\end{figure}


The way in which ``light'' arises in three-dimensional quantum 
ice and quantum dimer models is particularly simple.
%
The ice-rules constraint Eq.~(\ref{eq:ice-rules}) is most conveniently 
resolved as 
\begin{eqnarray}
\mathbf{\mathcal B}({\bf r}) = \nabla \times \mathbf{\mathcal A}({\bf r})  \, .
\end{eqnarray}
The new feature which enters where there is tunnelling between ice configurations 
is the fluctuation in time of the gauge field~$\mathbf{\mathcal A}({\bf r})$.
In conventional electromagnetism, this gives rise to an electric field 
\begin{eqnarray}
\mathbf{\mathcal E}({\bf r}) = -\frac{\partial \mathbf{\mathcal A}({\bf r}) }{\partial t} \, .
\end{eqnarray}
Then, following the heuristic arguments of Moessner and Sondhi~\cite{moessner03}
--- or the more microscopic derivation of Hermele {\it et al.}~\cite{hermele04} ---
it is reasonable to suppose that a quantum liquid found bordering the 
RK point ($\mathcal{H}_{\mu}$~[Eq.~(\ref{eq:Hmu-symbolic})] with $\mu \lesssim 1$), 
would be governed by the Maxwell action
\begin{eqnarray}
\mathcal{S}_{\sf Maxwell}
  = \frac{1}{8\pi}\int dt d^3 \mathbf{r} 
   \bigg[ \mathbf{\mathcal E}({\bf r})^2 - c^2 \mathbf{\mathcal B}({\bf r})^2  \bigg]
   \label{eq:SMaxwell}
\end{eqnarray}
Any state described by $\mathcal{S}_{\sf Maxwell}$ [Eq.~(\ref{eq:SMaxwell})]
automatically supports linearly-dispersing transverse excitations of the gauge 
field $\mathbf{\mathcal A}({\bf r})$ ---  ``photons'', with a speed of ``light'' $c$.
On the lattice, such a magnetic photon would have a dispersion $\omega(\bf k)$ 
of the form illustrated in Fig.~\ref{fig:ghostlyphoton}.


Moreover, the fact that the spins now fluctuate in time, as well 
as space, introduces an additional power of k in energy-integrated (i.e. equal time)
spin correlations~\cite{hermele04,castro-neto06}, 
\begin{eqnarray}
 \langle S_{\mu} (-\mathbf{k})  S_{\nu} (\mathbf{k}) \rangle_{\sf quantum}  
 \propto 
 k
 \left( \delta_{\mu \nu} - \frac{k_{\mu} k_{\nu}}{k^2} \right) ,
\end{eqnarray}
which serves to eliminate the pinch points seen in quasi-elastic neutron scattering
experiments [Fig.~\ref{fig:pinchpoint}(b)]~\cite{shannon12} .
More formally, this theory is a compact, frustrated $U(1)$ gauge theory on a 
diamond lattice, and we will refer to the liquid state it describes as the quantum 
$U(1)$ liquid below. 

\begin{figure}
\centering
\includegraphics[width=0.4\textwidth]{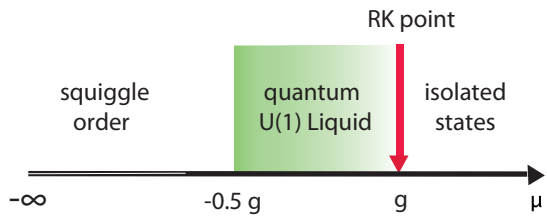}
\caption{
(Color online).
Zero-temperature phase diagram of the model of tunnelling between ice states 
${\mathcal H}_\mu$ [Eq.~\ref{eq:Hmu}], as determined by quantum Monte Carlo 
simulation in \onlinecite{shannon12}.   
The ``quantum ice'' point, $\mu=0$, lies deep within 
a quantum liquid phase with low-energy excitations described by 
a lattice analogue of quantum electromagnetism 
This extends from a ``squiggle'' ordered phase, found for $\mu < -0.5 g$, 
to the exactly-soluble RK point $\mu = g$.  
(Here $g$ is the strength of tunnelling between ice states).  
}
\label{fig:phase-diagram}
\end{figure}

The degree of fine-tuning in these arguments, and 
the need to introduce additional parameter $\mu$ [Eq.~(\ref{eq:deltaHmu})], 
might seem to render them of purely academic interest.
However the idea of a quantum $U(1)$ liquid found strong support in finite-temperature 
quantum Monte Carlo simulations of an ice-type model of frustrated charge order on the 
pyrochlore lattice~\cite{banerjee08}.
Subsequently, it has proved possible to determine the ground state phase diagrams of 
both the quantum dimer model of Moessner and Sondhi~\cite{moessner03}, 
and the quantum ice model of Hermele~{\it et al.}~\cite{hermele04}, 
from zero-temperature quantum Monte Carlo simulations~\cite{sikora09,sikora11,shannon12}.  
Both models contains extended regions of a quantum liquid phase, 
connecting to the RK point.
In both cases, this quantum liquid has low energy excitations which are described by a 
lattice analogue of quantum electromagnetism~\cite{sikora09,sikora11,shannon12}.
Significantly, in the case of the quantum ice model, this quantum liquid phase 
encompasses the ``physical'' point of the model $\mu=0$, and so does not 
require {\it any} fine-tuning [Fig.~\ref{fig:phase-diagram}]~\cite{shannon12}. 


The theoretical possibility of a three-dimensional spin-liquid state with excitations 
described by a lattice analogue of quantum electromagnetism is now well-established.  
What remains is to connect these ideas with experiments.
The purpose of this paper is therefore to set out predictions for the correlations
which would be measured in neutron scattering experiments, {\it if} such a state
were realised in a spin-ice material.
For concreteness, we work with the minimal lattice model introduced by 
Hermele {\it et al.}~\cite{hermele04}, transcribed to coordinates appropriate 
for a spin ice.
More realistic generalisations of this model will be considered elsewhere~\cite{olga-unpub}.


In Section~\ref{section:ice-to-em} of the paper, we develop the mathematical formalism 
needed to describe the spin correlations and low-energy spin excitations in a 
spin ice with a quantum $U(1)$-liquid ground state.
Using this theory, we make predictions for the photon dispersion $\omega(\mathbf{k})$ 
and dynamical structure factor $S^{\alpha\beta}(\mathbf{k}, \omega)$, which would be 
measured in neutron scattering experiments.


In Section~\ref{section:zero-T}, we make explicit comparison of the predictions of this 
theory with zero-temperature Quantum Monte Carlo 
simulations of the minimal, microscopic model of a quantum spin ice with tunnelling
between different ice configurations, ${\mathcal H}_\mu$ [Eq.~(\ref{eq:Hmu})].
We find essentially perfect, {\it quantitative} agreement between simulation 
results and the field theory solved on a finite-size lattice, for a range of parameters 
$0 \leq \mu \leq g$ which interpolate from the minimal model of a 
quantum spin ice ($\mu = 0$), to the classical correlations of the RK point ($\mu = g$). 
This analysis reinforces the conclusions reached in [\onlinecite{shannon12}]
about the existence of a quantum $U(1)$-liquid in this model, 
and puts the field-theory description on a quantitative footing.


In Section~\ref{section:finite-T} we make predictions for neutron scattering
experiments carried out at finite temperature.
In particular we analyse the way in which the characteristic ``pinch point'' structure in 
quasi-elastic scattering is lost as the system is cooled towards its zero-temperature ground state.
We conclude that the loss of the pinch points coincides with the progressive 
loss of the Pauling entropy as the system cools into a unique, quantum coherent, 
liquid ground state. 
Thus the signature features of the ice problem: pinch points and the Pauling entropy die 
together at low temperatures.
We also give a brief discussion of the uniform magnetic susceptibility and heat capacity, in the
low temperature quantum regime.


Finally, in Section~\ref{section:conclusions} we conclude with a discussion of some of 
the remaining issues relating to experiment.
As far as possible, each section of the paper is written so as to be 
self-contained.  
Readers uninterested in the mathematical development of the theory are therefore 
invited pass directly to Section~\ref{section:zero-T} and Section~\ref{section:finite-T}, 
referring to Section~\ref{section:ice-to-em} as required.


\section{From quantum ice to quantum electromagnetism}
\label{section:ice-to-em}


At first sight, an assembly of magnetic ions on a lattice does not look like a promising
place to search for a gauge theory which perfectly mimics quantum electromagnetism.
However in the simplest microscopic model for quantum mechanical tunnelling between 
spin configurations obeying the ``two in, two out'' ice rule, this is exactly what 
happens~\cite{hermele04,banerjee08,shannon12}.  
In what follows we retrace the steps which lead from a spin ice system to 
a theory of electromagnetism on a lattice.  


In Section~\ref{subsection:spins} we review the relevant microscopic models.
In Section~\ref{subsection:em}, we show how a lattice gauge theory resembling 
electromagnetism arises in these problems, recasting the earlier field-theoretical 
arguments of Hermele {\it et al.}~\cite{hermele04} in terms appropriate for a spin ice.  
In Section~\ref{subsection:photons} we explicitly construct the magnetic ``photon'' 
excitations of this lattice gauge theory.
In Section~\ref{subsection:Skomega-B} we use the mapping between spins and 
photons to calculate the correlations between spins in a quantum spin 
liquid described by this lattice gauge theory. 
Throughout this analysis we set $\hbar=k_B = 1$, restoring dimensional 
factors only where we quote a result for the speed of light.


\subsection{Spins on a pyrochlore lattice}
\label{subsection:spins}


\begin{figure*}
\centering
\includegraphics[width=0.8\textwidth]{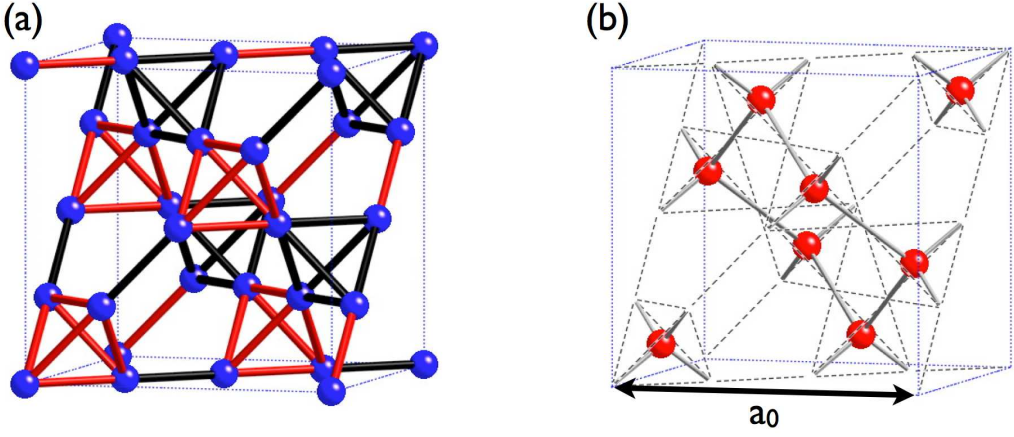}
\caption{
(Color online).
Structure of the pyrochlore lattice realised by the magnetic ions in spin-ice materials.
a) The lattice is built of corner sharing tetrahedra, and can be decomposed into 
a set of A-sublattice tetrahedra (here coloured red) and B--sublattice tetrahedra 
(here coloured black), each of which forms an FCC lattice in its own right.
The primitive unit of the pyrchlore lattice consists of a single tetrahedron with 
4 lattice sites.  
However it is also possible to define a cubic unit cell, of side $a_0$, containing
16 lattice sites.
b) Bipartite, diamond lattice formed by the centres of tetrahedra which make 
up the pyrochlore lattice. 
The bonds of this diamond lattice define the easy axes for spins in a spin ice, 
and play an important role in the lattice gauge theory of its excitations.
}
\label{fig:lattice}
\end{figure*}


The materials which we will seek to describe have magnetic ions which 
a) have a crystal-field ground state which is a doublet, and 
b) occupy the sites of the pyrochlore lattice shown in Fig.~\ref{fig:lattice}.   
In the case of the spin ices Ho$_2$Ti$_2$O$_7$ and 
Dy$_2$Ti$_2$O$_7$, this doublet has Ising character (rare-earth moments 
point into, or out of, the tetrahedra which make up the lattice), and the 
dominant interactions between these Ising spins are dipolar~\cite{siddharthan99}.
However, since these dipolar interactions are effectively 
self-screened, the correlations present in spin-ice are extremely well described 
by models with only nearest-neighbour interactions between 
spins~\cite{harris97,denHertog00,gingras01,melko04,isakov05}.  
This approximation gains further justification in ``quantum spin ice'' materials
such as Yb$_2$Ti$_2$O$_7$, where magnetic moments are smaller than for 
Ho$_2$Ti$_2$O$_7$ and Dy$_2$Ti$_2$O$_7$, and exchange interactions 
play a much larger role.  
%


As a starting point, we can therefore consider the Hamiltonian for 
a (pseudo) spin-1/2 degree of freedom on a pyrochlore lattice, with the most general 
nearest-neighbour exchange interactions allowed by 
symmetry~\cite{curnoe08} 
\begin{eqnarray}
  \mathcal{H}_{\sf S=1/2} 
 & = & \sum_{\langle ij\rangle} \Big\{ J_{zz} \mathsf{S}_i^z \mathsf{S}_j^z - J_{\pm}
  (\mathsf{S}_i^+ \mathsf{S}_j^- + \mathsf{S}_i^- \mathsf{S}_j^+) 
\nonumber \\ && 
 + J_{\pm\pm} \left[\gamma_{ij} \mathsf{S}_i^+ \mathsf{S}_j^+ + \gamma_{ij}^*
    \mathsf{S}_i^-\mathsf{S}_j^-\right] 
\nonumber \\ && 
+ J_{z\pm}\left[ \mathsf{S}_i^z (\zeta_{ij} \mathsf{S}_j^+ + \zeta^*_{ij} \mathsf{S}_j^-) +
  {i\leftrightarrow j}\right]\Big\}
  \label{eq:Hross}
\end{eqnarray}
Here we have followed the notation of Ross {\it et al.}~\cite{ross11-PRX}, in which 
the $\mathsf{S}_i^z$ is aligned with the local trigonal 
axes of the pyrochlore lattice on each site~$i$, and $\gamma_{ij}$ and 
$\zeta_{ij} $ are $4 \times 4$ complex unimodular matrices encoding the 
rotations between these local coordinate frames.
In the ``quantum spin ice'' Yb$_2$Ti$_2$O$_7$, where the ground state doublet of
Yb has XY-character~\cite{hodges01,cao09,onoda11}, Eq.~(\ref{eq:Hross}) gives a good account of diffuse 
structure observed in neutron scattering experiments {\it provided} that the exchange 
interactions $J_{\pm}$, $J_{z\pm}$ and $J_{\pm\pm}$  are taken into 
account~\cite{thompson11,chang-arXiv}.  
It also gives an excellent description of spin wave excitations about the saturated state 
of Yb$_2$Ti$_2$O$_7$ in applied magnetic field, with parameters 
$J_{zz} = 0.17 \pm 0.04$~meV, 
$J_{\pm} = 0.05 \pm 0.01$~meV, 
$J_{z\pm} =- 0.14 \pm 0.01$~meV and
$J_{\pm\pm} = 0.05 \pm 0.01$~meV 
obtained from fits to data~\cite{ross11-PRX}.
The phase diagram associated with $\mathcal{H}_{\sf S=1/2}$ [Eq.~(\ref{eq:Hross})] is 
explored in [\onlinecite{ross11-PRX,savary12,lee-arXiv}].  


We can further simplify the problem by setting $J_{\pm\pm}=0$, and focusing on the limit 
$J_{zz} >> J_{\pm}, J_{z\pm} > 0$.  
In this limit, the role of $J_{zz}$ is to enforce  the ``ice rules'' constraint,
while $J_{\pm}$ generates dynamics, and $J_{z\pm}$ 
lifts the degeneracy of ice-rule obeying states.
Performing degenerate perturbation theory in the basis of (spin) ice 
configurations, and dropping terms which lead only to a constant energy shift, 
leads to the effective Hamiltonian~\cite{ross11-PRX} 
\begin{eqnarray}
\label{eq:Heff}
\mathcal{H}_{\sf eff} =  \mathcal{H}_{\sf tunnelling} + \mathcal{H}_{\sf J_3} 
\end{eqnarray}
with
\begin{eqnarray}
\label{eq:Htunnelling}
\mathcal{H}_{\sf tunnelling} 
   = - g \sum_{\hexagon} 
        \left[ 
        \mathsf{S}^{+}_{1} \mathsf{S}^{-}_{2}\mathsf{S}^{+}_{3} 
        \mathsf{S}^{-}_{4}\mathsf{S}^{+}_{5} \mathsf{S}^{-}_{6} 
        + h.c. \right]
\end{eqnarray}
where $\sum_{\hexagon} $ runs over all hexagonal plaquettes in the pyorchlore 
lattice [cf. Fig~\ref{fig:tunnelling}] with
\begin{eqnarray}
g = \frac{12 J_{\pm}^3}{J_{zz}^2}  \, ,
\label{eq:g}
\end{eqnarray}
and
\begin{eqnarray}
\label{eq:HJ3}
\mathcal{H}_{\sf J_3}  
   = - J_3 \sum_{\langle ij \rangle_3} \mathsf{S}_i^z \mathsf{S}_j^z
\end{eqnarray}
where $\sum_{\langle ij \rangle_3}$ runs over third-neighbours bonds 
(parallel to the nearest-neighbour bonds), with 
\begin{eqnarray}
J_3 = \frac{3 J_{z\pm}^2}{J_{zz}} > 0
\end{eqnarray}


We note that, by construction, the Hamiltonian Eq.~(\ref{eq:Heff}) acts {\it only} on 
spin configurations satisfying the ice rules.
This implies that, in performing the degenerate perturbation theory, virtual excitations of
magnetic monopoles have been projected out of the problem.
This approximation will have little effect on the conclusions drawn in this paper, 
and could in principle be relaxed.  


It is also important to note that these spin ice configurations may possess 
a non-zero net magnetisation ${\bf M}$.
The tunnelling term $\mathcal{H}_{\sf tunnelling}$ [Eq.~(\ref{eq:Htunnelling})] generates 
dynamics by performing a cyclic exchange of spins on a hexagonal 
plaquette [cf. Fig.~\ref{fig:tunnelling}].  
This tunnelling process can be written symbolically as acting on a closed loop
of spins  [cf. Eq.~(\ref{eq:Htunnelling-symbolic})].  
Under these dynamics, the total magnetisation ${\bf M}$ is a conserved quantity.
%


We make the final simplification of neglecting $\mathcal{H}_{\sf J_3}$ [Eq.~(\ref{eq:HJ3})] 
and focusing exclusively on the spin-liquid favoured by the tunnelling term 
$\mathcal{H}_{\sf tunnelling}$ --- Eq.~(\ref{eq:Htunnelling}) 
or, symbolically,  Eq.~(\ref{eq:Htunnelling-symbolic}).
The neglected term $\mathcal{H}_{\sf J_3}$ favours the six ice states with 
the maximum possible magnetisation per site ${\bf m} = (\pm 1/\sqrt{3},0,0) \times S$, etc., 
where $S$ is the moment of the magnetic ion.
We have confirmed through zero-temperature quantum Monte Carlo simulation of 
$\mathcal{H}_{\sf eff}$ [Eq.~(\ref{eq:Heff})], that the system remains in quantum $U(1)$ 
liquid ground state up to a value of $J_3 \approx 0.27\, g$, at which point it 
undergoes first-order transition into this ordered, ferromagnetic state.
These results will be discussed elsewhere~\cite{olga-unpub}.
We note that a gauge mean-field theory for 
$\mathcal{H}_{\sf S=1/2} $ [Eq.~(\ref{eq:Hross})] predicts an intermediate
 ``Coulombic ferromagnet'' phase, in which the quantum $U(1)$ liquid 
spontaneously acquires a finite magnetisation 
for any finite $J_{z\pm}$~\cite{ross11-PRX,savary12}.
This does not appear to be a ground state of the 
effective model $\mathcal{H}_{\sf eff}$ [Eq.~(\ref{eq:Heff})].


%

Following Hermele~{\it et al.}~\cite{hermele04}, it is useful to augment the minimal 
model $\mathcal{H}_{\sf tunnelling}$
with an additional, artificial, interaction term $\delta{\mathcal H}_\mu$ [Eq.~(\ref{eq:deltaHmu})].  
This renders the model exactly soluble for $\mu = g$.  
Thus the most general microscopic model we consider in this paper 
can be written symbolically as 
\begin{eqnarray}
\label{eq:Hmu}
 \mathcal{H}_{\mu}
   &=&-g \sum_{\hexagon} 
   \big[
   |\! \circlearrowright \rangle\langle \circlearrowleft\! | + 
   |\! \circlearrowleft \rangle\langle \circlearrowright\! | 
   \big]
   \nonumber \\ && \quad 
+\mu  \sum_{\hexagon} 
   \big[
   |\! \circlearrowleft \rangle\langle \circlearrowleft\! | + 
   |\! \circlearrowright \rangle\langle \circlearrowright\! | 
   \big].
\end{eqnarray}
where $\mathcal{H}_{\mu}$ acts on the space of all possible (spin) ice configurations.  
This Hamiltonian is known to support a quantum $U(1)$ liquid ground state for 
$-0.5g < \mu \le g$~\cite{shannon12}.   


It is important to note that this effective description of tunnelling between ice configurations 
might equally have been derived for the model of hardcore bosons on the pyrochlore lattice
considered by Banerjee {\it et al.}~\cite{banerjee08}
\begin{eqnarray}
\mathcal{H}_{\sf charge-ice}
 &=& - t \sum_{\langle ij \rangle}  
     \left( b_i^{\dagger} b^{\phantom\dagger}_j 
    + b_j^{\dagger} b^{\phantom\dagger}_i \right) 
    \nonumber \\
    && \quad 
+ V \sum_{\langle ij \rangle} \left( n_i - \frac{1}{2} \right) \left( n_j - \frac{1}{2} \right)
\label{eq:HtV}
\end{eqnarray}
with $V \gg t$.    
At $1/2$-filling [$\langle n \rangle \equiv 1/2$],  
$V$ selects charge configurations with 
exactly two bosons in each tetrahedron of the lattice, and $\mathcal{H}_{\sf charge-ice}$ 
is exactly equivalent to the \mbox{pseudospin-$1/2$} model Eq.~(\ref{eq:Heff}), in the case 
where \mbox{$J_{z\pm} = J_{\pm\pm} \equiv 0$}.
The leading tunnelling matrix element between different (charge) ice configurations
is then 
\begin{eqnarray}
g = \frac{12 t^3}{V^2}  
\end{eqnarray}
We will return to this model below in the context of predictions for experiment and 
simulations performed at finite temperature~\cite{banerjee08}.  


The manifold of ice configurations on the pyrochlore lattice is equivalent to the 
set of possible close-packed loop coverings of the diamond lattice~\cite{jaubert11}.  
Exactly parallel arguments, leading to a formally identical Hamiltonian, 
can also be constructed for the closely related quantum dimer model on the 
diamond lattice~\cite{moessner03,bergman06}.
This model also exhibits a quantum $U(1)$ liquid ground states for a smaller --- but none the less 
finite --- range of parameters $0.75 g < \mu \le g$~[\onlinecite{sikora09,sikora11}].  


\subsection{Electromagnetism on a diamond lattice}
\label{subsection:em}

\begin{figure}[h!]
\centering
\includegraphics[width=0.5\textwidth]{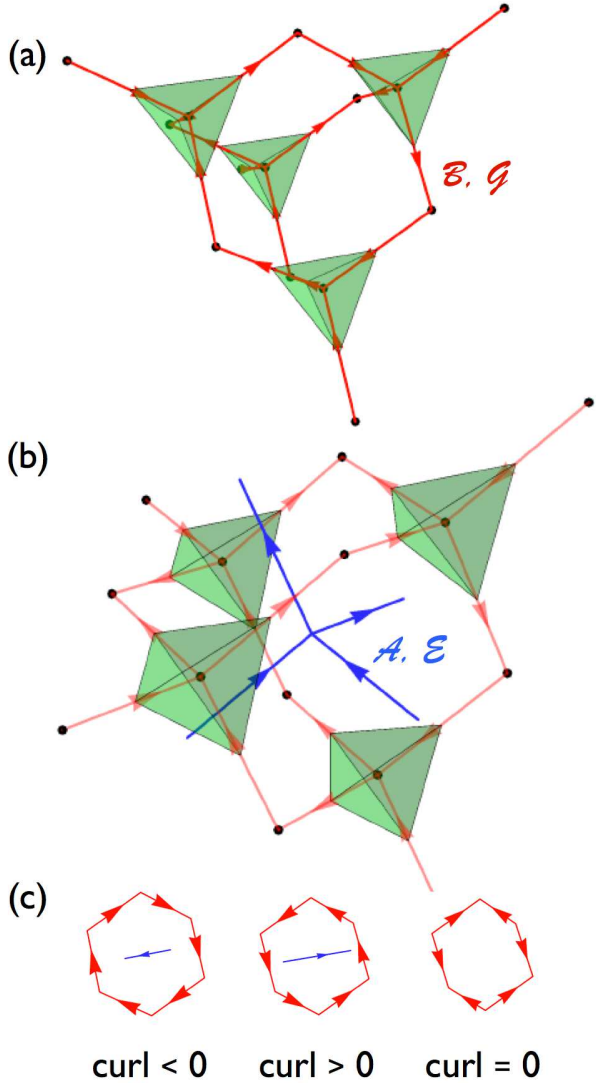}
\caption{
(Color online).  
The different fields used in Section~\ref{subsection:em} to construct lattice gauge theory 
of the spin liquid state, and the different lattices on which they are defined.  
(a) The ``magnetic'' field $\mathcal{B}_{\mathbf{r}\mathbf{r}'}$ [Eq.~(\ref{eq:B})], and its conjugate 
field $\mathcal{G}_{\mathbf{r}\mathbf{r}'}$ [Eq.~(\ref{eq:G})] are defined on the links 
of the diamond lattice, shown here in red.
Each link of this diamond lattice corresponds to a site of the original pyrochlore lattice, 
and $\mathcal{B}_{\mathbf{r}\mathbf{r}'}$ encodes the orientation of the spin on this site.  
(b) The compact $U(1)$ gauge field $\mathcal{A}_{\mathbf{s}\mathbf{s}'}$ and the conjugate 
``electric'' field $\mathcal{E}_{\mathbf{s}\mathbf{s}'}$ are defined on the links of a second, 
dual, diamond lattice, shown here in blue. 
The midpoints of these bonds also form a second, dual, pyrochlore lattice, corresponding to
the centres of hexagonal plaquettes in the original pyrochlore lattice.  
(c) An illustration of taking the lattice curl on the hexagonal plaquettes 
of the diamond lattice. 
The resulting vector lives on the links of the dual, diamond lattice. }
\label{fig:latticeEM}
\end{figure}

The mappings described in Section~\ref{subsection:spins} permit us to reduce 
complicated interactions between magnetic ions to a problem of tunnelling between 
spin configurations obeying the ``ice rules'' [cf. Fig.~{\ref{fig:tunnelling}}].  
If we think of these spins as field lines of a fictitious magnetic field ${\bf B}$, 
these rules can conveniently be written as 
$$
\nabla \cdot {\bf B} = 0 
$$
This naturally suggests an analogy with magnetostatics, with magnetic 
field lines constrained to lie on the bonds of a diamond lattice [cf. Fig.~\ref{fig:latticeEM}(a)].   
And in the presence of tunnelling between ice configurations, this analogy 
can be extended to a fully dynamical quantum electromagnetism.
Here we review the mapping from an ice with tunnelling, 
to a compact, $U(1)$ lattice gauge theory, before moving on to an analysis of its 
``photon'' excitations [Section~\ref{subsection:photons}] and spin correlations 
[Section~\ref{subsection:Skomega}] .  
In so doing we follow closely the arguments of Hermele {\it et al}~\cite{hermele04}, 
but recast the discussion in terms of the magnetic fields ${\bf B}$ usually associated 
with the spins of a spin ice.


We begin by transcribing the spin variables of ${\mathcal H}_{\sf tunnelling}$ 
[Eq.~(\ref{eq:Htunnelling})] in terms of a quantum rotor variable $\theta_i$, and 
its conjugate number operator $n_i$ 
\begin{eqnarray}
\label{eq:Sz}
{\mathsf S}^z_i &=& \left( n_i-\frac{1}{2} \right) \\
\label{eq:S+}
{\mathsf S}^{+}_i &=& \sqrt{n_i} \exp{[i \theta_i]} \sqrt{1-n_i} \\
\label{eq:S-}
{\mathsf S}^{-}_i &=& \sqrt{1-n_i} \exp{[-i \theta_i]} \sqrt{n_i}
\end{eqnarray}
where 
\begin{eqnarray}
\label{nphi0}
[\theta_i, n_j]=i \delta_{ij} .
\end{eqnarray}
The number operator $n_i$ could equally be associated with the density of 
(hard-core) bosons in a charge ice, and in order to remain in the physical subspace 
where $n_i = 0$ or $1$, we add the term 
\begin{eqnarray}
\label{hardcoreH}
\mathcal{H}_{\sf U}=\frac{U}{2} \sum_i (n_i  - 1/2)^2
\end{eqnarray}
to the Hamiltonian, subsequently taking the limit $U \to \infty$.   
With this restriction in place, the Hamiltonian becomes
\begin{eqnarray}
\label{eq:Hrotor}
\mathcal{H}_{\sf rotor} 
   &=& \frac{U}{2} \sum_i (n_i  - 1/2)^2  \nonumber \\
   &-& 2g \sum_{\hexagon}  
       \cos{\left( \theta_1 - \theta_2 + \theta_3 
                     - \theta_4 + \theta_5 - \theta_6 \right)} \nonumber \\
\end{eqnarray}
It is from this rotor form of the Hamiltonian that we will make the passage to 
a $U(1)$ gauge theory on the diamond lattice.


The site $i$ of the pyrochlore lattice can be thought of as the midpoint of the bond 
$\mathbf{r} \to \mathbf{r}'$ of a diamond lattice [cf. Fig.~\ref{fig:lattice}]. 
Since this diamond lattice is bipartite, it is possible to define directed variables 
on these bonds
\begin{eqnarray}
\mathcal{B}_{\mathbf{r} \mathbf{r}'}
    = - \mathcal{B}_{\mathbf{r}' \mathbf{r}} \quad \mathcal{G}_{\mathbf{r} \mathbf{r}'}
    = - \mathcal{G}_{\mathbf{r}' \mathbf{r}}
\end{eqnarray}
through the mapping
\begin{eqnarray}
\label{eq:B}
\mathcal{B}_{\mathbf{r} \mathbf{r}'} &=& \pm \left( \hat{n}_i-\frac{1}{2} \right) \\
\label{eq:G}
\mathcal{G}_{\mathbf{r} \mathbf{r}'} &=& \pm \theta_i
\end{eqnarray}
where the sign is taken to be positive if $\mathbf{r}$ belongs to the 
$A$-sublattice, and negative if $\mathbf{r}$ belongs to the $B$-sublattice.   
Taking this convention into account, we are left with a pair
of canonically conjugate variables 
\begin{eqnarray}
\label{BGcanon}
\left[\mathcal{G}_{\mathbf{r} \mathbf{r}'}, \mathcal{B}_{\mathbf{r}'' \mathbf{r}'''}\right] 
   = i \left(\delta_{\mathbf{r} \mathbf{r}''} \delta_{\mathbf{r}' \mathbf{r}'''} 
    -\delta_{\mathbf{r} \mathbf{r}'''} \delta_{\mathbf{r}' \mathbf{r}''}  \right).
\end{eqnarray}


The field, $\mathcal{B}_{\mathbf{r} \mathbf{r}'}$
will take on the role of a magnetic field in our lattice field theory.  
However in order to recreate ``electromagnetism'' we need also to discover an 
analogue to the electric field.  
The missing field, $\mathcal{E}_{\mathbf{s} \mathbf{s}'}$, inhabits
the bonds $\mathbf{s} \to \mathbf{s}'$ of a second diamond lattice, 
interpenetrating the first [Fig.~\ref{fig:latticeEM}(b)].
It is defined through a lattice curl 
\begin{eqnarray}
\mathcal{E}_{\mathbf{s} \mathbf{s}'}
= (\nabla_{\scriptsize\hexagon} \times \mathcal{G})_{\mathbf{s} \mathbf{s}'} 
= \sum_{\circlearrowleft} \mathcal{G}_{\mathbf{r} \mathbf{r}'}
\label{eq:Ecurl}
\end{eqnarray}
where the sum $\sum_{\circlearrowleft}$ is taken with {\it anticlockwise} sense
around the hexagonal plaquette of pyrochlore lattice sites encircling the bond 
$\mathbf{s} \to \mathbf{s}'$.   
It follows that $\mathcal{E}_{\mathbf{s} \mathbf{s}'}$ is also a directed variable 
\begin{eqnarray}
\mathcal{E}_{\mathbf{s} \mathbf{s}'}=-\mathcal{E}_{\mathbf{s}' \mathbf{s}} 
\end{eqnarray}
%


We are now in a position to transcribe $\mathcal{H}_{\sf rotor}$ completely  
in terms of ``electromagnetic'' fields
\begin{eqnarray}
\mathcal{H}_{\sf rotor} 
       &=& \frac{U}{2} \sum_{\langle \mathbf{r} \mathbf{r}' \rangle} 
                   \mathcal{B}_{\mathbf{r} \mathbf{r}'}^2 
          - 2g \sum_{\langle \mathbf{s} \mathbf{s}' \rangle}  
                \cos{(\mathcal{E}_{\mathbf{s} \mathbf{s}'})}
\label{eq:HcompactEM}
\end{eqnarray}
where the sum $\sum_{\langle \mathbf{r} \mathbf{r}' \rangle}$
runs over all bonds of the original diamond lattice, while the 
sum $\sum_{\langle \mathbf{s} \mathbf{s}' \rangle} $ runs over all 
bonds of the second, dual diamond lattice.  
The fact that the Hamiltonian is invariant under the transformation 
$\mathcal{E}_{\mathbf{s} \mathbf{s}'} \to \mathcal{E}_{\mathbf{s} \mathbf{s}'} + 2\pi$ 
makes it evident that this theory is compact.
It is also important to note that each of the components of 
the {\it total magnetic field} 
\begin{eqnarray}
  (\mathcal{B}_x, \mathcal{B}_y,\mathcal{B}_z)  
    &=& \sum_{\langle \mathbf{r} \mathbf{r}' \rangle} 
    \mathcal{B}_{\mathbf{r} \mathbf{r}'} 
    \hat{\mathbf{e}}_{\mathbf{r} \mathbf{r}'}
\end{eqnarray}
where $\hat{\mathbf{e}}_{\mathbf{r} \mathbf{r}'}$ is 
a unit vector directed from $\mathbf{r}$ to $\mathbf{r}'$, 
is a conserved quantity under the dynamics of 
$\mathcal{H}_{\sf rotor}$~[Eq.~(\ref{eq:HcompactEM})].  
More generally, reversing the sign of $\mathcal{B}_{\mathbf{r} \mathbf{r}'}$ 
on a closed loop of spins will tunnel one ice configuration to another, 
without changing the total magnetisation of the system.


In deriving Eq.~(\ref{eq:HcompactEM}), we have assumed that the ice rules hold, i.e.
\begin{eqnarray}
\label{eq:divB}
   (\nabla \cdot \mathcal{B})_\mathbf{r} 
     &=& \sum_{\langle \mathbf{r}' \rangle} \mathcal{B}_\mathbf{r r'} = 0
\end{eqnarray}
where the sum $\sum_{\langle \mathbf{r}' \rangle}$ runs over all sites
neighbouring $\mathbf{r}$.  
This condition is automatically satisfied if we write $\mathcal{B}_{\mathbf{r} \mathbf{r'}}$
as the lattice curl of a gauge field  $\mathcal{A}_{\mathbf{s} \mathbf{s'}}$.  
However we must also respect the requirement that the field $\mathcal{B}_{\mathbf{r} \mathbf{r}'}$ 
take on {\it half-integer} values [cf.  Eq.~(\ref{eq:B})].  
This can be accomplished by introducing a static background field $\mathcal{B}_{\mathbf{r} \mathbf{r'}}^0$, 
taken from {\it any} spin configuration which satisfies the ice rules, and writing 
\begin{eqnarray}
\left( \mathcal{B}_{\mathbf{r} \mathbf{r'}} - \mathcal{B}_{\mathbf{r} \mathbf{r'}}^0 \right)
   &=& \left( \nabla_{\scriptsize\hexagon} \times \mathcal{A} \right)_{\mathbf{r} \mathbf{r}'}
\label{eq:curlA}
\end{eqnarray}
to give
\begin{eqnarray}
\mathcal{H}_{\sf rotor} 
   &=& \frac{U}{2} \sum_{\langle \mathbf{r} \mathbf{r}' \rangle}
   \left[ \left(\nabla_{\scriptsize\hexagon} \times \mathcal{A}\right)_{\mathbf{r} \mathbf{r}'} 
    + \mathcal{B}_{\mathbf{r} \mathbf{r'}}^0 \right]^2  \nonumber\\
   && \quad - 2g \sum_{\langle \mathbf{s} \mathbf{s}' \rangle} 
    \cos{\left( \mathcal{E}_{\mathbf{s} \mathbf{s}'} \right)}
\label{eq:HemB0}
\end{eqnarray}
The fields $\mathcal{E}_{\mathbf{s} \mathbf{s}'}$ and $\mathcal{A}_{\mathbf{s} \mathbf{s}'}$ 
are canonically conjugate 
\begin{eqnarray}
\label{eq:EAcanon}
\left[ \mathcal{E}_{\mathbf{s} \mathbf{s}'}, \mathcal{A}_{\mathbf{s}'' \mathbf{s}'''} \right]
   = i \left( \delta_{\mathbf{s} \mathbf{s}''} \delta_{\mathbf{s}' \mathbf{s}'''} 
    - \delta_{\mathbf{s} \mathbf{s}'''} \delta_{\mathbf{s}' \mathbf{s}''}  \right)
\end{eqnarray}
Moreover, the theory has a local gauge symmetry since one can 
make the transformation
\begin{eqnarray}
\mathcal{A}_{\mathbf{s} \mathbf{s}'} 
   \to \mathcal{A}_{\mathbf{s} \mathbf{s}'} 
   + \lambda_{\mathbf{s}} -  \lambda_{\mathbf{s}'}
\end{eqnarray}
on any bond without changing the value of 
$(\nabla_{\scriptsize\hexagon} \times \mathcal{A})_{\mathbf{r} \mathbf{r}'}$ 
--- each value of $\lambda_{\mathbf{s}}$ 
occurs twice, with opposite signs.
The situation now bears more than a passing resemblance to 
quantum electromagnetism.


At this point a subtlety enters the problem.   
In passing from Eq.~(\ref{eq:Htunnelling}) to Eq.~(\ref{eq:HemB0}), we have performed 
a series of changes of variable without making any new approximations.  
However it still remains to take the limit $U \to \infty$.   
If the ``magnetic'' field $\mathcal{B}_{\mathbf{r}\mathbf{r}'}$ were an integer variable, 
it could be eliminated from the problem by setting $\mathcal{B}_{\mathbf{r}\mathbf{r}'} = 0$ 
on all bonds.   
This would be energetically favourable at large $U/g$, and would imply a phase transition
from a spin liquid phase at small $U/g$ , into a phase in which spinon excitations 
(magnetic monopoles) were confined at large $U/g$ [cf. \onlinecite{guth80}].   
However the fact that $\mathcal{B}_{\mathbf{r}\mathbf{r}'}$ takes on half-integer values
``frustrates'' the lattice theory, and makes it possible for a spin liquid phase to survive 
in the limit $U \to \infty$.    


Keeping this in mind, we now follow Hermele~{\it et al.}~\cite{hermele04} in assuming that 
an average over fast fluctuations of the gauge field a) softens the restriction 
that $\mathcal{B}_{\mathbf{r}\mathbf{r}'}$ take on half-integer values and, 
b) restricts $\mathcal{E}_{\mathbf{s}\mathbf{s}'}$ to small values.  
Provided that both of these assumptions hold true, we can drop the reference field 
$\mathcal{B}_{\mathbf{r}\mathbf{r}'}^0$ and expand the cosine in Eq.~(\ref{eq:HemB0}), 
to obtain
\begin{eqnarray}
\label{eq:Hem}
\mathcal{H}_{\sf U(1)} 
   = \frac{\mathcal{U}}{2} \sum_{\langle \mathbf{r} \mathbf{r}' \rangle}
   \left[ \left( \nabla_{\scriptsize\hexagon} \times \mathcal{A} \right)_{\mathbf{r} \mathbf{r}'} \right]^2 
   + \frac{\mathcal{K}}{2} \sum_{\langle \mathbf{s} 
   \mathbf{s}' \rangle} \mathcal{E}_{\mathbf{s} \mathbf{s}'}^2
\end{eqnarray}
where both the normalisation of the field $\mathcal{B}_{\mathbf{r}\mathbf{r}'}$, 
and the parameters of $\mathcal{H}_{U(1)}$ may be renormalized 
from their bare values $|\mathcal{B}_{\mathbf{r}\mathbf{r}'}| \sim 1/2$, 
$\mathcal{U} \sim U$, $\mathcal{K} \sim g$.
This, finally, is the Hamiltonian for non-compact quantum electromagnetism 
on a diamond lattice.


At first sight, the final step of this derivation might seem to involve an uncomfortably 
large leap of faith~\cite{henley10}.  
However this will be justified {\it a posteriori} in Section~\ref{subsection:QMC-zeroT} 
by the excellent, quantitative, agreement of the predictions of 
$\mathcal{H}_{U(1)}$ [Eq.~(\ref{eq:Hem})] 
with quantum Monte Carlo simulation of the microscopic model 
${\mathcal H}_\mu$ [Eq.~(\ref{eq:Hmu})].  
In order to extend this comparison to finite values of the control parameter 
$\mu$, we will augment $\mathcal{H}_{U(1)}$ with a term 
\begin{eqnarray}
\label{eq:deltaHem}
\delta\mathcal{H}_{\sf U(1)} 
   = \frac{\mathcal{W}}{2} \sum_{\langle \mathbf{s} \mathbf{s}' \rangle}
   \left[ \left( \nabla_{\scriptsize\hexagon} \times \left( \nabla_{\scriptsize\hexagon} 
   \times \mathcal{A} \right) \right)_{\mathbf{s} \mathbf{s}'} \right]^2
\end{eqnarray}
which mimics the effect of the ``RK'' potential [Eq.~(\ref{eq:deltaHmu})].  
Since this term is permitted by the gauge symmetry, in principle it might also be generated 
dynamically by an average over fast fluctuations of~$\mathcal{A}_{\mathbf{s}\mathbf{s}'}$.


\subsection{Constructing the photon}
\label{subsection:photons}


The lattice gauge theory described in Section~\ref{subsection:em} 
supports three types of excitation : 
magnetic charges (point sources of $\mathcal{B}$) and 
electric charges (point sources of $\mathcal{E}$), 
together with photons (transverse excitations of $\mathcal{A}$) 
which mediate Coulomb interactions between these emergent charges
[\onlinecite{moessner03,hermele04}].  


The magnetic charges are the magnetic monopoles of the classical 
theory~\cite{castelnovo08}, now quantised and endowed with 
dynamics~\cite{fulde02,nussinov07,wan12}.  
They correspond to the ``spinon'' excitations of the spin liquid.
Since they involve spin configurations lying outside the ice manifold, 
they have an energy gap $$2 \Delta_{\mathcal{B}} \sim  J^{zz}$$ [cf.~Eq~(\ref{eq:Hross})]. 
The electric charges are gapped, topological excitations which can be 
constructed as a wave packet of ice configurations with suitably 
chosen phases~\cite{moessner03,hermele04}.
These also have an energy gap 
$$\Delta_{\mathcal{E}} \sim {\mathcal K} \sim g = 12 J_{\pm}^3/J_{zz}^2$$ [cf.~Eq~(\ref{eq:g})].   


However the energy of the photons vanishes linearly 
at small wave vector
$$
\omega(\mathbf{k} \to 0) = c |\mathbf{k}|.
$$
and being gapless, the photons will control the low energy and low temperature 
properties of the system.  
We therefore concentrate on exploring the consequences of the photons 
in this paper, leaving other excitations for future work.


In what follows, we will explicitly construct a photon basis for the lattice gauge theory
developed in Section~\ref{subsection:em}, with a view to calculating the spin-spin 
correlation functions of the original model of a quantum spin ice.
We take as a starting point 
\begin{eqnarray}
\mathcal{H}^\prime_{\sf U(1)} 
   &=& \frac{\mathcal{U}}{2} \sum_{\mathbf{r} \in A, n}
    \left[ \left(\nabla_{\scriptsize\hexagon} \times \mathcal{A}\right)_{(\mathbf{r}, n)} \right]^2 
    \nonumber  \\
   &&
   + \frac{1}{2\mathcal{K}} \sum_{\mathbf{s}  \in A', m} 
      \left[ \frac{\partial \mathcal{A}_{(\mathbf{s}, m)}}{\partial t} \right]^2
    \nonumber  \\
   && + \frac{\mathcal{W}}{2} \sum_{\mathbf{s}  \in A', m} 
   \left[ \left( \nabla_{\scriptsize\hexagon} \times \nabla_{\scriptsize\hexagon} \times \mathcal{A} \right)_{(\mathbf{s}, m)} \right]^2
\label{eq:HA}
\end{eqnarray}
where we have used the fact that, in the absence of electric charges
\begin{eqnarray}
\mathcal{E}_{(\mathbf{s},m)}
    &=& - \frac{1}{\mathcal{K}} \frac{\partial \mathcal{A}_{(\mathbf{s}, m)}}{\partial t}
\end{eqnarray}
To avoid double counting of bonds, the sums over diamond lattice sites $\{\mathbf{r}\}$ 
and $\{\mathbf{s}\}$ are restricted to a single sublattice, with bonds labelled 
$$
(\mathbf{r}, n) = (\mathbf{r}, \mathbf{r}+\mathbf{e}_n) 
\quad , \quad 
(\mathbf{s}, m)=(\mathbf{s}, \mathbf{s}+\mathbf{e}_m)
$$
where
\begin{eqnarray}
\mathbf{e}_0 &=& \frac{a_0}{4} \left( 1, 1, 1 \right) \nonumber \\
\mathbf{e}_1 &= & \frac{a_0}{4} \left( 1, -1, -1 \right) \nonumber \\
\mathbf{e}_2 &= & \frac{a_0}{4} \left( -1, 1, -1 \right) \nonumber \\
\mathbf{e}_3 &= & \frac{a_0}{4} \left(-1, -1, 1 \right)
\label{eq:en}
\end{eqnarray}
and $a_0$ is the linear dimension of the cubic unit cell of the lattice, shown in Fig.~\ref{fig:lattice}.  


We proceed to quantise $\mathcal{A}_{\mathbf{s} m}$ by analogy with conventional 
electromagnetism, introducing a Bose operator 
$$
\big[ a^{\phantom\dagger}_{\lambda}, a^{\dagger}_{\lambda'} \big] 
   = \delta_{\lambda \lambda'}
$$
where the four sites of the tetrahedron in the primitive unit cell of the 
pyrochlore lattice translate into four bands $\lambda=1 \ldots 4$. 
We write
\begin{eqnarray}
\label{eq:Aquantised}
\mathcal{A}_{(\mathbf{s}, m)}
   &=&\sqrt{\frac{2}{N}} \sum_{\mathbf{k}} \sum_{\lambda=1}^{4} 
           \sqrt{\frac{\mathcal{K}}{\omega_{\lambda}(\mathbf{k})}} \nonumber \\
   && \times\left( \exp\left[-i \mathbf{k} \cdot (\mathbf{s}+\mathbf{e}_m/2)\right] 
          \eta_{m \lambda}(\mathbf{k}) a^{\phantom\dagger}_{\lambda}(\mathbf{k}) \right. \nonumber \\
   && \quad + \left. \exp\left[i \mathbf{k} \cdot (\mathbf{s}+\mathbf{e}_m/2)\right] 
             \eta^{\ast}_{\lambda m}(\mathbf{k}) a^{\dagger}_{\lambda}(\mathbf{k}) \right) \nonumber \\
\end{eqnarray}
where the sum $\sum_{\lambda=1}^{4}$ runs over all four branches of photons
and $\uuline{\eta}({\mathbf k})$ is a unitary, $4\times4$ matrix 
whose columns, $\eta_{\lambda}({\mathbf k})$, play the same role as 
the polarisation vector in conventional electromagnetism.  
%
%
By obvious extension
\begin{eqnarray}
\label{eq:Equantised}
\mathcal{E}_{(\mathbf{s}, m)}
    &=& i  \sqrt{\frac{2}{N}} \sum_{\mathbf{k}} \sum_{\lambda=1}^{4} 
            \sqrt{\frac{\omega_{\lambda}(\mathbf{k})}{\mathcal{K}}} \nonumber \\
   && \times \left( \exp\left[-i \mathbf{k} \cdot (\mathbf{s}+\mathbf{e}_m/2)\right] 
           \eta_{m \lambda}(\mathbf{k}) a^{\phantom\dagger}_{\lambda}(\mathbf{k}) \right. \nonumber \\
   && \quad - \left. \exp\left[i \mathbf{k} \cdot (\mathbf{s}+\mathbf{e}_m/2)\right] 
           \eta^{\ast}_{\lambda m}(\mathbf{k}) a^{\dagger}_{\lambda}(\mathbf{k}) \right) \nonumber \\
\end{eqnarray}


The Hamiltionian (Eq.~\ref{eq:HA}) is already quadratic in $a^{\phantom\dagger}_\lambda$.
What remains is to eliminate all terms which do not conserve photon number, by constructing 
a suitable matrix $\eta^{\ast}_{\lambda m}(\mathbf{k})$.   
To do this, we need to evaluate the Fourier transform of the lattice curl 
\mbox{$(\nabla_{\scriptsize\hexagon} \times A)_{(\mathbf{r}, n)}$}.
This operator is defined on a six-bond plaquette, composed of pairs of bonds 
which enter with opposite signs in the directed sum around the plaquette
[Fig.~\ref{fig:tunnelling}, Fig.~\ref{fig:latticeEM}].   
These bonds have midpoints located at 
$$
\mathbf{r} - \mathbf{e}_n/2 \pm \mathbf{h}_{nm}
$$
where
\begin{eqnarray}
\mathbf{h}_{nm} 
    \equiv \frac{a_0}{\sqrt{8}} \frac{\hat{\mathbf{e}}_n 
     \times \hat{\mathbf{e}}_m}{|\hat{\mathbf{e}}_n 
     \times \hat{\mathbf{e}}_m |}
\label{eq:hnm}
\end{eqnarray}
Hence 
\begin{eqnarray}
 && \left( \nabla_{\scriptsize\hexagon} \times \mathcal{A} \right)_{(\mathbf{r}, n)}  
=\sqrt{\frac{2}{N}} \sum_{\mathbf{k}} \sum_{\lambda=1}^{4} 
    \sqrt{\frac{\mathcal{K}}{\omega_{\lambda}(\mathbf{k})}} 
   \nonumber \\
&& \qquad  \times \bigg\{ \exp[-i \mathbf{k} \cdot (\mathbf{r}-\mathbf{e}_n/2)] a^{\phantom\dagger}_{\lambda}(\mathbf{k}) 
   \nonumber \\
&&  \qquad  \times \sum_{m} (-2 i \sin(\mathbf{k} \cdot \mathbf{h}_{nm})) \eta_{m \lambda}(\mathbf{k})  
  \nonumber \\
&& \qquad + \exp\left[ i \mathbf{k} \cdot (\mathbf{r}-\mathbf{e}_n/2) \right] a^{\dagger}_{\lambda}(\mathbf{k}) 
  \nonumber \\
&& \qquad \times \sum_{m} \left( 2 i \sin(\mathbf{k} \cdot \mathbf{h}_{nm}) \right) 
       \eta^{\ast}_{\lambda m}(\mathbf{k}) \bigg\}
\label{eq:curl1}
\end{eqnarray}
where, by inspection, $\mathbf{h}_{nn} \equiv 0$. 


We can rewrite the sum $\sum_{m}$ in Eq.~(\ref{eq:curl1}) in a more convenient form 
by introducing an Hermitian, anti-symmetric matrix 
\begin{eqnarray}
\label{eq:Z}
&&\uuline{Z}(\mathbf{k})= -2i  \times  \nonumber \\
&&
  \begin{pmatrix} 
  0 & \sin(\mathbf{k} \cdot \mathbf{h}_{01}) & \sin(\mathbf{k} \cdot \mathbf{h}_{02}) & \sin(\mathbf{k} \cdot \mathbf{h}_{03}) \\  
 -\sin(\mathbf{k} \cdot \mathbf{h}_{01}) & 0 & \sin(\mathbf{k} \cdot \mathbf{h}_{12}) & \sin(\mathbf{k} \cdot \mathbf{h}_{13})  \\
 -\sin(\mathbf{k} \cdot \mathbf{h}_{02}) & -\sin(\mathbf{k} \cdot \mathbf{h}_{12}) & 0 & \sin(\mathbf{k} \cdot \mathbf{h}_{23}) \\
 -\sin(\mathbf{k} \cdot \mathbf{h}_{03}) & -\sin(\mathbf{k} \cdot \mathbf{h}_{13}) & -\sin(\mathbf{k} \cdot \mathbf{h}_{23}) & 0  
  \end{pmatrix}\nonumber \\
  \end{eqnarray}
acting on the four component vectors $\eta_{\lambda}(\mathbf{k})$.


Since $\uuline{Z}({\mathbf k})$ is Hermitian, we are free to construct the matrix 
$\uuline{\eta}({\mathbf k})$ from the 
eigenvectors of $\uuline{Z}({\mathbf k})$, such that
\begin{eqnarray}
\uuline{Z}(\mathbf{k}) 
   \cdot 
   \begin{pmatrix} \eta_{\lambda 0} \\ \eta_{\lambda 1} \\ \eta_{\lambda 2} \\ \eta_{\lambda 3} \end{pmatrix} 
       &=& \zeta_{\lambda}(\mathbf{k}) 
   \begin{pmatrix} \eta_{\lambda 0} \\ \eta_{\lambda 1} \\ \eta_{\lambda 2} \\ \eta_{\lambda 3} \end{pmatrix} 
\label{eq:Zeigenvectors}
\end{eqnarray}
A specific choice of $\uuline{\eta}({\mathbf k})$ corresponds to a choice of gauge, since 
using Eq.~(\ref{eq:Aquantised}), the divergence of $\mathcal{A}_{{\bf s} {\bf s'}}$ is now fixed. 
The choice here, which is made for maximum
convenience in constructing the photon dispersion, is the radiation (or Coulomb) gauge
\begin{eqnarray}
\nabla \cdot \mathcal{A}=0.
\end{eqnarray}
It follows from Eqs.~(\ref{eq:curl1}) and~(\ref{eq:Zeigenvectors}) that
\begin{eqnarray}
&&(\nabla_{\scriptsize\hexagon} \times \mathcal{A})_{(\mathbf{r}, n)}
   = \sqrt{\frac{2}{N}} \sum_{\mathbf{k}} \sum_{\lambda=1}^{4} 
   \sqrt{\frac{\mathcal{K}}{\omega_{\lambda}(\mathbf{k})}} 
   \nonumber \\
&& \times \bigg( 
   \exp[ -i \mathbf{k} \cdot (\mathbf{r}-\mathbf{e}_n/2) ] 
   a_{\lambda}(\mathbf{k}) \zeta_{\lambda}(\mathbf{k}) 
   \eta_{n \lambda}(\mathbf{k})  
   \nonumber \\
&& \quad + \exp[i \mathbf{k} \cdot (\mathbf{r}-\mathbf{e}_n/2)] 
    a^{\dagger}_{\lambda}(\mathbf{k})\zeta_{\lambda}(\mathbf{k}) 
   \eta_{\lambda n}^{\ast}(\mathbf{k}) 
   \bigg). 
\label{eq:curl2}
\end{eqnarray}
Squaring and summing over $\mathbf{r}$ and $n$, we arrive at
\begin{eqnarray}
&&\sum_{(\mathbf{r}, n)}  (\nabla_{\scriptsize\hexagon} \times \mathcal{A})_{(\mathbf{r}, n)}^2 
   = \frac{1}{2}  \sum_{\mathbf{k}} \sum_{\lambda=1}^{4}  \sum_{\lambda'=1}^{4}  
   \nonumber \\
&& \times \sqrt{\frac{\mathcal{K}}{\omega_{\lambda}(\mathbf{k})}}  
  \sqrt{\frac{\mathcal{K}}{\omega_{\lambda'}(\mathbf{k})}} \zeta_{\lambda}(\mathbf{k}) \zeta_{\lambda'}(\mathbf{k}')  
   \nonumber \\
&&  \times \bigg\{ 
   a_{\lambda}(\mathbf{k}) a_{\lambda'}(-\mathbf{k}) \zeta_{\lambda}(\mathbf{k}) \zeta_{\lambda'}(-\mathbf{k})  
   \sum_n \eta_{n \lambda}(\mathbf{k})  \eta_{n\lambda'}(-\mathbf{k}) 
   \nonumber \\
&& + a^{\dagger}_{\lambda}(\mathbf{k}) a^{\dagger}_{\lambda'}(\mathbf{k}) 
    \zeta_{\lambda}(\mathbf{k}) \zeta_{\lambda'}(-\mathbf{k}) 
    \sum_n \eta_{\lambda n}^{\ast}(\mathbf{k}) 
    \eta_{\lambda' n}^{\ast}(-\mathbf{k}) 
    \nonumber \\
&& +  a_{\lambda}(\mathbf{k}) a^{\dagger}_{\lambda'}(\mathbf{k}) 
   \zeta_{\lambda}(\mathbf{k}) \zeta_{\lambda'}(\mathbf{k}) 
   \sum_n \eta_{n \lambda}(\mathbf{k}) 
   \eta_{\lambda' n}^{\ast}(\mathbf{k}) 
   \nonumber \\
&& + a^{\dagger}_{\lambda}(\mathbf{k}) a_{\lambda'}(\mathbf{k})  
   \zeta_{\lambda}(\mathbf{k}) \zeta_{\lambda'}(\mathbf{k}) 
   \sum_n \eta_{\lambda n}^{\ast}(\mathbf{k}) 
   \eta_{n \lambda'}(\mathbf{k}) \bigg\}.
\end{eqnarray}
This rather dense expression can be simplified using the unitarity of $\uuline{\eta}(\mathbf{k})$
\begin{eqnarray}
 \sum_n \eta_{\lambda n}^{\ast}(\mathbf{k}) \eta_{n \lambda'}(\mathbf{k})=\delta_{\lambda \lambda'}.
\label{eq:unitarity}
\end{eqnarray}
and the fact that 
$$
\uuline{Z}(-\mathbf{k})=\uuline{Z}(\mathbf{k})^{\ast}
$$
from which it follows that
\begin{eqnarray}
\eta_{\lambda}(-\mathbf{k}) &=& \eta_{\lambda}^{\ast}(\mathbf{k}) \\
\zeta_{\lambda}(\mathbf{k}) &=& \zeta_{\lambda}(-\mathbf{k}).
\end{eqnarray}
Whence, 
\begin{eqnarray}
&& \sum_{(\mathbf{r}, n)} (\nabla_{\scriptsize\hexagon} \times \mathcal{A})_{(\mathbf{r}, n)}^2
  = \frac{\mathcal{K}}{2} \sum_{\mathbf{k}} 
  \sum_{\lambda=1}^{4} \frac{\zeta_{\lambda}(\mathbf{k})^2}{\omega_{\lambda}(\mathbf{k})} 
  \nonumber \\
&& \quad \times \bigg\{ 
a_{\lambda}(\mathbf{k}) a_{\lambda}(-\mathbf{k}) 
   + a_{\lambda}^{\dagger}(\mathbf{k})a_{\lambda}^{\dagger}(-\mathbf{k}) 
   \nonumber \\
&& \qquad + a_{\lambda}^{\dagger}(\mathbf{k}) a_{\lambda}(\mathbf{k}) 
   + a_{\lambda}(\mathbf{k})a_{\lambda}^{\dagger}(\mathbf{k})
    \bigg\}.
\end{eqnarray}


Applying the same procedure again to Eq.~(\ref{eq:curl2}), we find
\begin{eqnarray}
&&\sum_{(\mathbf{s}, m)}  (\nabla_{\scriptsize\hexagon} \times \nabla_{\scriptsize\hexagon} \times \mathcal{A})_{(\mathbf{s}, m)}^2 
    = \frac{\mathcal{K}}{2} \sum_{\mathbf{k}} \sum_{\lambda=1}^{4} 
    \frac{\zeta_{\lambda}(\mathbf{k})^4}{\omega_{\lambda}(\mathbf{k})}   
    \nonumber \\
&& \times \bigg\{ a_{\lambda}(\mathbf{k}) a_{\lambda}(-\mathbf{k}) 
   + a_{\lambda}^{\dagger}(\mathbf{k})a_{\lambda}^{\dagger}(-\mathbf{k}) 
   \nonumber \\
&& \qquad + a_{\lambda}^{\dagger}(\mathbf{k}) a_{\lambda}(\mathbf{k}) 
 + a_{\lambda}(\mathbf{k})a_{\lambda}^{\dagger}(\mathbf{k}) \bigg\}.
\end{eqnarray}
The remaining, electric field, term in $\mathcal{H}_0$ [Eq.~(\ref{eq:Hem})] yields 
\begin{eqnarray}
&\sum_{(\mathbf{s}, m)}  &
   \left(
      \frac{\partial \mathcal{A}_{(\mathbf{s}, m)}}{\partial t}
   \right)^2 
   = \frac{1}{2} \sum_{\mathbf{k}}  \sum_{\lambda=1}^{4} 
    \omega_{\lambda}(\mathbf{k})
    \nonumber \\
&& \times \bigg\{-a_{\lambda}(\mathbf{k}) a_{\lambda}(-\mathbf{k}) 
   - a_{\lambda}^{\dagger}(\mathbf{k})a_{\lambda}^{\dagger}(-\mathbf{k}) \nonumber \\
&& \qquad + a_{\lambda}^{\dagger}(\mathbf{k}) a_{\lambda}(\mathbf{k}) 
   + a_{\lambda}(\mathbf{k})a_{\lambda}^{\dagger}(\mathbf{k}) \bigg\}.
\end{eqnarray}
Inserting all of this into the Hamiltonian Eq.~(\ref{eq:HA}) gives
\begin{eqnarray}
&\mathcal{H}^\prime_{\sf U(1)}&
    = \sum_{\mathbf{k}} \sum_{\lambda=1}^{4} 
   \bigg[ 
   \bigg( 
   \frac{\mathcal{U} \mathcal{K} \zeta_{\lambda}(\mathbf{k})^2}{4 \omega_{\lambda}(\mathbf{k})} 
     + \frac{\mathcal{W} \mathcal{K} \zeta_{\lambda}(\mathbf{k})^4}{4 \omega_{\lambda}(\mathbf{k})} 
     + \frac{\omega_{\lambda}(\mathbf{k})}{4} 
    \bigg) 
    \nonumber \\
&& \qquad  \times \left( 
     a_{\lambda}(\mathbf{k}) a_{\lambda}^{\dagger}(\mathbf{k}) 
     + a_{\lambda}^{\dagger}(\mathbf{k}) a_{\lambda}(\mathbf{k}) 
    \right) 
    \nonumber \\ 
&& \qquad + \left( 
       \frac{\mathcal{U} \mathcal{K} \zeta_{\lambda}(\mathbf{k})^2}{4 \omega_{\lambda}(\mathbf{k})} 
       + \frac{\mathcal{W} \mathcal{K} \zeta_{\lambda}(\mathbf{k})^4}{4 \omega_{\lambda}(\mathbf{k})} 
       - \frac{\omega_{\lambda}(\mathbf{k})}{4} 
       \right)
    \nonumber \\
&& \qquad \times \left( 
     a_{\lambda}(\mathbf{k}) a_{\lambda}(-\mathbf{k}) 
     + a_{\lambda}^{\dagger}(\mathbf{k}) a_{\lambda}^{\dagger}(-\mathbf{k}) 
     \right)  
    \bigg] 
\end{eqnarray}
To diagonalize the Hamiltonian we require
\begin{eqnarray}
\frac{\mathcal{U} \mathcal{K} \zeta_{\lambda}(\mathbf{k})^2}{4 \omega_{\lambda}(\mathbf{k})} 
   + \frac{\mathcal{W} \mathcal{K} \zeta_{\lambda}(\mathbf{k})^4}{4 \omega_{\lambda}(\mathbf{k})}
   =  \frac{\omega_{\lambda}(\mathbf{k})}{4}.
\label{diagonal condition}
\end{eqnarray}
which implies 
\begin{eqnarray}
\mathcal{H}^\prime_{\sf U(1)} 
   = \sum_{\mathbf{k}} \sum_{\lambda=1}^{4} \omega_{\lambda}(\mathbf{k}) 
   \left( a^{\dagger}_{\lambda} (\mathbf{k}) a_{\lambda} (\mathbf{k})+\frac{1}{2} \right)
\label{diagonalH1}
\end{eqnarray}
with dispersion relation fixed by Eq. (\ref{diagonal condition})
\begin{eqnarray}
\omega_{\lambda}(\mathbf{k}) 
  = \mathcal{K} 
  \sqrt{\frac{\mathcal{U}}{\mathcal{K}} 
  \zeta_{\lambda}(\mathbf{k})^2
  + \frac{\mathcal{W}}{\mathcal{K}} 
  \zeta_{\lambda}(\mathbf{k})^4}.
  \label{eq:omegakzeta}
\end{eqnarray}


All that now remains is to determine the eigenvalues of the matrix 
$Z(\mathbf{k})$, $\zeta_{\lambda}(\mathbf{k})$. 
We find 
\begin{eqnarray}
\zeta_{1}(\mathbf{k}) &=& +\sqrt{2} \sqrt{\sum_{mn} \sin{(\mathbf{k} \cdot \mathbf{h}_{mn})}^2} \\
\label{eq:zeta1}
\zeta_{2}(\mathbf{k}) &=& -\sqrt{2} \sqrt{\sum_{mn} \sin{(\mathbf{k} \cdot \mathbf{h}_{mn})}^2} \\
\label{eq:zeta2}
\zeta_{3}(\mathbf{k}) &=& 0 \\
\zeta_{4}(\mathbf{k}) &=& 0.
\end{eqnarray}
It follows that the four bands of excitations  $\zeta_{\lambda}(\mathbf{k})$ correspond to two, degenerate, 
physical photon modes, and two unphysical, zero energy modes.  
The unphysical modes arise because of the gauge redundancy in $\mathcal{A}$ and make no 
contribution to either the Hamiltonian or to any gauge invariant correlation functions.  


Keeping only the physical photon modes from Eq.~(\ref{diagonalH1}), we finally arrive at
\begin{eqnarray}
\mathcal{H}^\prime_{\sf U(1)} 
  = \sum_{\mathbf{k}} \sum_{\lambda=1}^{2} 
  \omega(\mathbf{k}) 
  \left( 
  a^{\dagger}_{\lambda} (\mathbf{k}) a^{\phantom\dagger}_{\lambda} (\mathbf{k}) 
  + \frac{1}{2} 
  \right)
\label{eq:Hphoton}
\end{eqnarray}
where $\lambda$ now has the interpretation of the polarisation of the photon.
The photon dispersion $\omega(\mathbf{k})$ is independent of polarisation 
and can be written 
\begin{eqnarray}
&&\omega(\mathbf{k}) =  \mathcal{K} 
   \sqrt{\frac{\mathcal{U}}{\mathcal{K}} 
     \zeta(\mathbf{k}) 
   + \frac{\mathcal{W}}{\mathcal{K}} 
     \zeta(\mathbf{k})^2}
    \nonumber \\
\label{eq:omegak}
\end{eqnarray}
where
\begin{eqnarray}
\zeta(\mathbf{k})
   =\zeta_{1}(\mathbf{k}) 
   = -\zeta_{2}(\mathbf{k})
   = \sqrt{2} \sqrt{\sum_{mn} \sin{(\mathbf{k} \cdot \mathbf{h}_{mn})}^2}
\label{eq:zeta}   
\end{eqnarray}
with $\mathbf{h}_{mn}$ defined by Eq.~(\ref{eq:hnm}).  


For all $\ \mathcal{U}/\mathcal{K} > 0$ the photon dispersion is linear in the 
long-wavelength limit
\begin{eqnarray}
\omega(\mathbf{k}\approx \mathbf{0}) 
   \approx 
   \sqrt{\mathcal{U}\mathcal{K}} \ a_0 |\mathbf{k}|
\end{eqnarray}
This means that there is a well-defined speed of light
\begin{eqnarray}
   c = \sqrt{\mathcal{U}\mathcal{K}}\ a_0\ \hbar^{-1}
\end{eqnarray}
where we have restored the dimensional factor of $\hbar$.   

\begin{figure}
\includegraphics[width=0.45\textwidth]{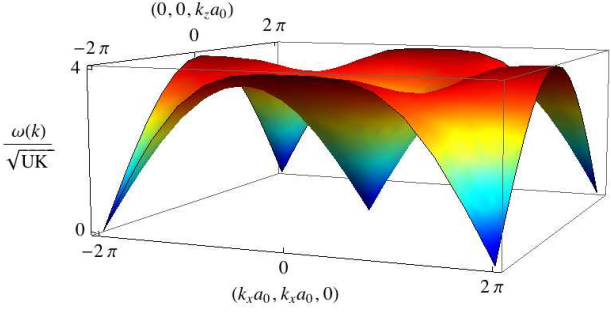}
\caption{
(Color online).  
Dispersion $\omega(\mathbf{k})$ of magnetic photon excitations, calculated for the lattice 
field theory Eq.~(\ref{eq:HA}) in the ``quantum ice'' limit $\mathcal{W}/\mathcal{K} \to 0$.
The dispersion is plotted in the $(h,h,l)$ plane, following Eq.~(\ref{eq:omegak}).  
The dispersion is linear in $|\mathbf{k}|$ in the long-wavelength limit, with a 
speed of light $c=\sqrt{\mathcal{U}\mathcal{K}}a_0$.
}
\label{fig:W=0-dispersion}
\end{figure}
\begin{figure}
\includegraphics[width=0.45\textwidth]{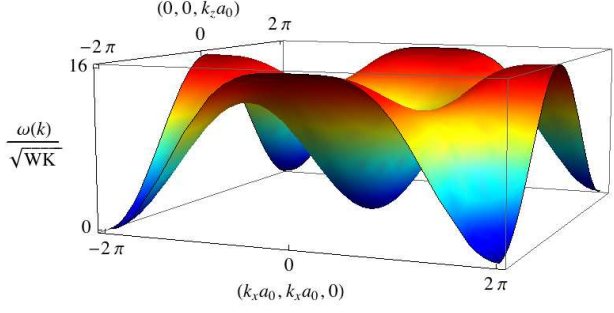}
\caption{
(Color online).  
Dispersion $\omega(\mathbf{k})$ of magnetic photon excitations, calculated for the lattice 
field theory Eq.~(\ref{eq:HA}) in the limit $\mathcal{U}/\mathcal{K} \to 0$.
The dispersion is plotted in the $(h,h,l)$ plane, following Eq.~(\ref{eq:omegak}).  
The dispersion is quadratic in $\mathbf{k}$ the long wavelength limit.  
This situation is realised in the microscopic ``quantum ice'' model 
${\mathcal H}_\mu$ [Eq.~(\ref{eq:Hmu})], at the RK point, $\mu=g$.
}
\label{U=0dispersion}
\end{figure}

However in the limiting case $\mathcal{U}/\mathcal{K} \to 0$, $c \to 0$, and 
the dispersion of the photon becomes quadratic in the long-wavelength limit
\begin{eqnarray}
\omega(\mathbf{k}) 
   \approx 
   \sqrt{\mathcal{W}\mathcal{K}} a_0^2 |\mathbf{k}|^2.
\end{eqnarray}
Precisely this limit is realised at the RK point $\mu=g$ of the quantum ice model 
${\mathcal H}_\mu$ [Eq.~(\ref{eq:Hmu})], and defines the boundary 
of the quantum $U(1)$ liquid phase~\cite{hermele04,shannon12}.  
The photon dispersion relations in the two extreme cases $\mathcal{U}/\mathcal{K}=0$ 
and $\mathcal{W}/\mathcal{K}=0$ are plotted in Fig.~\ref{fig:W=0-dispersion} and 
Fig.~\ref{U=0dispersion}.


\subsection{From photons to structure factors}
\label{subsection:Skomega-B}


Spin correlations in real materials can be measured directly by neutron scattering.  
Here we convert the analysis of photons in Section~\ref{subsection:photons}
into concrete predictions for the dynamical structure factors measured in such 
an experiment.  
We also consider the structure factors which might be measured in, e.g., X-ray 
scattering experiments on a charge ice of the type considered by 
Banerjee {\it et al.}~\cite{banerjee08}.
Specifically, we will consider 
\begin{eqnarray}
S_{\sf spin}^{\alpha\beta} (\mathbf{k}, \omega)
    &=& \int dt e^{-i \omega t}
    \langle {\mathsf S}^\alpha(-\mathbf{k}, t) {\mathsf S}^\beta(\mathbf{k}, 0) \rangle
\label{eq:Skomega-spin-definition}
\end{eqnarray}
and 
\begin{eqnarray}
S_{\sf charge} (\mathbf{k}, \omega)
    &=&  \int dt e^{-i \omega t}
    \langle n(-\mathbf{k}, t) n(\mathbf{k}, 0) \rangle
\label{eq:Skomega-charge-definition}
\end{eqnarray}
Possessing the full photon wave function [Eq.~(\ref{eq:Aquantised})] permits  
us to calculate these dynamical structure factors on a lattice, passing directly 
from the correlations of $\mathcal{A}_{(\mathbf{s},m)}$ to those of 
${\mathsf S}_{(\mathbf{r},n)}$ or $n_{(\mathbf{r},n)}$.  


We first consider the charge ice and, following \onlinecite{banerjee08}, 
introduce an additional (dimensionless) scale factor $\kappa \lesssim 1$ to take account 
of any renormalization of the field $\mathcal{B}$ when an average is taken over fast 
fluctuations of $\mathcal{A}_{(\mathbf{s},m)}$ [cf. Eq.~(\ref{eq:HemB0}) to Eq.~(\ref{eq:Hem})]
\begin{eqnarray}
S_{\sf charge} (\mathbf{k}, t)  &=& \kappa^2
   \sum_{mn} 
   \langle 
       \tilde{\mathcal{B}}_n(-\mathbf{k}, t)  \tilde{\mathcal{B}}_m(\mathbf{k}, 0)  
   \rangle 
\label{eq:Skomega-charge-definition}
\end{eqnarray}
where 
\begin{eqnarray}
\tilde{\mathcal{B}}_n(\mathbf{k},t) 
= \frac{1}{\sqrt{N}}  \sum_{\mathbf{r}} 
   \exp[-i \mathbf{k} \cdot (\mathbf{r} +\mathbf{e}_n/2)] 
   \mathcal{B}_n(\mathbf{r},t)
\end{eqnarray}
with
$$
\mathcal{B}_n(\mathbf{r}) \equiv \mathcal{B}(\mathbf{r}-\mathbf{e}_n/2) 
   = \left( \nabla_{\scriptsize\hexagon} \times \mathcal{A} \right)_{(\mathbf{r}, n)}
$$
The time evolution of 
$\mathcal{A}_{(\mathbf{s}, m)}$ follows directly from 
$\mathcal{H}^\prime_{\sf U(1)}$ [Eq.~(\ref{eq:Hphoton})]   
\begin{eqnarray}
\tilde{\mathcal{B}}_n(\mathbf{k}, t) 
&=& \frac{\sqrt{2}}{4} \sum_{\lambda=1}^{4}  
   \sqrt{\frac{\mathcal{K}}{\omega_{\lambda}(\mathbf{k})}} 
   \zeta_{\lambda}(\mathbf{k}) 
   \nonumber \\
&& \times 
   \left(
   \eta_{n \lambda}( -\mathbf{k}) 
   a_{\lambda}(-\mathbf{k}) 
   e^{-i \omega_{\lambda}(\mathbf{k}) t} 
   \right. \nonumber \\
&&  \left. \qquad + 
   \eta_{\lambda n}^{\ast}(\mathbf{k}) 
   a_{\lambda}^{\dagger}(\mathbf{k}) 
   e^{i \omega_{\lambda}(\mathbf{k}) t} 
   \right).
\label{eq:Bkt}
\end{eqnarray}
such that
\begin{eqnarray}
S_{\sf charge} (\mathbf{k}, \omega)
    &=& \frac{\kappa^2}{8} \sum_{mn} 
     \sum_{\lambda=1}^{4} \frac{\mathcal{K}}{\omega_{\lambda}(\mathbf{k})} 
     \zeta_{\lambda}(\mathbf{k})^2 \eta_{m \lambda}(\mathbf{k}) \eta_{\lambda n}^{\ast} 
     \nonumber \\
    && \quad \times \langle a^{\phantom\dagger}_{\lambda}(\mathbf{k}) a_{\lambda}^{\dagger}(\mathbf{k}) 
    + a_{\lambda}^{\dagger}(\mathbf{k}) a^{\phantom\dagger}_{\lambda}(\mathbf{k})\rangle
    \nonumber \\
    && \quad \times \delta \left(\omega - \omega_{\lambda}(\mathbf{k}) \right) 
\label{eq:Sq-charge}
\end{eqnarray}
where we have dropped all terms which fail to preserve photon number or polarisation.


The unphysical photon polarisations $\lambda=3$,$4$ do {\it not} contribute to 
Eq.~(\ref{eq:Sq-charge}), since 
\begin{eqnarray}
\zeta_\lambda(\mathbf{k})^2/\omega_\lambda(\mathbf{k})|_{\lambda=3,4} \equiv 0
\label{eq:zero-modes}
\end{eqnarray}
For the physical polarisations $\lambda=1$,$2$ 
\begin{eqnarray}
\langle a_{\lambda}^{\dagger}(\mathbf{k}) a^{\phantom\dagger}_{\lambda}(\mathbf{k})\rangle
    = \frac{1}{e^{\frac{\omega (\mathbf{k})}{T}}-1} \equiv n_B (\mathbf{k})
\end{eqnarray}
since the photons are bosons.   
Noting that 
\begin{eqnarray}
\sum_{\lambda=1}^{4} \zeta_{\lambda}(\mathbf{k})^2 
\eta_{m \lambda}(\mathbf{k}) \eta_{\lambda n}^{\ast} =  \sum_{mn} (\uuline{Z}(\mathbf{k})^2)_{mn}
\label{eq:spectral-decomp-Z}
\end{eqnarray}
we arrive at a result for the dynamical structure factor of a quantum charge ice 
\begin{eqnarray}
S_{\sf charge} (\mathbf{k},\omega) 
   && = \frac{\kappa^2}{2} \frac{\mathcal{K}}{\omega(\mathbf{k})} 
 \sum_{mn} \sum_l 
   \sin{( \mathbf{k} \cdot \mathbf{h}_{ml} )} 
   \sin{( \mathbf{k} \cdot \mathbf{h}_{nl} )}
       \nonumber \\
    && \quad \times \bigg( \delta \left(\omega - \omega_{\lambda}(\mathbf{k}) \right) (1+n_B(\mathbf{k}))
    \nonumber \\
    && \quad + \  \delta \left(\omega + \omega_{\lambda}(\mathbf{k}) \right) n_B(\mathbf{k}) \bigg)
\label{eq:Skomega-charge}
\end{eqnarray}
where the vectors $\mathbf{h}_{nm}$ are defined by Eq.~(\ref{eq:hnm}). 


In comparing with quantum Monte Carlo simulation, we 
will also make extensive use of the zero-temperature, equal-time (i.e. energy-integrated) 
structure factor
\begin{eqnarray}
S_{\sf charge} (\mathbf{k},t=0)_{T=0}  = \int d\omega \ S_{\sf charge} (\mathbf{k},\omega)_{T=0}  
 \end{eqnarray}
This can be written as a function of just two, dimensionless, ratios of parameters, 
$\overline{\mathcal{U}}$ and $\overline{\mathcal{W}}$
\begin{eqnarray}
S_{\sf charge} (\mathbf{k},t=0)_{T=0} 
= \frac{
    \overline{S}_{\sf 0}(\mathbf{k})
    }{
    \sqrt{
    \overline{\mathcal{U}} 
    \zeta(\mathbf{k})^2
    + 
    \overline{\mathcal{W}} 
    \zeta(\mathbf{k})^4
    }
    }
\label{eq:Sk-charge}    
\end{eqnarray}
where $\zeta(\mathbf{k})$ is defined by Eq.~(\ref{eq:zeta}),  
\begin{eqnarray}
\overline{S}_{\sf 0}(\mathbf{k})
   = \sum_{mn} 
   \sum_l \sin{(\mathbf{k}\cdot\mathbf{h}_{ml})}  
   \sin{(\mathbf{k}\cdot\mathbf{h}_{nl})} \, ,
\end{eqnarray}
and the dimensionless ratios of parameters are given by 
\begin{eqnarray}
&&\overline{\mathcal{U}} = \frac{\mathcal{U}}{\mathcal{K} \kappa^4} 
   \quad , \quad
\overline{\mathcal{W}} =\frac{\mathcal{W}}{\mathcal{K} \kappa^4} \, .
\label{eq:dimensionless-parameters}
\end{eqnarray}
It is this form of the result, evaluated at the discrete set of wave vectors 
$\{\mathbf{k} \}$ appropriate for a finite-size cluster with given boundary
conditions, which we will fit to simulation results in Section~\ref{subsection:QMC-zeroT}.


Calculating the dynamical structure factor $S_{\sf spin}^{\alpha\beta} (\mathbf{k}, \omega)$ 
for a spin ice means generalising Eq.~(\ref{eq:Skomega-charge}) to take account of neutron 
polarisation, and the local easy axes of spins in a spin ice~\cite{henley05}.   
However the underlying field-theoretical description of the problem ${\mathcal H}^\prime_{\sf U(1)}$ 
[Eq.~(\ref{eq:HA})] is unchanged, and the two results differ only in the way in which the 
contraction of fields on different sublattices
$\langle \mathcal{B}_n(-\mathbf{k}) \mathcal{B}_m(\mathbf{k}) \rangle$ 
contribute to correlation functions.
In a charge ice we simply sum over $m$, $n$ as in Eq.~(\ref{eq:Skomega-charge-definition}). 
In a spin ice we must account for the easy axes which lie along the vectors $\hat{\mathbf{e}}_n$
[Eq.~(\ref{eq:en}] and then calculate the projection of the spin along the axis of interest~\cite{henley05}. 
Thus, the equal time structure factor is
\begin{eqnarray}
&& S^{\alpha \beta}_{\sf spin}(\mathbf{k}, t=0) =  \nonumber \\
&&\quad  \kappa^2 \sum_{mn}\bigg( 
     \hat{\mathbf{e}}_m \cdot \hat{{\bf \alpha}}     \bigg) 
     \bigg( 
     \hat{\mathbf{e}}_n \cdot  \hat{{\bf \beta}}
     \bigg)  \langle \mathcal{B}_n(-\mathbf{k}) \mathcal{B}_m(\mathbf{k}) \rangle 
\end{eqnarray}
where $\hat{{\bf \alpha}}$ and $\hat{{\bf \beta}}$ are unit vectors in the $\alpha$ and $\beta$ 
directions.


Following the same procedure as described above for the charge ice we 
come to the general result for the dynamical structure factor in a spin ice
\begin{eqnarray}
&& S^{\alpha \beta}_{\sf spin}(\mathbf{k}, \omega)  =\frac{\kappa^2}{2} 
   \frac{\mathcal{K}}{\omega(\mathbf{k})} 
  \sum_{mn} \sum_l \sin{(\mathbf{k}\cdot\mathbf{h}_{ml})} 
      \sin{(\mathbf{k}\cdot\mathbf{h}_{nl})} 
      \nonumber \\
&& \quad \times 
      \bigg( 
     \hat{\mathbf{e}}_m \cdot \hat{{\bf \alpha}}     \bigg) 
     \bigg( 
     \hat{\mathbf{e}}_n \cdot  \hat{{\bf \beta}}
     \bigg) 
  \bigg( \delta \left(\omega - \omega_{\lambda}(\mathbf{k}) \right) (1+n_B(\mathbf{k}))
    \nonumber \\
    && \quad + \  \delta \left(\omega + \omega_{\lambda}(\mathbf{k}) \right) n_B(\mathbf{k}) \bigg)
\label{eq:Skomega-general}
\end{eqnarray}


For concreteness, where we come to plot results, we will follow 
the conventions of Fennell {\it et al.}~\cite{fennell09}, who used neutrons 
with polarisation parallel to
\begin{eqnarray}
\bf{n}_{\nu}=(1, -1, 0) \nonumber
\end{eqnarray}
to measure the energy-integrated structure factor 
\mbox{$S_{\sf spin}^{\alpha\beta} (\mathbf{k}, t=0)$}, for 
transfered momentum $\mathbf{k}$ in the $(h,h,l)$ plane. 
We also follow the conventions of Ref.~\onlinecite{fennell09} 
in choosing a coordinate system in which
\begin{eqnarray}
{\bf x} \parallel \mathbf{k} \ , \ {\bf y} \parallel {\bf n}_\nu \times \mathbf{k} \ , \ {\bf z} \parallel {\bf n}_{\nu}, 
\label{eq:axes-definition}
\end{eqnarray}
and consider the ``spin-flip'' channel $S_{\sf spin}^{yy}(\mathbf{k}, \omega)$.
In this convention, the non spin-flip channel measures 
$S_{\sf spin}^{zz}(\mathbf{k}, \omega)$. 


It follows from Eq.~(\ref{eq:Skomega-general}) that the dynamical structure factor in 
the spin-flip channel is given by 
\begin{eqnarray}
&& S_{\sf spin}^{yy} (\mathbf{k}, \omega)
   =\frac{\kappa^2}{2} 
   \frac{\mathcal{K}}{\omega(\mathbf{k})} 
 \sum_{mn} \sum_l \sin{(\mathbf{k}\cdot\mathbf{h}_{ml})} 
      \sin{(\mathbf{k}\cdot\mathbf{h}_{nl})} 
      \nonumber \\
&& \quad \times 
      \left( 
        \frac{\hat{\mathbf{e}}_m \cdot (\mathbf{n}_\nu \times \mathbf{k})}{|(\mathbf{n}_\nu \times \mathbf{k})|} 
     \right) 
     \left( 
         \frac{\hat{\mathbf{e}}_n \cdot (\mathbf{n}_\nu \times \mathbf{k})}{|(\mathbf{n}_\nu \times \mathbf{k})|} 
     \right) 
     \nonumber \\
&& \quad  \times  \big( \delta \left(\omega - \omega_{\lambda}(\mathbf{k}) \right) (1+n_B(\mathbf{k}))
    \nonumber +  \delta \left(\omega + \omega_{\lambda}(\mathbf{k}) \right) n_B(\mathbf{k}) \big)
\label{eq:Skomega-spin}
\end{eqnarray}
and the corresponding zero-temperature, energy-integrated (i.e. equal-time) structure factor is
\begin{eqnarray}
&& S_{\sf spin}^{yy} (\mathbf{k}, t=0)_{T=0}
   =\frac{\kappa^2}{2} 
   \frac{\mathcal{K}}{\omega(\mathbf{k})} 
   \nonumber \\
&& \quad \times \sum_{mn} \sum_l \sin{(\mathbf{k}\cdot\mathbf{h}_{ml})} 
      \sin{(\mathbf{k}\cdot\mathbf{h}_{nl})} 
      \nonumber \\
&& \quad \times 
      \left( 
        \frac{\hat{\mathbf{e}}_m \cdot (\mathbf{n}_\nu \times \mathbf{k})}{|(\mathbf{n}_\nu \times \mathbf{k})|} 
     \right) 
     \left( 
         \frac{\hat{\mathbf{e}}_n \cdot (\mathbf{n}_\nu \times \mathbf{k})}{|(\mathbf{n}_\nu \times \mathbf{k})|} 
     \right) 
\label{eq:Sk-spin}
\end{eqnarray}
Once again, we will make extensive use of this result when comparing with quantum
Monte Carlo simulation. 
At finite temperature we obtain, for the energy integrated structure factor in SF channel
\begin{eqnarray}
&& S_{\sf spin}^{yy} (\mathbf{k}, t=0)
   =\frac{\kappa^2}{2} 
   \frac{\mathcal{K}}{\omega(\mathbf{k})}  \coth{\left( \frac{\omega(\mathbf{k})}{2T} \right)}
   \nonumber \\
&& \quad \times \sum_{mn} \sum_l \sin{(\mathbf{k}\cdot\mathbf{h}_{ml})} 
      \sin{(\mathbf{k}\cdot\mathbf{h}_{nl})} 
      \nonumber \\
&& \quad \times 
      \left( 
        \frac{\hat{\mathbf{e}}_m \cdot (\mathbf{n}_\nu \times \mathbf{k})}{|(\mathbf{n}_\nu \times \mathbf{k})|} 
     \right) 
     \left( 
         \frac{\hat{\mathbf{e}}_n \cdot (\mathbf{n}_\nu \times \mathbf{k})}{|(\mathbf{n}_\nu \times \mathbf{k})|} 
     \right).
\label{eq:Sk-spin}
\end{eqnarray}


Where neutron scattering is performed with unpolarized neutrons, experiments 
measure an average over different components of the dynamical structure factor 
\begin{eqnarray}
I ({\bf k}, \omega) 
   \propto \sum_{\alpha \beta} 
  \left( \delta_{\alpha \beta}-\frac{k_{\alpha} k_{\beta}}{k^2} \right)
  S^{\alpha \beta}_{\sf spin} (\mathbf{k}, \omega)  
\label{eq:Ikomega-unpolarized}
\end{eqnarray}
This is the result plotted where we illustrate photon dispersions in
Fig.~\ref{fig:ghostlyphoton} and Fig.~\ref{fig:inelastic-unpolarised}.
The corresponding quasi-elastic (energy integrated) form 
of Eq.~(\ref{eq:Ikomega-unpolarized}) is given by
\begin{eqnarray}
I ({\bf k}) 
    \propto \sum_{\alpha \beta} 
    \left( \delta_{\alpha \beta}-\frac{k_{\alpha} k_{\beta}}{k^2} \right)
    S^{\alpha \beta}_{\sf spin} (\mathbf{k}, t=0) .
\label{eq:Ik-unpolarized-quasielastic}
\end{eqnarray}


It is important to note the predictions for quasi-elastic neutron scattering 
$S_{\sf spin}^{yy} (\mathbf{k}, \omega)$~[Eq.~(\ref{eq:Skomega-spin})]
and 
$I ({\bf k}, \omega)$~[Eq.~(\ref{eq:Ikomega-unpolarized})], 
and their energy-integrated counterparts
\mbox{$S_{\sf spin}^{yy} (\mathbf{k}, t=0)$~[Eq.~(\ref{eq:Sk-spin})]}
and 
$I ({\bf k})$~[Eq.~(\ref{eq:Ik-unpolarized-quasielastic})] 
{\it only} include contributions from the low-energy photon excitations of a quantum spin ice.
In a real quantum spin ice material, their would also be contributions at higher energy 
from gapped ``electric charges'', and magnetic monopole excitations.
These are not treated in the present theory.


\section{ ``Electromagnetism'' in a Quantum Spin Ice at $T=0$}
\label{section:zero-T}


The arguments presented in Section~\ref{subsection:em} explain how a spin liquid state 
with correlations described by an effective electromagnetism can arise in a quantum 
spin ice, but stop short of offering proof that this happens in any real material or 
microscopic model.  
In what follows, we validate our use of Gaussian 
electromagnetism $\mathcal{H}_{\sf U(1)}$ [Eq.~(\ref{eq:Hem})] as a description 
of the quantum ice model $\mathcal{H}_{\mu}$ [Eq.~(\ref{eq:Hmu})], 
by making explicit comparison with the results of zero-temperature 
quantum Monte Carlo simulation.  


However before considering results on a lattice, it is useful to ask how 
correlations in quantum spin ice might differ from those in a classical spin ice, 
within a simple continuum field theory.
This is considered in Section~\ref{subsection:continuum-zeroT}.  
We then turn to simulation of the lattice model $\mathcal{H}_{\mu}$ [Eq.~(\ref{eq:Hmu})] 
in Section~\ref{subsection:QMC-zeroT}, demonstrating that the lattice field theory 
$\mathcal{H}_{\sf U(1)}$ [Eq.~(\ref{eq:Hem})] provides an excellent {\it quantitative} 
description of the results for $S_{\sf spin}^{\alpha\beta} (\mathbf{k}, t=0)$.
In Section~\ref{subsection:Skomega} we use the same lattice field theory 
to make predictions for the magnetic photon excitations which could be observed in 
inelastic neutron scattering experiments.
Finally in Section~\ref{subsection:c} we use the finite-size scaling 
of ground state energies in simulation to put an absolute scale on the speed of 
light $c$ associated with these magnetic photons.  
Throughout this analysis we set $\hbar=1$, restoring dimensional factors of $\hbar$ 
only where we quote results for the speed of light.


\begin{figure*}
\centering
\includegraphics[width=0.9\textwidth]{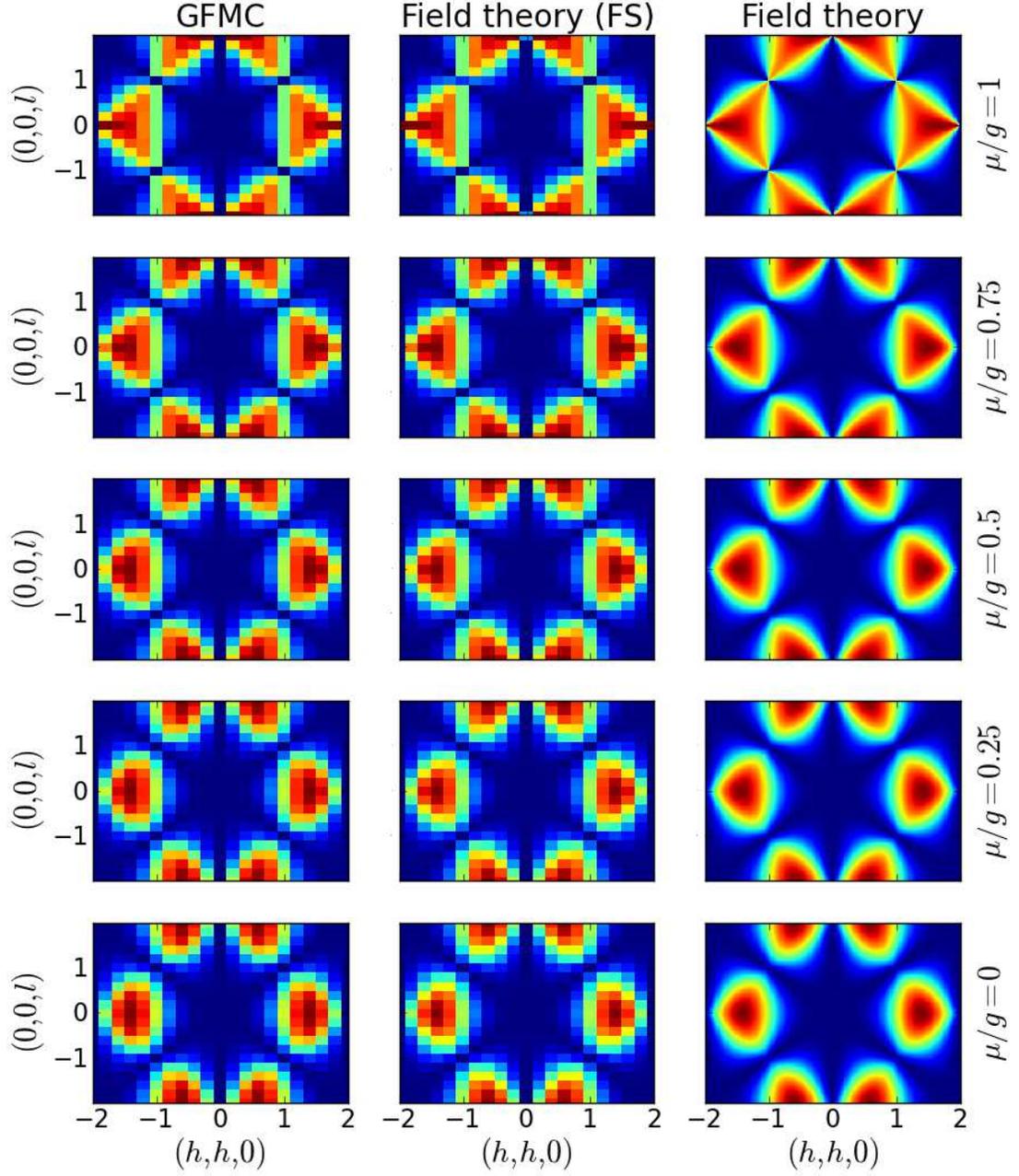}
\caption{
(Color online).  
Comparison between the predictions of the lattice field theory 
${\mathcal H}^\prime_{\sf U(1)}$~[Eq.~(\ref{eq:HA})]
and quantum Monte Carlo simulation of the microscopic model 
${\mathcal H}_\mu$~[Eq.~(\ref{eq:Hmu})], 
for a quantum charge ice at $T=0$.  
{\bf First column~:} 
equal-time structure factor $S_{\sf charge} ({\mathbf k}, t=0)$ calculated using Green's function 
Monte Carlo (GFMC) simulation of a 2000-site cubic cluster, for a range of $\mu$ ranging from 
$\mu/g = 1$ (RK point) to $\mu/g = 0$ (quantum ice).
{\bf Second column~:} 
best fit of the finite-size (FS) prediction of the lattice field theory to simulation, 
following Eq.~(\ref{eq:Sk-charge}).
There is excellent, quantitative, agreement between theory and simulation for
all values of $\mu/g$.
{\bf Third column~:}
prediction of lattice field theory in the thermodynamic limit, for parameters obtained
from fits to simulation.
}
\label{fig:evolutionSk-charge}
\end{figure*}


\subsection{Structure factors within continuum theory}
\label{subsection:continuum-zeroT}


The long-wavelength properties of a quantum $U(1)$ liquid are well-described by a 
continuum field theory of the form considered in Ref.~\onlinecite{moessner03}
\begin{eqnarray}
\mathcal{S}_{\sf eff} 
   = \frac{1}{8\pi}\int dt d^3 \mathbf{r} 
   &\bigg[& \mathbf{\mathcal E}({\bf r})^2 - c^2 \mathbf{\mathcal B}({\bf r})^2 
   \nonumber \\
   && \quad - \rho_{\sf c}\ \bigg( \nabla  \times \mathbf{\mathcal B}({\bf r}) \bigg)^2 \bigg]
\label{eq:Seff}
\end{eqnarray}
This therefore provides a convenient starting point for discussing the evolution of 
spin correlations in quantum spin ice.   
We emphasise that such a theory can be derived as a 
continuum limit of ${\mathcal H}^\prime_{\sf U(1)}$ [Eq.~(\ref{eq:HA})]~\cite{hermele04}.  
And where we go on to make comparison with quantum Monte Carlo simulation in 
Section~\ref{subsection:QMC-zeroT}, we will use the appropriate results on a lattice, 
i.e. Eq.~(\ref{eq:Sk-charge}) and Eq.~(\ref{eq:Sk-spin}).


For $\rho_{\sf c} = 0$, $\mathcal{S}_{\sf eff}$ reduces to the familiar 
Maxwell action of quantum electromagnetism.
Crucially, this action supports photon excitations with dispersion $\omega(\mathbf{k}) = c |\mathbf{k}|$.   
The additional term $\rho_{\sf c}\ ( \nabla  \times \mathbf{\mathcal B}({\bf r})  )^2$ is invariant 
under gauge transformations  
$ \mathbf{\mathcal A}({\bf r}) \to  \mathbf{\mathcal A}({\bf r}) + \nabla \phi ({\bf r})$, 
and is an {\it irrelevant perturbation} in the RG sense~\cite{hermele04}.
However it introduces a new length scale into the problem 
\begin{eqnarray}
\lambda_{\sf c} = 2 \pi \frac{\sqrt{\rho_{\sf c}}}{c} \, ,
\label{eq:lambda}
\end{eqnarray}
which controls the curvature of the photon dispersion
\begin{eqnarray}
   \omega(\mathbf{k})=c |\mathbf{k}| \sqrt{1 +\left( \frac{\lambda_c}{2 \pi} \right)^2 |\mathbf{k}|^2} \, ,
\end{eqnarray}
and has an important impact on how correlations evolve as a function of distance.


The role of $\lambda_c$ can most easily be understood in the limit $c \to 0$, 
where correlations of $\mathbf{\mathcal B}({\bf r})$ are controlled entirely by~$\rho_{\sf c}$.
Precisely this limit is realised in the microscopic model $\mathcal{H}_{\mu}$ 
[Eq.~(\ref{eq:Hmu})] at the exactly soluble ``RK'' point \mbox{$\mu = g$}. 
At the RK point, {\it all} ice configurations are degenerate, and 
the photons have dispersion 
$\omega(\mathbf{k}) = \sqrt{\rho_{\sf c}} |\mathbf{k}|^2$~[\onlinecite{moessner03,hermele04}].  
Correlations of the magnetic field 
\begin{eqnarray}
   C^{\mathbf{\mathcal B}}_{\mu \nu}(\mathbf{k}) 
   = \langle 
   \mathbf{\mathcal B}_{\mu} (-\mathbf{k})  
   \mathbf{\mathcal B}_{\nu} (\mathbf{k}) 
   \rangle
\end{eqnarray}
can be calculated from Eq.~(\ref{eq:Seff}), and for $c=0$ these 
behave as
\begin{eqnarray}
C^{\mathbf{\mathcal B}}_{\mu \nu}(\mathbf{k}) 
   \approx 
   \frac{8 \pi^4}{\sqrt{\rho_{\sf c}}} 
   \left( 
   \delta_{\mu \nu} - \frac{k_{\mu} k_{\nu}}{k^2} 
   \right)
\label{eq:pinch-point}
\end{eqnarray}
exhibiting the pinch-point singularities characteristic of the 
``Coulombic'', classical $U(1)$ liquid phase~\cite{huse03,henley05,youngblood80}.  
On Fourier transform,  Eq.~(\ref{eq:pinch-point}) corresponds to dipolar 
correlations in a three-dimensional space
\begin{eqnarray}
C^{\mathbf{\mathcal B}}_{\mu \nu}(\mathbf{r}) 
\propto \frac{3 r_\mu r_\nu/r^2 -  \delta_{\mu \nu}}{r^3}
\label{eq:3Ddipolar}
\end{eqnarray}


The quantum $U(1)$ liquid phase, with its linearly dispersing photons, is 
stabilised by the emergence of finite value of the speed of light $c$ for 
$\mu < g$~[\onlinecite{moessner03,hermele04,shannon12}].  
In this case, we find
\begin{eqnarray}
 C^{\mathbf{\mathcal B}}_{\mu \nu}(\mathbf{k})
   = \frac{8 \pi^4 k}{c \sqrt{ 1+\left( \frac{\lambda_{\sf c} k}{2 \pi} \right)^2}} 
       \left( \delta_{\mu \nu} - \frac{k_{\mu} k_{\nu}}{k^2} \right).
\label{eq:no-pinch-point}
\end{eqnarray}
[cf. \onlinecite{hermele04,castro-neto06}].  
For wavelengths $\lambda \ll \lambda_{\sf c}$, Eq.~(\ref{eq:no-pinch-point})
reduces to Eq.~(\ref{eq:pinch-point}), and the system exhibits ``classical'' 
dipolar correlations of the form Eq.~(\ref{eq:3Ddipolar}).
However for long wavelengths  $\lambda \gg \lambda_c$ the additional 
factor of $k$ in the numerator of Eq.~(\ref{eq:no-pinch-point}) ``hollows out'' 
the pinch point singularities.  
In this limit, $r \gg \lambda_{\sf c}$, Eq.~(\ref{eq:no-pinch-point}) corresponds 
to dipolar correlations in a {\it four-dimensional} space
\begin{eqnarray}
C^{\mathbf{\mathcal B}}_{\mu \nu}(\mathbf{r}) 
\propto \frac{2 r_\mu r_\nu/r^2 -  \delta_{\mu \nu}}{r^4}  \, ,
\label{eq:4Ddipolar}
\end{eqnarray}
the additional dimension arising because of fluctuations in time~\cite{hermele04, castro-neto06}.
We therefore associate $\lambda_{\sf c}$ with the length-scale over which
the system crosses over from ``classical'' ice correlations, decaying as 
$1/r^3$, to ``quantum'' ice correlations decaying as $1/r^4$.


The length-scale $\lambda_{\sf c}$ will also play an important role where we compare the 
predictions of field theory with simulation of  microscopic model 
${\mathcal H}_\mu$~[Eq.~(\ref{eq:Hmu})] as a function of $\mu$.
We can gain some insight into the $\mu$ dependence of $\lambda_{\sf c}$, from 
degenerate perturbation theory about the RK 
point~\cite{hermele04,sikora09,sikora11,shannon12}. 
We find that $c^2 \sim (g-\mu)$, while $\rho_c \approx const.$, and   
it follows from Eq.~(\ref{eq:lambda}) that $\lambda_c$ diverges as 
$\lambda_c \sim 1/\sqrt{g-\mu}$.
Exactly at the RK point, where $g=\mu$, $\lambda_c$ is infinite and correlations have 
the classical form Eq.~(\ref{eq:3Ddipolar}) at {\it all} length scales, as expected.
However, as we move away from the RK point into the quantum liquid phase
for $\mu/g < 1$, there will be a progressive evolution of correlations from 
classical (pinch points) at short distances to quantum (no pinch points)
at long distances.  
This expectation is born out by quantum Monte Carlo simulations, 
described below. 


For the purposes of these simulations, $\lambda_{\sf c}$ also sets the minimum 
size of cluster which is needed to capture quantum effects at a given $\mu$. 
At \mbox{$\mu=0$} we find that \mbox{$\lambda_{\sf c} \approx 0.8 a_0$}, and 
hence a cluster of linear dimension \mbox{$L=5 a_0 \ (N=2000)$} is comfortably 
big enough to observe the quantum spin liquid phase~\cite{shannon12}.


\subsection{Comparison with Quantum Monte Carlo simulation}
\label{subsection:QMC-zeroT}



\begin{figure}
\includegraphics[width=0.4\textwidth]{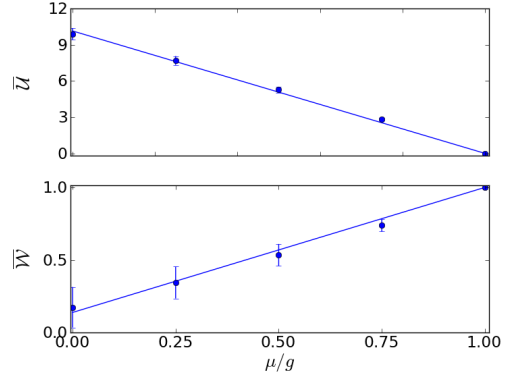}
\caption{
Parameters of lattice field theory ${\mathcal H}^\prime_{\sf U(1)}$~[Eq.~(\ref{eq:HA})] 
as a function of $\mu/g$, from comparison with quantum 
Monte Carlo simulation of spin correlations in the microscopic model 
${\mathcal H}_\mu$~[Eq.~(\ref{eq:Hmu})]
at $T=0$.
The dimensionless combinations of parameters 
$\overline{\mathcal U}$~and~$\overline{\mathcal W}$~[Eq.~(\ref{eq:dimensionless-parameters})]
were obtained by fitting the predictions of the lattice field theory 
$S_{\sf charge} (\mathbf{k},t=0)_{T=0}$~[Eq.~(\ref{eq:Sk-charge})] 
to simulation results at fixed $\mu/g$ [cf. Fig~\ref{fig:evolutionSk-charge}].}
\label{fig:fit-parameters}
\end{figure}


\begin{figure*}
\centering
\includegraphics[width=0.9\textwidth]{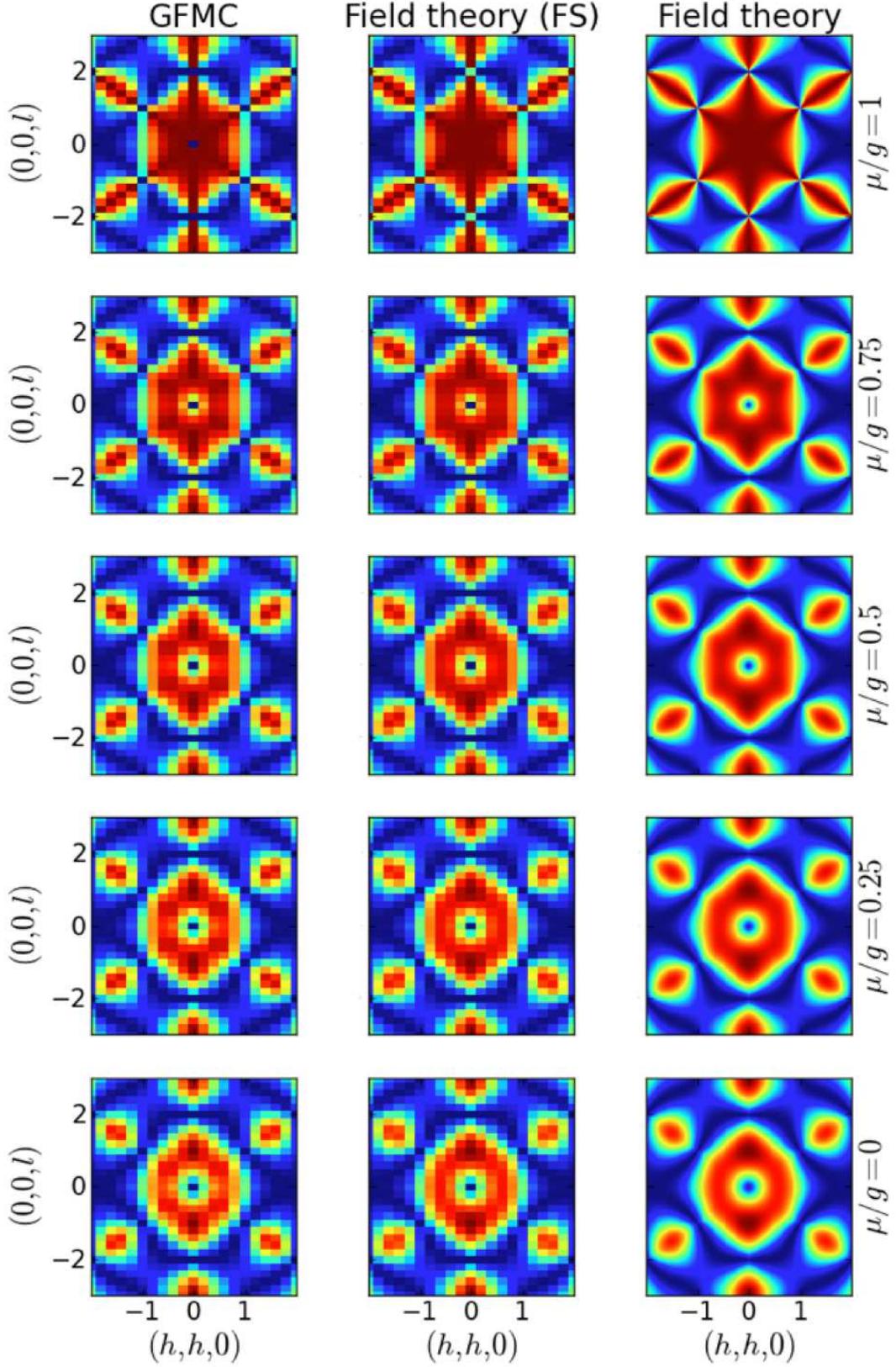}
\caption{
(Color online).  
Comparison between the predictions of the lattice field theory 
${\mathcal H}^\prime_{\sf U(1)}$~[Eq.~(\ref{eq:HA})]
and quantum Monte Carlo simulation of the microscopic model 
${\mathcal H}_\mu$~[Eq.~(\ref{eq:Hmu})], 
for a quantum spin ice at $T=0$.  
First column~: 
equal-time structure factor $S_{\sf spin}^{yy} ({\mathbf k}, t=0)$, as measured in neutron
scattering by Fennell~{\it et al.}~\cite{fennell09}, calculated using Green's function 
Monte Carlo (GFMC) simulation of a 2000-site cubic cluster for parameters ranging from $\mu/g = 1$ 
(RK point) to $\mu/g = 0$ (quantum ice).
Second column~: 
best fit of the finite-size (FS) prediction of the lattice field theory to simulation, 
following Eq.~(\ref{eq:Sk-spin}).
There is excellent, quantitative, agreement between theory and simulation for
all values of $\mu/g$.
Third column~: 
prediction of lattice field theory in the thermodynamic limit, for parameters obtained
from fits to simulation.
}
\label{fig:evolutionSk-spin}
\end{figure*}


We now turn to zero-temperature quantum Monte Carlo simulation of
the microscopic model ${\mathcal H}_\mu$ [Eq.~(\ref{eq:Hmu})].  
We have previously argued that this model supports a quantum 
$U(1)$ liquid ground state for a range of parameters 
$-0.5g < \mu < g$ --- cf. Fig.~\ref{fig:phase-diagram} 
and \onlinecite{shannon12}. 
In this earlier work, evidence for the ground state phase diagram was taken 
from the finite-size scaling of energy spectra.  
Our main tool here will be the equal time structure factor $S (\mathbf{k},t=0)$, 
calculated from simulation, and from the lattice field theory 
${\mathcal H}^\prime_{\sf U(1)}$ [Eq.~(\ref{eq:HA})].  
These two independent calculations are found to be in excellent, quantitative
agreement, confirming the conclusions of \onlinecite{shannon12}.
Making a direct comparison between the field theory and simulation also serves 
to put the field theory on a quantitative footing, providing information about the 
evolution of the parameters of the field theory as a function of the microscopic 
parameter $\mu$.


Simulations were performed using a Green's Function Monte Carlo (GFMC) 
technique based on the statistical sampling of ice configurations.
This sampling is weighted using a variational estimate of the ground state wave 
function, which is optimised in a separate variational Monte Carlo (VMC) 
calculation.
In this sense, GMFC can be thought of a systematic method of improving
upon variational calculations. 
There is no sign problem associated with ${\mathcal H}_\mu$, since {\it all} 
of its off-diagonal matrix elements are equal to $0$ or $-g$, with $g > 0$.
Where simulations converge, the results obtained are numerically exact.
Our implementation of VMC and GFMC calculations for quantum 
ice~\cite{shannon12} exactly parallels our earlier work on the quantum dimer 
model on a diamond lattice~\cite{sikora09,sikora11}, with correlation functions
calculated using techniques described in Ref. \onlinecite{calandra98}.  
We refer the interested reader to these papers for further details of the method.


In the left-hand column of Fig.~\ref{fig:evolutionSk-charge}, we present GFMC 
simulation results for the equal-time correlations in a quantum charge ice 
$$
S_{\sf charge} ({\mathbf k}, t=0)_{T=0} = \langle n ({\mathbf k}) n (-{\mathbf k}) \rangle_{T=0}
$$
Simulations were performed for a 2000-site cubic cluster possessing the 
full symmetry of the lattice, for parameters 
\mbox{$\mu/g = 1$, $0.75$, $0.5$, $0.25$, $0$}.


The classical, dipolar correlations at the RK point $\mu/g = 1$ are clearly 
visible as sharp ``bow-tie'' motifs in \mbox{$S_{\sf charge} ({\mathbf k}, t=0)$}, 
centred on pinch-points at ${\mathbf k} = (1,1,1)$, etc.
As expected, these pinch points are progressively eliminated as $\mu/g \to 0$, and 
quantum effects come to dominate the long lengthscale physics of the problem.
This erosion of the pinch points is accompanied by a gradual redistribution 
of spectral weight, with high intensity regions evolving from a triangular into an 
oval shape.


In the central column of Fig.~\ref{fig:evolutionSk-charge}, we present the best fit
to simulation results obtained from the lattice field theory.
Fits were made using the result 
$S_{\sf charge} ({\mathbf k}, t=0)_{T=0}$~[Eq.~(\ref{eq:Sk-charge})], 
evaluated for the same 2000-site cluster, as a function of the two dimensionless 
parameters $\overline{\mathcal{U}}$ and  $\overline{\mathcal{W}}$  
[Eq.~(\ref{eq:dimensionless-parameters})].
The two results are indistinguishable by eye, and differ maximally by a few 
percent, for values of ${\mathbf k}$ close to the Brillouin zone boundary.   
The quality of these fits implies that they can be used to accurately parameterize the 
lattice field theory ${\mathcal H}^\prime_{\sf U(1)}$ [Eq.~(\ref{eq:HA})], 
and the values of $\overline{\mathcal{U}}$ and  $\overline{\mathcal{W}}$ obtained 
are shown in Fig.~\ref{fig:fit-parameters}.    
We note that the values obtained at the RK point,  
$\overline{\mathcal{U}} =0$ and $\overline{\mathcal W} = 1$, 
are uniquely determined by the known form of correlations 
within the classical ice states~\cite{henley05}. 
A separate evaluation of the speed of light $c \propto \sqrt{\mathcal UK}$
from finite size scaling of the ground state energy is given in 
Section~\ref{subsection:c} below.


In Fig.~\ref{fig:evolutionSk-spin} we show equivalent results for the equal-time 
structure factor of a spin ice
$$
S_{\sf spin}^{yy} (\mathbf{k}, t=0)_{T=0} 
   = \langle {\mathsf S}^y (\mathbf{k}) {\mathsf S}^y (\mathbf{-k}) \rangle_{T=0} 
$$
in the spin-flip channel considered by Fennell {\it et al.}~\cite{fennell09}.
Superficially, these results look very different to those presented in 
Fig.~\ref{fig:evolutionSk-charge}.   
This is because the local easy axis is different for each of the four sublattices, leading to 
a staggering of correlations not present in the charge ice problem. 
However the information content of the two structure factors is exactly the same.


At the RK point $\mu/g=1$, correlations are classical, and $S_{\sf spin}^{yy} (\mathbf{k}, t=0)$
exhibits a characteristic ``snow flake'' motif in the $(h,h,l)$ plane, also seen in neutron 
scattering experiments on Ho$_2$Ti$_2$O$_7$ [\onlinecite{fennell09}].  
Pinch point singularities are clearly visible at the reciprocal lattice vectors 
${\mathbf k} = (1,1,1)$, etc.


Once again, these pinch points are progressively eroded as the system is tuned 
away from the RK point into the quantum spin-liquid regime for $\mu/g < 1$.  
Probably the most striking change, however, occurs at $\mathbf{k} = (0,0,0)$.  
Here, for a classical spin ice 
$$
S_{\sf spin}^{yy} (\mathbf{k}\to \mathbf{0}, t=0)_{T=0} \to const.
$$
However, in a quantum spin ice,
$$
S_{\sf spin}^{yy} (\mathbf{k}=\mathbf{0}, t=0)_{T=0} \equiv 0 \, ,
$$
and spectral weight is progressively excavated from the region of 
reciprocal space around $\mathbf{k} = (0,0,0)$ for $\mu/g < 1$.
This has important consequences for the evolution of correlations 
at finite temperature, discussed in Section~\ref{subsection:SkT}
and for the uniform magnetic susceptibility, 
discussed in Section~\ref{subsection:thermodynamic}.  


We wish to emphasise that the results shown in Fig.~\ref{fig:evolutionSk-spin} 
are {\it not} the outcome of separate simulations of a quantum spin ice.
They are taken from the same simulations of the quantum ice model 
${\mathcal H}_\mu$ [Eq.~(\ref{eq:Hmu})], recast in the coordinates 
appropriate for a spin ice.
It follows that the parameters obtained from fits to field theory at finite size 
are exactly the same as those for a charge ice, given in Fig.~\ref{fig:fit-parameters}.    


\subsection{Seeing the light : photons and inelastic neutron scattering}
\label{subsection:Skomega}



\begin{figure}
\includegraphics[width=0.4\textwidth]{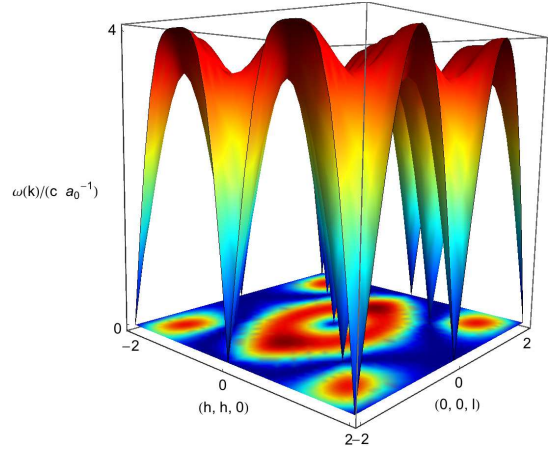}
\caption{
(Color online).  
Relationship between the dispersion of the magnetic photon excitation 
$\omega(\mathbf{k})$ [Eq.~(\ref{eq:omegak})], and the equal time structure factor 
$S^{yy}_{\sf spin}(\mathbf{k}, t=0)$ [Eq.~(\ref{eq:Sk-spin})] in a quantum 
spin ice.  
The photon dispersion $\omega(\mathbf{k})$ in the $(h,h,l)$ plane is plotted above the 
corresponding equal-time structure factor, demonstrating 
how the photon disperses out of the (suppressed) pinch points 
at reciprocal lattice vectors.  
Note that the intensity of the scattering $S^{yy}_{\sf spin}(\mathbf{k}, t=0) \to 0$ 
where $\omega(\mathbf{k}) \to 0$ [Eq.~(\ref{eq:vanishingSkomega})].  
Results were calculated within the lattice field theory [Eq.~(\ref{eq:HA})]
for $\mathcal{W}=0$, with energy measured in units such that $\hbar=1$.
}
\label{fig:dispersion+S(q)}
\end{figure}


\begin{figure*}
\centering
\includegraphics[width=0.9\textwidth]{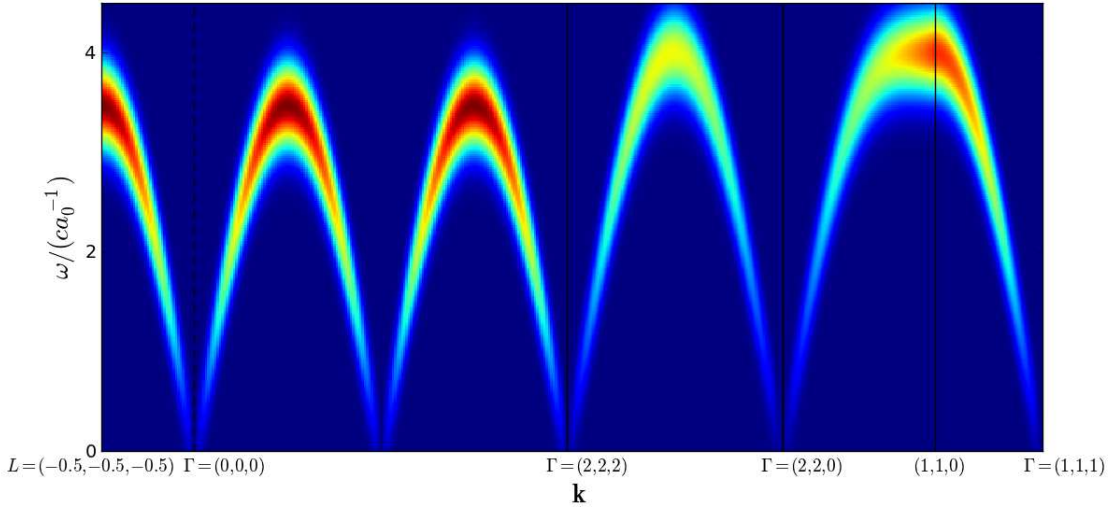}
\caption{
(Color online).  
Ghostly magnetic ``photon'' excitation as it might appear in an inelastic 
neutron scattering experiment on a quantum spin ice realising a quantum 
ice ground state, for a series of cuts along high symmetry directions in reciprocal space.  
The prediction of the lattice field theory ${\mathcal H}^\prime_{\sf U(1)}$~[Eq.~(\ref{eq:HA})] 
for inelastic scattering by unpolarized neutrons, $I(\mathbf{k},\omega)$ [Eq.~(\ref{eq:Ikomega-unpolarized})]
has been convoluted with a Gaussian of variance $0.3 \ c\ a_0^{-1}$ to represent the 
finite energy resolution of the instrument. 
The intensity of scattering vanishes as $\omega \to 0$, and is strongest at high energies.
Energy is measured in units such that $\hbar=1$, and the photon dispersion 
calculated for $\mathcal{W} = 0$.  
}
\label{fig:inelastic-unpolarised}
\end{figure*}


\begin{figure}
\centering
\includegraphics[width=0.9\columnwidth]{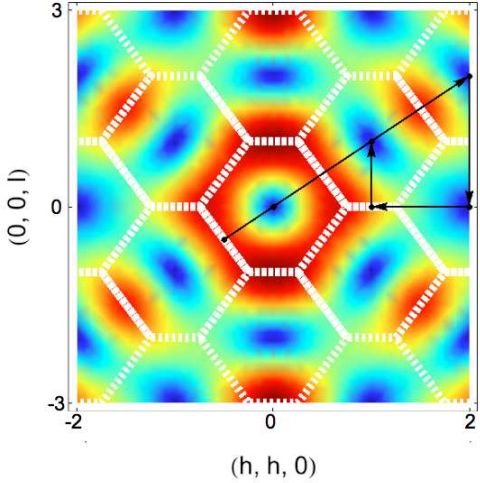}
\caption{
(Color online).  
Prediction of the lattice field theory ${\mathcal H}^\prime_{\sf U(1)}$~[Eq.~(\ref{eq:HA})]
for quasi-elastic neutron scattering performed using unpolarised neutrons, for comparison 
with Fig.~\ref{fig:inelastic-unpolarised}. 
Results for $I(\mathbf{k})$ are taken from Eq.~(\ref{eq:Ik-unpolarized-quasielastic}), 
and calculated for $\mathcal{W}=0$.  
The path within the $[h,h,l]$ plane used for plotting the photon dispersion 
in Fig.~\ref{fig:inelastic-unpolarised} is shown using unbroken black arrows, 
with Brillouin zone boundaries marked as dashed white lines. 
}
\label{fig:quasi-elastic-unpolarised}
\end{figure}


Inelastic neutron scattering provides a direct method of measuring the dynamical 
structure factor $S_{\sf spin}^{\alpha\beta}(\mathbf{k}, \omega)$, 
and so of resolving photon excitations in a quantum spin ice.  
These photons disperse linearly out of those reciprocal lattice vectors
where pinch points are observed in quasi-elastic scattering experiments.
However, since these experiments measure the energy integral of the dynamical 
structure factor,  the suppression of pinch points in a quantum spin ice at $T=0$ has
important implications for the observation of its photon excitations.
Specifically, for non-interacting photons, the suppression of energy-integrated structure 
factor must imply the suppression of the weight in the photon peak itself.
This is illustrated in Fig.~\ref{fig:dispersion+S(q)}.


To see how this works, we consider the result for the dynamical structure factor 
in a quantum spin ice $S_{\sf spin}^{\alpha\beta}(\mathbf{k}, \omega)$~[Eq.~(\ref{eq:Skomega-general})], 
in the (physically relevant) limit where $\mathcal{W}=0$.
In this case weight in the photon peak is determined by the ratio
$$ 
\frac{\overline{S}^{\alpha\beta}_0(\mathbf{k})}{\omega(\mathbf{k})}
$$
where, 
\begin{eqnarray}
\overline{S}^{\alpha\beta}_0(\mathbf{k}) &=& 
      \sum_{mn} \sum_l \sin{(\mathbf{k}\cdot\mathbf{h}_{ml})} 
      \sin{(\mathbf{k}\cdot\mathbf{h}_{nl})} 
         \nonumber \\ 
&& \quad  \times \bigg( 
     \hat{\mathbf{e}}_m \cdot \hat{{\bf \alpha}} \bigg) 
     \bigg( 
     \hat{\mathbf{e}}_n \cdot  \hat{{\bf \beta}} 
     \bigg) 
 \label{eq:S0alphabeta}
\end{eqnarray}
and
\begin{eqnarray}
\omega_{\lambda}(\mathbf{k})=\sqrt{\mathcal{U} \mathcal{K}} \zeta_{\lambda}(\mathbf{k}).
\end{eqnarray}
We can use the spectral representation of $\uuline{Z}(\mathbf{k})$ [Eq.~(\ref{eq:spectral-decomp-Z})] 
to write 
\begin{eqnarray}
&& \sin{(\mathbf{k}\cdot\mathbf{h}_{ml})} 
      \sin{(\mathbf{k}\cdot\mathbf{h}_{nl})} 
      \nonumber\\
&& \qquad =  \frac{1}{4} \sum_{\lambda=1}^{4} 
      \frac{\omega_{\lambda}(\mathbf{k})^2}{\mathcal{KU}} 
      \ \eta_{m \lambda} (\mathbf{k}) \eta_{\lambda n}^{\ast}(\mathbf{k}) 
      \label{eq:sinsin}
\end{eqnarray}
Since the only contributions to the RHS of Eq.~(\ref{eq:sinsin})  
come from the two dispersing modes $\lambda =1$, $2$, 
[cf.~Eq.~(\ref{eq:zero-modes})], Eq.~(\ref{eq:S0alphabeta}) simplifies to
\begin{eqnarray}
\overline{S}^{\alpha\beta}_0(\mathbf{k}) 
&=&  \frac{1}{4}  \frac{\omega(\mathbf{k})^2 }{\mathcal{KU}} 
      \sum_{\lambda=1}^{2} \sum_{mn}
      \ \eta_{m \lambda} (\mathbf{k}) \eta_{\lambda n}^{\ast}(\mathbf{k}) 
      \nonumber \\
&&  \quad  \times  
     \bigg( 
     \hat{\mathbf{e}}_m \cdot \hat{{\bf \alpha}}     
     \bigg) 
     \bigg( 
     \hat{\mathbf{e}}_n \cdot  \hat{{\bf \beta}} 
     \bigg) 
\label{eq:Sknumerator}
\end{eqnarray}
Expanding in the first Brillouin zone, for $\mathbf{k} \approx 0$, we find  
$$
\sum_{mn} \eta_{m \lambda} (\mathbf{k}) \eta_{\lambda n}^{\ast}(\mathbf{k}) 
\bigg( 
     \hat{\mathbf{e}}_m \cdot \hat{{\bf \alpha}}     
     \bigg) 
     \bigg( 
     \hat{\mathbf{e}}_n \cdot  \hat{{\bf \beta}} 
     \bigg) 
\approx \frac{1}{3}
$$
for $\alpha = \beta = y,z$ and zero otherwise.  
It follows that 
\begin{eqnarray}
&& S^{yy}_{\sf spin}(\mathbf{k} \approx \mathbf{0}, \omega \approx 0) \nonumber \\
&& \quad = S^{zz}_{\sf spin}(\mathbf{k} \approx \mathbf{0}, \omega \approx 0) \nonumber \\
   && \qquad \propto \omega(\mathbf{k})  \ \delta(\omega-\omega(\mathbf{k})).
\label{eq:vanishingSkomega}
\end{eqnarray}


Therefore at low energies, in the first Brillouin zone, inelastic neutron scattering experiments will 
resolve the magnetic photon excitation as a ghostly, linearly dispersing peak, with 
intensity vanishing as $I \propto \omega(\mathbf{k})$, as noted in [\onlinecite{savary12}].   
However at higher energies and in other Brillouin zones, the momentum dependence of
$\eta_{m \lambda} (\mathbf{k}) \eta_{\lambda n}^{\ast}(\mathbf{k})$ in Eq.~(\ref{eq:Sknumerator})
will lead to a significant variation in the intensity of the signal at fixed $\omega$. 
This behaviour is illustrated in Fig.~\ref{fig:inelastic-unpolarised}, where we have plotted
the intensity of scattering $I(\mathbf{k},\omega)$ [Eq.~(\ref{eq:Ikomega-unpolarized})] 
for an experiment performed using {\it unpolarised} neutrons.
The corresponding quasi elastic scattering, and the path within the $[h,h,l]$ plane, 
are shown in Fig.~(\ref{fig:quasi-elastic-unpolarised}).


The phenomenology of this photon excitation stands in stark contrast to 
conventional antiferromagnets, whose linearly-dispersing spin-wave excitations have the 
greatest intensity approaching the zero-energy magnetic Bragg peak associated with magnetic order.
The difference between these two problems stems from the fact that the photon
is a quantised excitation of $\mathcal{A}$, while neutron scattering measures
correlations of $\mathcal{B}$.
The lattice curl needed to relate one to the other introduces additional factors 
of $\zeta_{\lambda}(\mathbf{k})$ in 
$S_{\sf spin}^{\alpha\beta}(\mathbf{k}, \omega)$~[Eq.~(\ref{eq:Skomega-general})], 
which leads to the suppression of spectral weight at low energies.


A better point of comparison, in fact, is the scattering of neutrons by {\it real} photons in
models of a hot early universe\cite{gould93}.
In both cases, photons are associated with a periodically fluctuating magnetic field, transverse 
to the direction of their propagation.
And in both cases neutrons scatter inelastically from these locally fluctuating 
magnetic fields.
In a spin ice, this scattering can occur in both the spin-flip (SF) channel, in which case 
there is a transfer of angular momentum to the sample, and in the non spin-flip (NSF)
channel (cf. Fig.~\ref{fig:inelastic-unpolarised}, Fig.~\ref{fig:quasi-elastic-unpolarised} 
and Fig.~\ref{fig:SkT}).  
It is also interesting to note that the same phenomenology of linearly-dispersing excitations, 
with a vanishing spectral weight at long wavelengths, is encountered 
in quantum spin nematics~\cite{penc11}.
In this case, low-energy spin fluctuations are controlled by a time derivative
of the underlying nematic order parameter~\cite{smerald-preprint}, and so vanish for $\omega \to 0$. 


\subsection{Estimating the speed of light}
\label{subsection:c}

\begin{figure}
\includegraphics[width=0.43\textwidth]{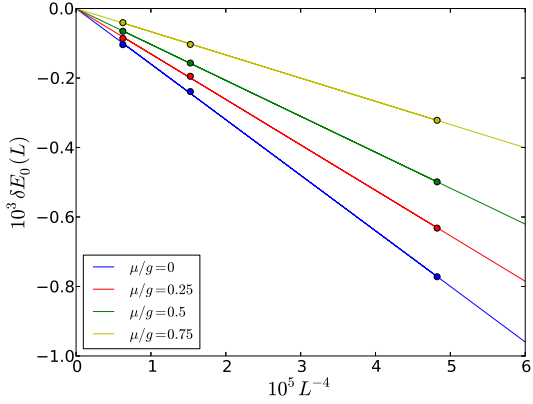}
\caption{
(Color online).
Finite-size scaling of the finite-size correction to the ground state energy per site 
$\delta E_0/N$ found in quantum Monte Carlo simulations of the quantum ice model 
${\mathcal H}_\mu$ [Eq.~(\ref{eq:Hmu})].
Results are shown for cubic clusters of $N=432$, $1024$ and $2000$ sites, for
parameters $\mu/g=0$, $0.25$, $0.5$, $0.75$, as function of the linear dimension
of the system $L = (a_0/2) (N/2)^\frac{1}{3}$.  
The fact that $\delta E_0/N \sim 1/L^4$ implies the existence of a linearly dispersing 
excitation --- the ``photon'' of the underlying lattice gauge theory.
}
\label{fig:deltaE0}
\end{figure}

The signal feature of the quantum $U(1)$ liquid is its photon excitations. 
One important consequence of these, so far as the simulation of finite-size systems 
is concerned, is a characteristic finite-size correction to the ground state energy 
per site $E_0/N$, coming from the zero-point energy of the photons
\begin{eqnarray}
\frac{\delta E_0(L)}{N} = \frac{1}{N}\left[ E_0(L) - E_0(\infty) \right] =  x_1 L^{-4} + \ldots
\label{Escale}
\end{eqnarray}
where $L \sim N^\frac{1}{3}$ is the linear dimension of the cluster,
and the coefficient $x_1$ is proportional to the speed of light $c$ 
[cf. \onlinecite{sikora11}].
This means that it is possible to extract the speed of light from the finite-size 
scaling of the ground state energy found in simulations of 
${\mathcal H}_\mu$~[Eq.~(\ref{eq:Hmu})], shown in Fig.~\ref{fig:deltaE0}.  


Approaching this problem from the lattice field theory 
${\mathcal H}^\prime_{\sf U(1)}$~[Eq.~(\ref{eq:HA})], 
we know that 
\begin{eqnarray}
c = \sqrt{\mathcal{U} \mathcal{K}} a_0 
= \kappa^2 \mathcal{K} \sqrt{\overline{\mathcal{U}}} a_0.
\label{eq:c}
\end{eqnarray}
where the dimensionless parameter 
$\overline{\mathcal{U}} = \frac{\mathcal{U}}{\mathcal{K} \kappa^4}$ can be determined separately 
from fits to structure factors (cf Fig.~\ref{fig:fit-parameters}).
We also have enough information from the fits to the structure factor to evaluate the sum
\begin{eqnarray}
\frac{1}{N}\sum_{\mathbf{k}} \frac{\omega(\mathbf{k})}{\kappa^2 \mathcal{K}}
 = \frac{1}{\kappa^2 \mathcal{K}}\left(\frac{E_0}{N} + const \right)
\label{eq:owen}
\end{eqnarray}
where $\frac{E_0}{N}$ is the ground state energy per site found from 
Monte Carlo simulations.  


For $\overline{\mathcal{U}}=0$ the LHS of Eq.~(\ref{eq:owen}) does not 
depend on $L$.
This is consistent with simulations of the microscopic model at $\mu=g$.  
For $\overline{\mathcal{U}}>0$ we expect a scaling law $\sim \frac{1}{L^4}$ for large~$L$.  
We write
\begin{eqnarray}
\epsilon(L) 
\equiv \frac{1}{N}\sum_{\mathbf{k}} \frac{\omega(\mathbf{k})}{\kappa^2 \mathcal{K}}
=\epsilon(\infty)-x_2 L^{-4}
\label{FTscaling}
\end{eqnarray}
and it follows that
\begin{eqnarray}
\frac{x_1}{x_2}=\kappa^2 \mathcal{K}
\end{eqnarray}
with
\begin{eqnarray}
c=\frac{x_2}{x_1} \sqrt{\overline{\mathcal{U}}} a_0.
\end{eqnarray}
where the coefficients $x_1$ and $x_2$ can be found from the finite-size scaling
of the ground state energy in simulation [Fig.~\ref{fig:deltaE0}], and through the 
numerical evaluation of the $\sum_{\mathbf{k}}$ in Eq.~(\ref{eq:owen}) for a 
finite-size system.


We find that, for $0 \le \mu \le 1$, the evolution of the speed of light as a function of $\mu$ 
is well-described by 
\begin{eqnarray}
c^2 = \alpha \delta \overline{\mu} + \beta \delta \overline{\mu}^2 + {\mathcal O}(\delta \overline{\mu}^3)
\end{eqnarray}
where
\begin{eqnarray}
\delta \overline{\mu}  &=& 1 - \mu/g \\
\alpha &=& 0.22\ g^2\ a_0^2 \\
\beta &=& 0.13\ g^2\ a_0^2
\end{eqnarray}
In particular, for $\mu=0$, the physical point of our model, we find
\begin{eqnarray}
c=(0.6 \pm 0.1) \ g \ a_0 \ \hbar^{-1}
\label{eq:speedoflight}
\end{eqnarray}
where we have restored the dimensional factor of $\hbar$.   
We have also calculated an upper bound on $c$ from a single mode approximation, 
in the spirit of Ref. \onlinecite{moessner03}.  
We find 
\begin{eqnarray}
c  \le (0.6 \pm 0.1) \ g\ a_0 \ \hbar^{-1}
\end{eqnarray}
Within errors, the two numbers are indistinguishable.


It is interesting to use this result to make an order of magnitude estimate 
of the speed of light in a quantum spin ice material.
Considering Yb$_2$Ti$_2$O$_7$, as (presently) the best-characterised material, 
and inserting the exchange parameters obtained by Ross {\it et al.}~\cite{ross11-PRX} 
into the expression for the tunnelling matrix element [Eq.~(\ref{eq:g})], 
we obtain 
\begin{eqnarray}
   g_{\text{\tiny Yb$_2$Ti$_2$O$_7$}} \approx 0.05 \ \text{meV}.
\end{eqnarray}
From Eq.~(\ref{eq:speedoflight}), and the known size of the unit cell
\mbox{$a_0 = 10.026$}~\AA\ [\onlinecite{thompson11}], we find a speed of light 
\begin{eqnarray}
   c \sim 0.3 \ \text{meV \AA} \sim 50 \ \text{ms$^{-1}$}
\label{eq:cYTO}
\end{eqnarray}
which implies a photon bandwidth
\begin{eqnarray}
   \Delta \omega \sim 0.1 \ \text{meV} \, ,
\label{eq:bandwidthYTO}
\end{eqnarray}
within the range accessible to modern inelastic neutron scattering 
experiments~\cite{coldea10}.


The accuracy of this estimate is limited by the approximations made in setting up 
the minimal model of a quantum spin ice 
${\mathcal H}_{\sf tunnelling}$ [Eq.~(\ref{eq:Htunnelling})], 
and so it should only be regarded as a ``ballpark'' figure.  
It should also be remembered that Yb$_2$Ti$_2$O$_7$ is believed to 
order ferromagnetically at the lowest temperatures~\cite{chang-arXiv,savary12}. 
However as long as a given system remains an ``ice'', 
the inclusion of further tunnelling processes beyond 
${\mathcal H}_{\sf tunnelling}$ should only increase the speed of light.


\section{``Electromagnetism'' in a Quantum Spin Ice at finite temperature}
\label{section:finite-T}


\begin{figure*}
\centering
\includegraphics[width=1.0\textwidth]{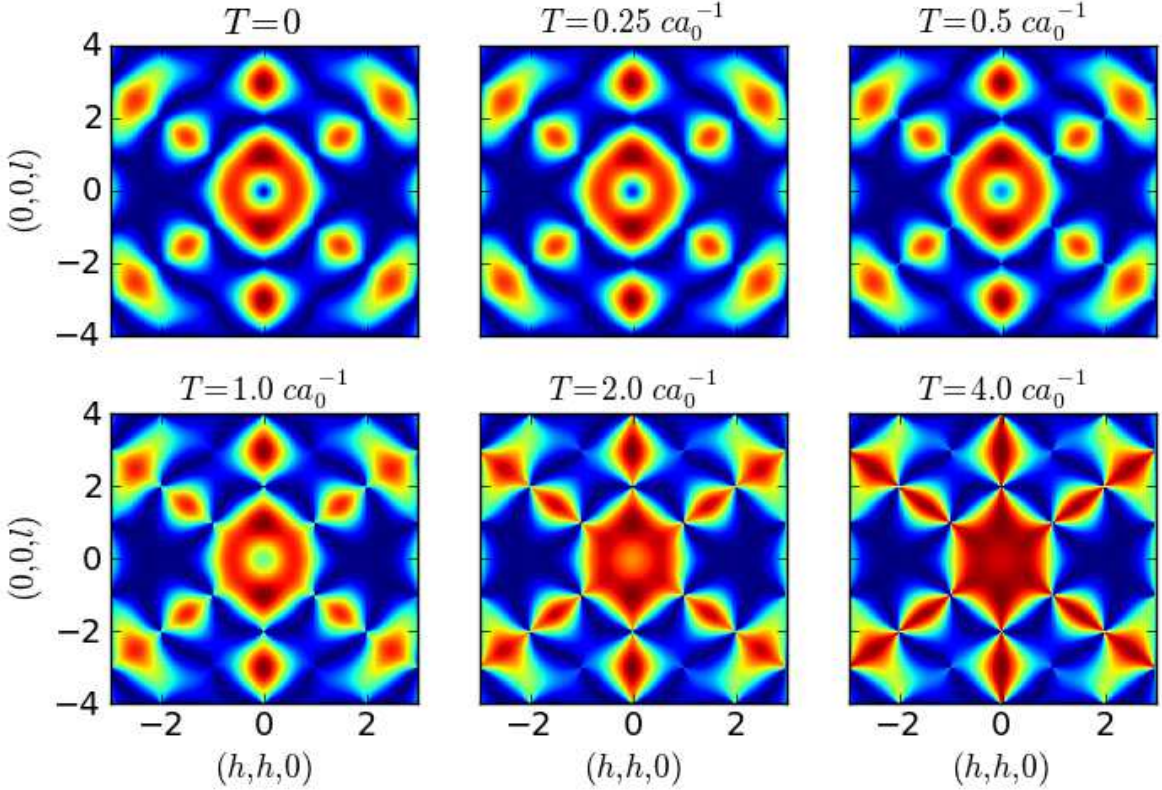}
\caption{
(Color online).
Slow, cold death of pinch points in a quantum ice.
Equal-time structure factor $S^{yy}_{\sf spin} (\mathbf{k}, t=0)$ [Eq.~(\ref{eq:Sk-spin})]
in the spin-flip channel measured by Fennell {\it et al.}~\cite{fennell09}, calculated from the 
lattice field theory ${\mathcal H}^\prime_{\sf U(1)}$ [Eq.~(\ref{eq:HA})], for comparison 
with neutron scattering experiments on a quantum spin ice.
Results are plotted for temperatures ranging from $T=0$ to \mbox{$T= c \ a_0^{-1}$}, 
where $c$ is the speed of light, and $a_0$ the linear dimension of the cubic unit cell, 
with temperature measured in units such that $\hbar = k_B = 1$.  
The pinch-point structure observed at finite temperature is progressively ``hollowed out'' 
as the system is cooled towards its zero-temperature ground state.
}
\label{fig:SkT}
\end{figure*}

In Section~\ref{subsection:QMC-zeroT} we have demonstrated that the field theory 
${\mathcal H}^\prime_{\sf U(1)}$~[Eq.~(\ref{eq:HA})] --- quantum electromagnetism on a 
pyrochlore lattice --- gives an excellent account of the results of zero-temperature quantum 
Monte Carlo simulations of the minimal microscopic model of a quantum spin ice, 
${\mathcal H}_\mu$~[Eq.~(\ref{eq:Hmu})].
These results confirm the conjecture that this model could 
support a spin-liquid phase, down to $T=0$.  
Encouraged by this, we now use the same field theory to explore how 
correlations in this spin liquid state develop at finite temperature.


In Section~\ref{subsection:SkT} we assess how the thermal 
excitation of magnetic photons changes the temperature dependence of the energy-integrated 
structure factors measured in quasi-elastic scattering.
We find that pinch-points eliminated by quantum fluctuations at zero temperature, 
are progressively restored as the temperature of the spin liquid is raised.


In Section~\ref{subsection:QMC-finiteT} we compare the results of the lattice field
theory with published results for quantum Monte Carlo simulations of quantum
charge ice at finite temperature~\cite{banerjee08}.
We find that both the form and the temperature dependence of the correlations are 
well described by the lattice field theory.


Finally, in Section~\ref{subsection:thermodynamic} we conclude with a few remarks
about the finite temperature behaviour of the heat capacity and uniform magnetic 
susceptibility in a quantum spin ice.


Throughout this analysis we set $\hbar=k_B = 1$, restoring dimensional factors of $\hbar$ and
$k_B$ only where we quote results for the coefficient of the heat capacity
associated with photons.


\subsection{Temperature dependence of structure factors}
\label{subsection:SkT}


The qualitative changes in correlations between spins at finite temperature can most
easily be understood within the continuum field theory ${\mathcal S}_{\sf eff}$ [Eq.~(\ref{eq:Seff})].
The thermal excitation of photons enhances correlations of the magnetic field ${\mathcal B}$
at small~$|\mathbf{k}|$ 
\begin{eqnarray}
C^{\mathcal B}_{\mu \nu}(\mathbf{k})
   &=& \frac{8 \pi^4 k}{c \sqrt{ 1+\left( \frac{\lambda_c k}{2 \pi} \right)^2}}
     \left( \delta_{\mu \nu} - \frac{k_{\mu} k_{\nu}}{k^2} \right) \nonumber \\
    && \times \coth{\left( \frac{c k \sqrt{ 1+\left( \frac{\lambda_c k}{2 \pi} \right)^2}}{2T} \right)}.
\label{eq:SkT-continuum}
\end{eqnarray}
and introduces a thermal de Broglie wavelength for the photons.  
\begin{eqnarray}
\lambda_{\sf T}=\frac{\pi c}{T}
\end{eqnarray}
Over sufficiently long distances, this enhancement of correlations
exactly cancels their suppression by quantum fluctuations.
Assuming that $\lambda_{\sf c} \ll \lambda_{\sf T}$, and expanding 
Eq.~(\ref{eq:SkT-continuum}) for small wave number, 
we find 
\begin{eqnarray}
\label{B finite T 2}
C^{\mathcal B}_{\mu \nu}(|\mathbf{k}| \ll 2\pi/\lambda_{\sf T}) 
= \frac{16 \pi^4 T}{c^2} \left( \delta_{\mu \nu} - \frac{k_{\mu} k_{\nu}}{k^2} \right)
+ \ldots
\end{eqnarray}
This implies that, for these small wave vectors, the pinch point is restored,  
but with a prefactor that depends linearly on temperature.


This result has very simple interpretation.  
At finite temperature photons are only coherent quantum excitations over a length scale 
$\lambda_{\sf T}$.
Therefore, while correlations in a quantum spin ice may decay as $1/r^4$ over distances 
$\lambda_{\sf c} \ll r \ll \lambda_{\sf T}$,  at long distances \mbox{for $r \gg \lambda_{\sf T}$} 
the classical $1/r^3$ decay of the spin correlations is restored.


\begin{figure}
\centering
\includegraphics[width=0.45\textwidth]{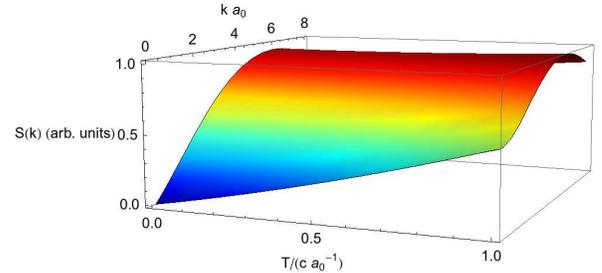}
\caption{
(Color online)
Angle-integrated scattering intensity $I (k \approx 0, T)$ [Eq.~(\ref{eq:SkT-integrated})] 
calculated from the lattice field theory ${\mathcal H}^\prime_{\sf U(1)}$ [Eq.~(\ref{eq:HA})], 
for comparison with neutron scattering experiments on a powder sample of 
a quantum spin ice.
Results are plotted for temperatures ranging from $T=0$ to \mbox{$T= 1.0 c a_0^{-1}$}, 
where $c$ is the speed of light, and $a_0$ the linear dimension of the cubic unit cell, with 
temperature measured in units such that $\hbar = k_B = 1$.  
The progressive elimination of pinch points as the sample is cooled 
manifests itself as a steady loss of scattering for $|\mathbf{k}| \to 0$.
}
\label{fig:SkT-integrated}
\end{figure}


All of these arguments generalise to the lattice field theory ${\mathcal H}^\prime_{\sf U(1)}$
[Eq.~(\ref{eq:HA})], and to expressions for the equal-time structure factor at finite temperatures
derived from $S_{\sf spin}^{\alpha\beta} (\mathbf{k}, \omega)$ [Eq.~(\ref{eq:Skomega-spin})].  
Thus we anticipate that they will apply equally to a quantum spin ice at finite temperatures.
This suggests a simple diagnostic for a quantum spin ice in quasi-elastic neutron scattering
experiments --- as the sample is cooled, and photons become coherent over longer 
length scales, the pinch points observed at reciprocal lattice vectors are progressively
``bleached out''.
This slow, cold, death of pinch points is illustrated in Fig.~\ref{fig:SkT}.   


Since there is also a characteristic loss of spectral weight in 
$S_{\sf spin}^{\alpha\beta} (\mathbf{k}, t=0)$ for $\mathbf{k} \approx \mathbf{0}$, 
exactly the same process could  be seen in the angle integrated structure factor 
measured in neutron scattering experiments on powder samples.
In this case, the intensity of scattering is given by 
\begin{eqnarray}
I (k, T) 
\propto \sum_{\alpha\beta} \int d\mathbf{\Omega} 
   \left( \delta_{\alpha \beta}-\frac{k_{\alpha} k_{\beta}}{k^2} \right)
      \ S_{\sf spin}^{\alpha\beta} (\mathbf{k}, t=0) 
   \nonumber\\      
   \label{eq:SkT-integrated}
\end{eqnarray}
For classical spin ice, or a quantum spin ice at sufficiently high temperature, 
$$ I (k \approx 0, T) \approx const. $$
However, as a quantum spin ice is cooled to zero temperature, the growing 
coherence of photons will manifest itself as a progressive loss of
spectral weight at small $k$, 
$$ I (k = 0, T) \sim T $$
until, for $T=0$, spectral weight at $k=0$ is eliminated entirely 
$$ I (k \approx 0, T=0) \propto k $$
This progression is illustrated in Fig.~\ref{fig:SkT-integrated}.  


\subsection{Comparison with quantum Monte Carlo simulation}
\label{subsection:QMC-finiteT}


It is also interesting to compare the predictions of the lattice field theory 
${\mathcal H}^\prime_{\sf U(1)}$ [Eq.~(\ref{eq:HA})], with the results of finite-temperature 
quantum Monte Carlo simulations of a quantum charge ice described by 
${\mathcal H}_{\sf t-V}$ [Eq.~(\ref{eq:HtV})], as published by Banerjee~{\it et al.}~\cite{banerjee08}.   
Banerjee {\it et al.} performed their simulations for hard-core bosons on a pyrochlore 
lattice at half filling, with hopping integral $t=1$, and nearest-neighbour repulsion $V=19.4$, 
at temperatures $T = 1.05 g$ and $T = 1.57 g$,  where $g = 12 t^3/V^2$ is the size of the 
leading tunnelling matrix element between different charge ice configurations.

\begin{figure}[h!]
\centering
\includegraphics[width=0.45\textwidth]{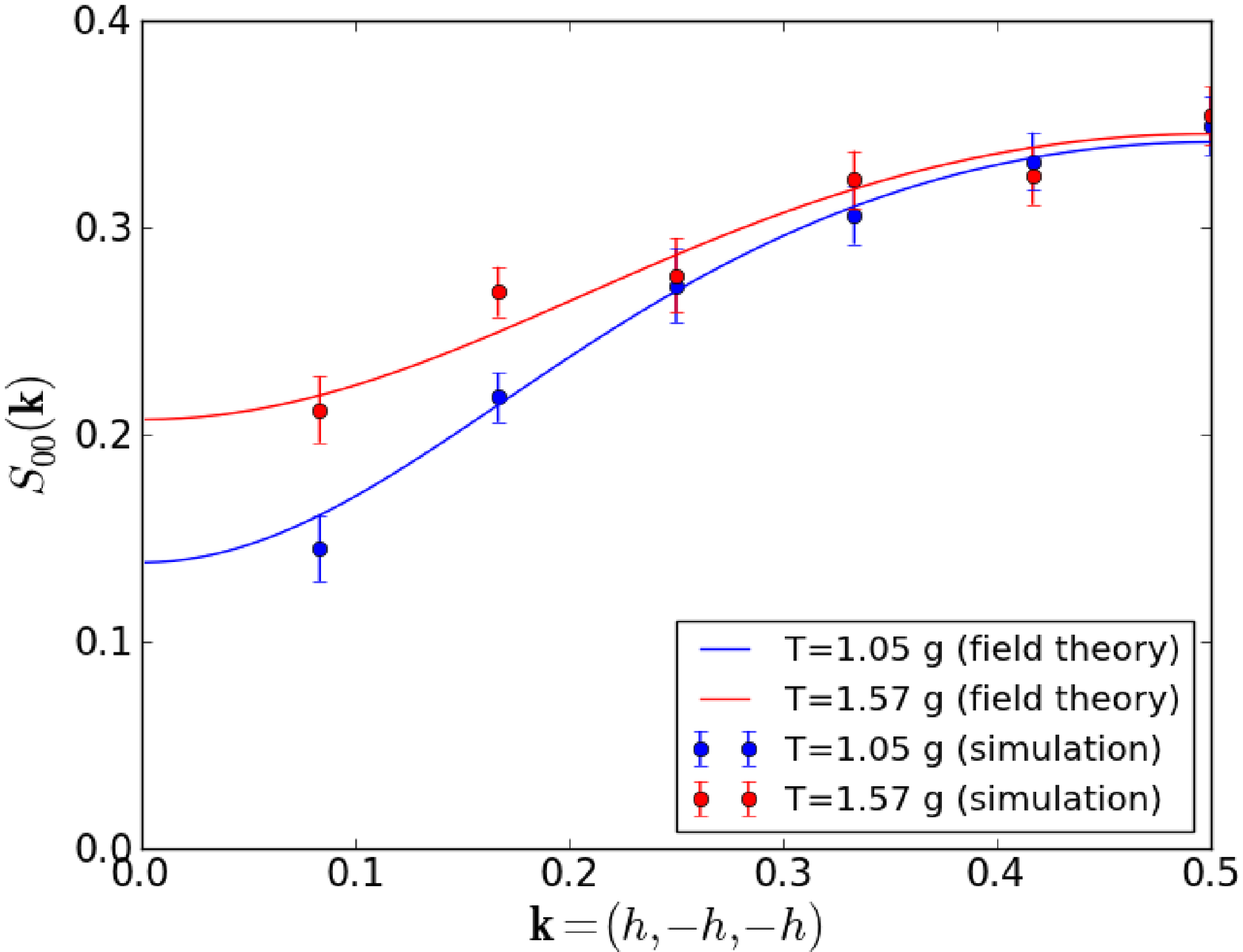}
\caption{
(Color online).
Comparison of the predictions of the lattice field theory ${\mathcal H}^\prime_{\sf U(1)}$ 
[Eq.~(\ref{eq:HA})] with the results of finite-temperature quantum Monte Carlo simulations 
of a quantum charge ice described by ${\mathcal H}_{\sf t-V}$ [Eq.~(\ref{eq:HtV})].
Results are shown for the equal time, on-sublattice structure factor
\mbox{$S_{00}(\mathbf{k})=\langle n_0 (\mathbf{-k}) n_0(\mathbf{k}) \rangle$}
Simulations are taken from Banerjee~{\it et al.}~\cite{banerjee08}, 
and were performed at temperatures $T = 1.05 g$ and $T = 1.57 g$,   
where $g = 12 t^3/V^2$ is the size of the leading tunnelling matrix 
element between different charge ice configurations.
The temperature dependence of the spin correlations makes it possible 
to estimate the speed of light $c \approx 1.8 \ g \ a_0 \ \hbar^{-1}$.  
}
\label{fig:banerjee}
\end{figure}

In Fig.~\ref{fig:banerjee} we plot simulation results for $S_{\sf charge} (\mathbf{k}, t=0)$ 
at these temperatures, calculated within a single sublattice of pyrochlore lattice sites, together 
with the best fit to Eq.~(\ref{eq:Sk-charge}), projected onto a single sublattice.
We assume that the parameters of the field theory depend relatively weakly 
on temperature, and attribute the temperature dependence of correlations
entirely to the thermal excitations of photons.
Under these assumptions, the lattice field theory gives a good account of both the form 
and the temperature dependence of $S_{\sf charge} (\mathbf{k}, t=0)$, within the error 
bars on points taken from simulation.


These fits suggest a speed of light
\begin{eqnarray}
c=(1.8 \pm 0.2) \ g \ a_0 \ \hbar^{-1}
\end{eqnarray}
which is $\sim 3$ times larger than that found in Section~\ref{subsection:c} from finite 
size scaling of the ground state energy of ${\mathcal H}_\mu$ [Eq.~(\ref{eq:Hmu})].   
This discrepancy can probably be attributed to the fact that the simulations 
of Banerjee~{\it et al.} were performed close to the melting point of the charge 
ice~\cite{banerjee08}, where both interactions between photons, and tunnelling processes 
involving more than six lattice sites, are likely to play an important role.
Since all of these processes will contribute to the rate at which the gauge field fluctuates
in time, they can be expected to increase the speed of light.


\subsection{Heat capacity and magnetic susceptibility at low temperatures}
\label{subsection:thermodynamic}


Neutron scattering experiments have the potential to give decisive information
about emergent electromagnetism in a quantum spin ice.  
However these experiments are expensive and difficult to perform, and depend critically
on the size and quality of available samples.
We therefore conclude with a few brief remarks on potential signatures of 
a quantum $U(1)$ liquid in thermodynamic quantities.   
The results given will hold in the low-temperature regime where the physics 
of a quantum spin-ice can be described as a gas of photons. 
At higher temperatures the thermal excitation of the gapped spinons (monopoles)
and electric charges also play an important role.


We have seen in Section~\ref{subsection:SkT} how quantum fluctuations lead to an 
equal-time structure factor which, in the limit \mbox{$k \to 0$}, vanishes at low 
temperatures as
\begin{eqnarray}
\lim_{k \to 0}  S(k, T) \propto T
\end{eqnarray}
This in turn implies a bulk magnetic susceptibility $\chi(T)$ which is 
independent of temperature at low temperatures
\begin{eqnarray}
\chi^{-1}
(T \ll g)  =  \frac{3\mathcal{U}}{\kappa^2}.
\label{eq:chi}
\end{eqnarray}
where $\mathcal{U}$ is the coefficient of $\mathcal{B}^2$ in the effective 
Hamiltonian $\mathcal{H}_{\sf U(1)}$ [Eq.~(\ref{eq:Hem})], and 
$\kappa \approx 1$ is the dimensionless scale factor introduced in 
Eq.~(\ref{eq:Skomega-charge-definition}).


This result provides another means of parameterizing the lattice field theory.
It is also a potentially useful diagnostic for experiment, since, a classical spin ice 
which remains in thermodynamic equilibrium at low temperatures, 
will exhibit an effective Curie law~\cite{isakov04,ryzhkin05} 
\begin{eqnarray}
\chi^{-1}
(T) \sim T
\end{eqnarray}
This result follow directly from the fact that there are more spin 
ice configurations with vanishing magnetisation $\mathbf{M} = 0$ 
than with any finite magnetisation $\mathbf{M} \ne 0$, and so, 
in the absence of any other considerations, a state with 
$\mathbf{M} = 0$ is selected by an entropic term 
$\delta {\mathcal F} = T \delta {\mathcal S}\ \sim T\ \mathbf{M}^2$ in the free energy~\cite{henley05}.    
Nonetheless, any comparison with a classical spin ice 
should be approached with some caution, as these systems need not remain in 
equilibrium at low temperatures~\cite{matsuhira00,snyder01},
and the character of the spin fluctuations which control $\chi(T)$ changes 
as a function of temperature~\cite{matsuhira01, jaubert09}.


As noted elsewhere~\cite{hermele04,savary12}, the fact that photons are linearly 
dispersing excitations implies that they must make a $T^3$ contribution to the heat 
capacity at low temperatures. 
While this contribution has exactly the same temperature dependence as that from 
acoustic phonons, the large amount of entropy available in ice states, and low 
speed of light [cf. Section~\ref{subsection:c}], mean that the heat capacity at low 
temperatures will be dominated by photons.
The photon contribution to the heat capacity per mole is
\begin{eqnarray}
C_{\sf photon}[mole]= B T^3
\label{eq:cV}
\end{eqnarray}
with the coefficient B given by
\begin{eqnarray}
B=   \left( \frac{\pi^2}{30}\right) R \left( \frac{k_B a_0}{\hbar c}\right)^3.
\end{eqnarray}
From the characterisation of Yb$_2$Ti$_2$O$_7$ by Ross {\it et al.}~\cite{ross11-PRX}, 
and the analysis of the speed of light in Section~\ref{subsection:c}, we estimate 
\begin{eqnarray}
B \approx 65\ \text{J}\ \text{mol}^{-1}\ \text{K}^{-4}
\end{eqnarray}
which is several orders of magnitude larger than the expected phonon contribution.
This should be compared with the value
$$
1\ \text{J}\ \text{mol}^{-1}\ \text{K}^{-4} 
$$
obtained in Ref.~\onlinecite{savary12}.  
We note that, since the photons are magnetic excitations, measurements of the heat capacity
in an applied magnetic field may also prove instructive.
 

\section{Discussion and Conclusions}
\label{section:conclusions}


In this paper we have developed a detailed theory for the simplest 
microscopic model which could describe quantum tunnelling between different
spin ice configurations [Eq.~(\ref{eq:Htunnelling})].
The striking claim that this type of model could support a spin liquid phase 
which perfectly mimics quantum electromagnetism~\cite{hermele04}
has been verified by quantum Monte Carlo 
simulations~\cite{banerjee08,shannon12}.
Here we have explored how such a quantum spin liquid might manifest 
itself in experiment, parameterizing an ``electromagnetic'' lattice gauge theory 
from quantum Monte Carlo simulations at zero temperature, and using this 
to calculate the dynamical structure factor $S^{\alpha\beta}(\mathbf{k},\omega)$ 
[Eq.~(\ref{eq:Skomega-spin})] which would be measured in neutron 
scattering experiments at finite temperature.


We find that a key signature of the emergent electromagnetism is the 
suppression of pinch points singularities in the energy-integrated structure 
factor $S^{\alpha\beta}(\mathbf{k}, t=0)$ as the system is cooled to its 
zero-temperature ground state [Fig.~\ref{fig:SkT}].
This will coincide with the appearance of a gapless, linearly dispersing, 
mode --- the photon of the lattice gauge theory ---- in inelastic neutron scattering 
[Fig.~\ref{fig:inelastic-unpolarised}].  
In sharp contrast with a conventional antiferromagnet, the dispersing feature 
associated with this photon vanishes as $\omega \to 0$.   
These photons will also strongly influence the low-temperature
thermodynamic properties of the system, giving rise to a temperature-independent 
contribution to the magnetic susceptibility [Eq.~(\ref{eq:chi})] and an anomalously 
large $T^3$ contribution to specific heat [Eq.~(\ref{eq:cV})]~\cite{savary12}.  


Neither the idea of ``artificial light''~\cite{foerster80,motrunich02,wen02-PRL, 
wen02-PRB, wen03,motrunich05}, 
nor the observation that there could be quantum tunnelling between different spin ice 
configurations~\cite{bramwell98}, is new.  
However the possibility that one might lead to another is both new and exciting, 
and adds to the general frisson surrounding pyrochlore magnets.  
Without attempting to review all of this fast-developing field --- but with the possibility 
of observing photons in mind --- it is interesting to ask whether any of the materials 
currently studied ``fit the bill''.


The most widely studied example of a three-dimensional spin liquid is the insulating
pyrochlore oxide Tb$_2$Ti$_2$O$_7$~\cite{gardner10}. 
Tb$_2$Ti$_2$O$_7$ does not order down to $50$~mK [\onlinecite{gardner03}], despite having 
a Curie-Weiss temperature \mbox{$\theta_{\sf CW} \sim 14$~K} [\onlinecite{gingras00}], 
and a strong tendency to order under magnetic field or pressure~\cite{mirebeau02,mirebeau04}.   
In a series of papers, Gingras and coauthors have argued that Tb$_2$Ti$_2$O$_7$ 
is a ``quantum spin ice'', in which spins fluctuate strongly about the crystallographic $[111]$ 
axes.
These claims were made on the basis of a characteristic checkerboard structure observed 
in diffuse neutron scattering experiments at high temperatures~\cite{kao03, enjalran04}, 
and a subsequent microscopic analysis of crystal field levels~\cite{molavian07,molavian-arXiv}, 
and find support in the recent observation of partial magnetisation
plateau for magnetic field applied along a $[111]$ axis~\cite{baker-arXiv,molavian09}.


Within this framework, the field at which the plateau is observed implies  
that the energy scale relevant for ``quantum spin ice'' behaviour in Tb$_2$Ti$_2$O$_7$ 
is \mbox{$J_{\sf eff} \approx 0.2 K$}~\cite{baker-arXiv}.  
Unfortunately, the interpretation of experiment at these low temperatures 
is muddied by questions of sample quality, with inconsistent results for spin-freezing
obtained by different authors~\cite{gardner99,luo01,yasui02,gardner03,hamaguchi04}.  
Published thermodynamic data at low temperatures is also  
less than conclusive, showing hints of a saturation of 
$\chi(T)$ at low temperatures, but strong sample 
dependence~\cite{luo01,gardner03,hamaguchi04,ofer07,chapuis10}.
And the picture is further complicated by strong fluctuations of the 
lattice~\cite{ruff07, ruff-arXiv}, with alternative theories of Tb$_2$Ti$_2$O$_7$ 
building on lattice effects~\cite{bonville11, bonville-arXiv, gaulin11}


At present there is no published neutron scattering data for Tb$_2$Ti$_2$O$_7$ 
with the combination of k-resolution, energy resolution and low temperature needed 
to compare with the predictions in Section~\ref{subsection:Skomega-B} 
and Section~\ref{subsection:Skomega} of this paper.
However recent evidence of ``pinch-point'' structure in quasi-elastic neutron 
scattering on single crystals of Tb$_2$Ti$_2$O$_7$~\cite{fennel12}, taken together 
with inelastic neutron scattering experiments on powder samples~\cite{takatsu12}, 
suggest that the comparison might be interesting.
The latter find evidence of a quasi-elastic feature evolving into two bands of excitations 
at temperatures $T < 0.4$K~\cite{takatsu12}.
If --- and it remains a big IF --- the behaviour of Tb$_2$Ti$_2$O$_7$ is connected 
with the physics of the quantum ice described in this paper, it would be tempting 
to identify these bands with the excitations of electromagnetism on a lattice 
--- gapless photons, together with gapped ``electric''  and ``magnetic'' charges (spinons).  
But more, and more delicate, experiments will be needed to determine whether
this is indeed the case.
And ultimately, Tb$_2$Ti$_2$O$_7$ will remain a fascinating system to study.   
regardless of whether or not it is a quantum spin ice.


Recently, there has also been intense experimental and theoretical interest in the 
closely-related Yb pyrochlore, Yb$_2$Ti$_2$O$_7$.  
Originally identified in the pioneering survey of Bl\"ote {\it et al.}~\cite{bloete69} as a ferromagnet 
with \mbox{$T_c=0.21$K} and \mbox{$\theta_{\sf CW}=0.4$K}, 
Yb$_2$Ti$_2$O$_7$ differs from the classical spin ice materials 
Ho$_2$Ti$_2$O$_7$ and Dy$_2$Ti$_2$O$_7$ in that the 
lowest lying crystal field state is a Kramers doublet with easy-plane 
anisotropy~\cite{hodges01,cao09,onoda10}.
An XY ferromagnet on a pyrochlore lattice 
--- modern estimates suggest \mbox{$\theta_{\sf CW} \approx 0.65$K} 
for Yb$_2$Ti$_2$O$_7$ [\onlinecite{yasui03,hodges01}] ---
would naturally be expected to order ferromagnetically at low temperatures.
However Yb$_2$Ti$_2$O$_7$ exhibits a far more complicated phenomenology.


Neutron scattering experiments at temperatures below $10$K find diffuse liquid-like structure
 which offers evidence of anisotopic exchange interactions~\cite{thompson11,chang-arXiv}.
At temperature of order $1$K, rod like structure emerges, reminiscent of a dimensional 
crossover~\cite{bonville04,ross09,ross11-PRB,thompson11,ross-arXiv}.
Some authors have found evidence of a first order transition into a ferromagnetically 
ordered state at $T_c=0.24$K~\cite{yasui03,chang-arXiv}, although this has been 
contested, and may not occur in all samples~\cite{hodges01,gardner04, hodges02,bonville04,ross09}.


That Yb$_2$Ti$_2$O$_7$ orders ferromagnetically in applied magnetic field is, 
however, uncontroversial.
And this has made it possible for Ross {\it et al.}~\cite{ross11-PRX} to accurately 
characterise an exchange Hamiltonian for Yb$_2$Ti$_2$O$_7$ [Eq.~(\ref{eq:Hross})] 
from fits to spin wave excitations in the polarised state.  
The parameters obtained confirm that the dominant interactions in Yb$_2$Ti$_2$O$_7$
favour  ``ice'' states, but that these are complimented by terms which will drive significant 
fluctuations at low temperatures.


Reassuringly, this description of Yb$_2$Ti$_2$O$_7$ is also in quantitative
agreement with measurements of thermodynamic properties over a wide 
range of temperatures~\cite{applegate-arXiv}.
This makes Yb$_2$Ti$_2$O$_7$  the best-characterised ``quantum spin ice'',
and as such, it is a natural place to look for emergent electromagnetism.
However neutron scattering data with sufficient resolution to 
compare with the predictions of this paper are not, as yet, available.


Tb$_2$Ti$_2$O$_7$ and Yb$_2$Ti$_2$O$_7$ are by no means 
the only pyrochlore systems with spin-liquid properties~\cite{gardner10}, and some of
these other systems, notably Pr$_2$Sn$_2$O$_7$~\cite{matsuhira04,zhou08,onoda10}
and Pr$_2$Zr$_2$O$_7$~\cite{onoda10,onoda11-PRB,matsuhira09}
are also worth investigating as potential realisations of a quantum ice. 
It might also be interesting to revisit two-dimensional ice-type materials, such as the 
proton bonded ferroelectric copper formate tetrahydrate~\cite{youngblood80}.  
While two-dimensional quantum ice models are known to order at low 
temperatures~\cite{chakravarty02, shannon04, syljuasen06, pollmann06, poilblanc08, pollmann11}, 
they are described by the same class of lattice gauge theory,  
and possess the same spinon excitations as their three-dimensional 
counterparts~\cite{fulde02,shannon04,pollmann11}.
These excitations will be confined in the ordered state, but might be visible at finite energy, 
and above the ordering temperature.


Although the theoretical possibility of emergent electromagnetism in 
quantum ice~\cite{hermele04,castro-neto06,banerjee08,shannon12} and 
quantum dimer~\cite{moessner03,bergman06,sikora09,sikora11} 
models is now well-established, many theoretical questions remain open.
In this paper we have considered only the simplest microscopic model 
of a quantum spin ice [Eq.~(\ref{eq:Hmu})], and fully characterised 
only its photon excitations.
The study of more realistic models, and of other excitations, 
is still in its infancy~\cite{savary12,wan12,applegate-arXiv}.
We have also made no attempt to resolve the question of how 
the quantum ice state which we find at low temperatures,
becomes a classical ice at high temperatures.
All of these issues remain for future study.
But we believe that the best motivation for studying them is experiment, and 
hope that the results in this paper will encourage further experiments 
on spin liquid materials which might realise artificial light.


\section{Acknowledgements}


The authors are pleased to acknowledge helpful conversations with 
Steven Bramwell, 
John Chalker, 
Peter Fulde, 
Bruce Gaulin, 
Michel Gingras, 
Paul McClarty, 
Roderich Moessner,  
Karlo Penc, 
Frank Pollmann, 
Lucile Savary and 
Alan Tennant.
We are particularly grateful to Tom Fennell and Radu Coldea for critical readings
of the manuscript.  
This work was supported by EPSRC Grants EP/C539974/1 and EP/G031460/1.  
OS and NS gratefully acknowledge the hospitality of the guest program of 
MPI-PKS Dresden.

%
%

\bibliographystyle{apsrev}

\begin{thebibliography}{10}


\bibitem{anderson73}
P.~W.~Anderson, 
\newblock  Mat. Res. Bull. {\bf 8}, 153 (1973).

\bibitem{lee08}
P.~A.~Lee
\newblock  Science {\bf 321}, 1306, (2008).

\bibitem{balents10}
L.~Balents.
\newblock  Nature {\bf 464}, 199, (2010).


\bibitem{harris97}
M.~J.~Harris, S.~T.~Bramwell, D.~F.~McMorrow, T.~Zeiske, and K.~W.~Godfrey , 
\newblock  Phys. Rev. Lett. {\bf 79}, 2554 (1997)

\bibitem{bramwell01}
S.~T.~Bramwell and M.~J.~P.~Gingras,
\newblock  Science {\bf 294}, 1495 (2001).

\bibitem{gardner10}
J.~S.~Gardner, M.~J.~P.~Gingras, and J.~E.~Greedan, 
\newblock  Rev. Mod. Phys. {\bf 82}, 53 (2010).


\bibitem{huse03} 
D.~A.~Huse, W.~Krauth, R.~Moessner and S.~L.~Sondhi,
\newblock  Phys. Rev. Lett. {\bf 91}, 167004 (2003).

\bibitem{henley05}
C.~L.~Henley, 
\newblock  Phys. Rev. B {\bf 71}, 014424 (2005).

\bibitem{fennell09}
T.~Fennell, P.~P.~Deen ,A.~R.~Wildes, K.~Schmalzl, D.~Prabhakaran,  A.~T.~Boothroyd, 
R.~J.~Aldus, D.~F.~McMorrow and S.~T.~Bramwell,
\newblock  Science {\bf 326}, 415 (2009).

\bibitem{henley10}
C.~L.~Henley, 
\newblock  Annu. Rev. Condens. Matter Phys. {\bf 1}, 179 (2010).


\bibitem{ryzhkin05}
I.~Ryzhkin.
\newblock JETP {\bf 101}, 481, (2005).
%
\bibitem{castelnovo08}
C.~Castelnovo, R.~Moessner and S.~L. Sondhi,
\newblock Nature {\bf 451}, 42, (2008).

\bibitem{bramwell09}
S.~T.~Bramwell, S.~R.~Giblin, S.~Calder, R.~Aldus, D.~Prabhakaran and T.~Fennell,
\newblock Nature {\bf 461}, 956, (2009).

\bibitem{morris09}  
D.~J.~P.~Morris, D. A. Tennant, S. A. Grigera, B.~Klemke, C.~Castelnovo, R.~Moessner, C.~Czternasty, M.~Meissner, K.~C.~Rule, 
J.~U.~Hoffmann, K.~Kiefer, S.~Gerischer, D.~Slobinsky and R.~S.~Perry,
Science {\bf  16}, 411 (2009).

\bibitem{kadowaki09}
H.~Kadowaki, N.~Doi, Y.~Aoki, Y.~Tabata, T.~J.~Sato, J.~W.~Lynn, K.~Matsuhira, Z.~Hiroi
J. Phys. Soc. Jpn. {\bf  78}. 103706 (2009).

\bibitem{jaubert09}  
L.~D.~C.~Jaubert and P.~C.~W.~Holdsworth, 
Nature Phys. {\bf 5}, 258 (2009).

\bibitem{giblin11}
S. R. Giblin, S. T. Bramwell, P. C. W. Holdsworth, D. Prabhakaran and I.~Terry,
\newblock Nature Physics, {\bf 7}, 252, (2011).


\bibitem{gingras00}
M.~J.~P.~Gingras, B. C. den Hertog, M.Faucher, J.~S.~Gardner, S.~R.~Dunsiger, L.~.J.~Chang, B.~D.~Gaulin,
N.~P.~Raju and J.~E.~Greedan,
\newblock Phys. Rev. B {\bf  62}, 6496 (2000).

\bibitem{enjalran04}
M.~Enjalran and M.~J.~P.~Gingras, 
\newblock Phys. Rev. B {\bf  70}, 174426 (2004).

\bibitem{molavian07}
H.~R.~Molavian, M.~J.~P.~Gingras, and B.~Canals, 
\newblock Phys. Rev. Lett. {\bf 98}, 157204 (2007).

\bibitem{molavian09}
H.~R.~Molavian and M.~J.~P.~Gingras
\newblock J. Phys.: Condens. Matter {\bf 21} 172201 (2009)  

\bibitem{gardner01}
J.~S.~Gardner, B.~D.~Gaulin, A.J.~Berlinsky, P.~Waldron, S.~R.~Dunsiger, 
N.~P.~Raju and J.~E.~Greedan, 
\newblock Phys. Rev. B {\bf  64}, 224416 (2001).

\bibitem{gardner03}
J.~S.~Gardner, A.~Keren, G.~Ehlers, C.~Stock,
E.~Segal,  J.~M.~Roper, B.~F\r{a}k, 
M.~B.~Stone, P.~R.~Hammar, D.~H.~Reich and B.~D.~Gaulin,
\newblock Phys.~Rev. B {\bf  68}, 180401(R) (2003).

\bibitem{gardner99}
J.~S.~Gardner, S.~R.~Dunsiger, B.~D.~Gaulin, M.~J.~P.~Gingras, J.~E.~Greedan, 
R.~F.~Kiefl, M.D.~Lumsden, W.~A.~MacFarlane, N.~P.~Raju, J.~E.~Sonier, 
I.~Swainson and Z.~Tun, 
\newblock Phys. Rev. Lett. {\bf  82}, 1012 (1999).

\bibitem{fennel12}
T.~Fennell, M.~Kenzelmann, B.~Roessli, M.K.~Haas, and R.J.~Cava
\newblock Phys. Rev. Lett. {\bf  109}, 017201 (2012).


\bibitem{thompson11}
J.~D.~Thompson, P.~A.~McClarty, H.~M.~Ronnow, L.~ P.~ Regnault, A.~Sorge and M.~J.~P.~Gingras,
\newblock Phys. Rev. Lett. {\bf 106}, 187202 (2011).

\bibitem{ross11-PRX}
K.~A. Ross, L.~Savary, B.~D.~Gaulin and L.~Balents,
\newblock Phys. Rev. X {\bf 1}, 021002 (2011).

\bibitem{ross11-PRB}
K.~A.~Ross, L.~R.~Yaraskavitch, M.~Laver, J.~S.~Gardner, J.~A.~Quilliam, 
S.~Meng, J.~B.~Kycia, D.~K.~Singh, T.~Proffen, H.~A.~Dabkowska and B.~D.~Gaulin
\newblock Phys. Rev. B {\bf 84}, 174442 (2011).

\bibitem{chang-arXiv}
L.-J.~Chang, S.~Onoda, Y.~Su, Y.-J.~Kao, K.-D.~Tsuei, Y.~Yasui, K.~Kakurai and M.~R.~Lees, 
\newblock  arXiv:1111.5406.

\bibitem{applegate-arXiv}
R.~Applegate, N.~R.~Hayre, R.~R.~P.~Singh, T.~Lin, A.~G.~R.~Day and M.~J.~P.~Gingras, 
\newblock  arXiv:1203.4569


\bibitem{zhou08}
H.~D.~Zhou, C.~R.~Wiebe, J.~A.~Janik, L.~Balicas, Y.~J.~Yo, Y.~Qiu, J.~R.~D.~Copley
and J.~S.~Gardner
\newblock Phys. Rev. Lett. {\bf  101}, 227204 (2008).

\bibitem{onoda10}
S. Onoda and Y. Tanaka, 
\newblock  Phys. Rev. Lett. {\bf 105}, 047201(2010).

\bibitem{onoda11-PRB}
S. Onoda and Y. Tanaka, 
\newblock  Phys. Rev. B {\bf 83}, 094411(2011).


\bibitem{matsuhira09}
K.~Matsuhira, C.~Sekine, C.~Paulsen, M.~Wakeshima, Y.~Hinatsu, T.~Kitazawa, 
Y.~Kiuchi, Z.~Hiroi and S.~Takagi,
\newblock  J. Phys. Conf. Ser. {\bf 145}, 012031 (2009).


\bibitem{lee-arXiv}
S.-B.~Lee, S.~Onoda and L.~Balents
\newblock arXiv:1204.2268v2.


\bibitem{gingras01}
M.~J.~P.~Gingras and B.~C.~den~Hertog, 
\newblock  Can. J. Phys. {\bf 79},1339 (2001)

\bibitem{melko04}
R.~G.~Melko, and M.~J.~P.~Gingras, 
\newblock  J. Phys.: Condens. Matter {\bf 16}, R1277 (2004).


\bibitem{bernal33} 
J.~D.~Bernal  and R.~H.~Fowler, 
\newblock  J. Chem. Phys. {\bf 1}, 515 (1933).


\bibitem{anderson58} 
P.~W.~Anderson, 
\newblock  Phys. Rev. {\bf 102}, 1008 (1956).

\bibitem{fulde02} 
P.~Fulde, K.~Penc, N.~Shannon,
\newblock  Ann. Phys. (Leipzig) {\bf 11}, 892  (2002).

\bibitem{youngblood80} 
R.~W.~Youngblood, J.~D.~Axe, and B.~M.~McCoy, 
\newblock  Phys. Rev. B {\bf 29}, 5212 (1980).

\bibitem{kondev98} 
J.~Kondev and J.~L.~Jacobsen,  
\newblock  Phys. Rev. Lett. {\bf 81}, 2922 (1998).


\bibitem{pauling35} 
L.~Pauling, 
\newblock  J. Am. Chem. Soc. {\bf 57}, 2680 (1935).

\bibitem{giauque36}
W.~F.~Giauque and J.~W.~Stout, 
\newblock  Am. J. Chem. Phys {\bf 58}, 58 (1936).

\bibitem{ramirez99}
A.~P.~Ramirez, A.~Hayashi, R.~J.~Cava, R.~Siddharthan and B.~S.~Shastry
\newblock  Nature {\bf 399}, 333 (1999).


\bibitem{moessner03} 
R.~Moessner and S.L.~Sondhi, 
\newblock  Phys. Rev. B {\bf 68},184512 (2003). 

\bibitem{hermele04} 
M.~Hermele, M.P.A.~Fisher, and L.~Balents, 
\newblock  Phys. Rev. B {\bf 69}, 064404 (2004).

\bibitem{castro-neto06} 
A.H.~Castro-Neto, P.~Pujol and E.~Fradkin,
\newblock  Phys. Rev. B {\bf 74}, 024302 (2006).


\bibitem{rokhsar88} 
D.S.~Rokhsar and  S.A.~Kivelson, 
\newblock  Phys. Rev. Lett. {\bf 61},2376 (1988).


\bibitem{foerster80}
D.~Foerster, H.~B.~Nielsen, and M.~Ninomiya, 
\newblock  Phys. Lett. {\bf 94}, 135 (1980), and references therein.

\bibitem{wen02-PRL}
X.-G.~Wen, 
\newblock  Phys. Rev. Lett. {\bf 88}, 011602 (2001)

\bibitem{wen02-PRB}
X.-G.~Wen, 
\newblock  Phys. Rev. B {\bf 65}, 165113 (2002).

\bibitem{motrunich02}
O.~I.~Motrunich and T.~Senthil, 
\newblock  Phys. Rev. Lett. 89, 277004 (2002).

\bibitem{wen03}
X.-G.~Wen, 
\newblock  Phys. Rev. B {\bf 68}, 115413 (2003).

\bibitem{motrunich05}
O.~I.~Motrunich and T.~Senthil, 
\newblock  Phys. Rev. B {\bf 71}, 125102 (2005).

\bibitem{motrunich04}
O.I.~Motrunich and A.~Vishwanath, 
\newblock  Phys. Rev. B {\bf 70}, 075104 (2004).

\bibitem{levin05}
M.~Levin and X.-G.~Wen, 
\newblock  Rev. Mod. Phys. {\bf 77}, 871 (2005).


\bibitem{shannon12}
N.~Shannon, O.~Sikora, F.~Pollmann, K.~Penc and P.~Fulde,
\newblock  Phys. Rev. Lett. {\bf 108}, 067204 (2012).

\bibitem{banerjee08} 
A.~Banerjee, S.~V.~Isakov, K.~Damle and Y.-B.~Kim,
\newblock  Phys. Rev. Lett. {\bf 100}, 047208 (2008).

\bibitem{sikora09}  
O.~Sikora, F.~Pollmann, N.~Shannon, K.~Penc and P.~Fulde,
\newblock  Phys. Rev. Lett. {\bf 103}, 247001 (2009).

\bibitem{sikora11}  
O.~Sikora, N.~Shannon, F.~Pollmann, K.~Penc and P.~Fulde,
\newblock  Phys. Rev. B {\bf 84}, 115129 (2011).


\bibitem{olga-unpub}
O.~Sikora, O.~Benton, P. McClarty, F.~Pollman, K.~Penc, R.~Moessner 
and N.~ Shannon, {\it in preparation}.


\bibitem{siddharthan99}
R.~Siddharthan, B.~S.~Shastry, A.~P.~Ramirez, A.~Hayashi, R.~J.~Cava, and S.~Rosenkranz, 
\newblock  Phys. Rev. Lett. {\bf 83}, 1854 (1999)

\bibitem{denHertog00}
B.~C.~den~Hertog and M.~J.~P.~Gingras, 
\newblock  Phys. Rev. Lett. {\bf 84}, 3430 (2000).

\bibitem{isakov05}
S.~V.~Isakov, R.~Moessner, and S.~L.~Sondhi, 
\newblock  Phys. Rev. Lett. {\bf 95}, 217201 (2005).


\bibitem{curnoe08}
S.~H.~Curnoe,  
\newblock  Phys. Rev. B. {\bf 78}. 094418 (2008).


\bibitem{hodges01}
J.~A.~Hodges, P.~Bonville, A.~Forget, M.~Rams, K.~Krolas and G.~Dhalenne, 
\newblock  J. Phys.: Condens. Matter {\bf 13}, 93010 (2001).

\bibitem{cao09}
H.~Cao, A.~Gukasov, I.~Mirebeau, P.~Bonville, C.~Decorse and G.~Dhalenne, 
\newblock  Phys. Rev. Lett. {\bf 103}, 056402 (2009)

\bibitem{onoda11}
S.~Onoda,
\newblock  J. Phys. Conf. Ser. {\bf 320}, 012065 (2011).


\bibitem{savary12}
L.~Savary and L.~Balents.
\newblock  Phys. Rev. Lett. {\bf 108}, 037202, (2012).

%



\bibitem{jaubert11}
L.~D.~C.~Jaubert, M.~Haque and R.~Moessner,
\newblock  Phys. Rev. Lett. {\bf 107}, 177202 (2011).


\bibitem{bergman06}
D.~L.~Bergman, G.~A.~Fiete, and L.~Balents, 
\newblock  Phys. Rev. B {\bf 73}, 134402 (2006).


\bibitem{guth80}
A.~H. Guth
\newblock Phys. Rev. D {\bf 21}, 2291, (1980).


\bibitem{wan12}
Y.~Wan and O.~Tchernyshyov, 
\newblock Phys. Rev. Lett. {\bf 108} 247210, (2012).

\bibitem{nussinov07}
Z.~Nussinov, C.D.~Batista, B.~Normand, and S.A.~Trugman
\newblock  Phys. Rev. B {\bf 75}, 094411 (2007).


\bibitem{calandra98} 
M.~Calandra~Buonaura and S.~Sorella,
\newblock  Phys. Rev. B {\bf 57},11446 (1998).


\bibitem{gould93}
R.~J.~Gould,
\newblock Astrophys. J. {\bf 12}, 417, (1993).


\bibitem{penc11} 
K.~Penc and A.~Lauchli, 
\newblock  {\it ``Introduction to frustrated magnetism''}, Chapter 13, Springer, Berlin (2011)

\bibitem{smerald-preprint}
A.~Smerald and N.~Shannon, 
\newblock  preprint.


\bibitem{coldea10}
R.~Coldea, D.~A~Tennant, E.~M.~Wheeler, E.~Wawrzynska, D.~Prabhakaran, 
M.~Telling, K.~Habicht, P.~Smeibidl and  K.~Kiefer, 
\newblock  Science {\bf 327} 177 (2010)


\bibitem{isakov04}
S. V. Isakov, K. S. Raman, R. Moessner, and S. L. Sondhi, 
\newblock  Phys. Rev. B {\bf 70}, 104418 (2004).

\bibitem{matsuhira00}
K.~Matsuhira, Y.~Hinatsu, K.~Tenya and T.~Sakakibara,
\newblock  J. Phys.: Condens. Matter {\bf 12}, L649 (2000).

\bibitem{snyder01}
J.~Snyder, J.~S.~Slusky, R.~J.~Cava and P.~Schiffer, 
\newblock  Nature {\bf 413}, 48 (2001).

\bibitem{matsuhira01}
K.~Matsuhira, Y.~Hinatsu and T.~Sakakibara,
\newblock  J. Phys.: Condens. Matter {\bf 13}, L737 (2001).




\bibitem{bramwell98}
S.~T.~Bramwell and M.~J.~ Harris, 
\newblock J. Phys.: Condens. Matter {\bf 10}, L215 (1998)


\bibitem{mirebeau02}
I.~Mirebeau, I.N.~Goncharenko, P.~Cadavez-Peres, S.T.~Bramwell, M.J.P.~Gingras and J.S.~Gardner, 
Nature {\bf 420}, 54 (2002).

\bibitem{mirebeau04}
I.~Mirebeau, I.N.~Goncharenko, G.~Dhalenne and A.~Revcolevschi,
Phys. Rev. Lett. {\bf 93}, 187204 (2004).

\bibitem{kao03}
Y.-J.~Kao, M.~Enjalran, A.~Del~Maestro, H.~R.~Molavian and M.~J.~P.~Gingras,
Phys. Rev. B {\bf  68}, 172407 (2003).

\bibitem{molavian-arXiv}
H.~R.~Molavian, P.~A.~McClarty, and M.~J.~P.~Gingras,
arXiv:0912.2957v1.

\bibitem{baker-arXiv}
P.~J.~Baker, M.~J.~Matthews, S.~R.~Giblin, P.~Schiffer, C.~Baines and D.~Prabhakaran
\newblock arXiv:1105.2196v1

\bibitem{yasui02}
Y.~Yasui, M.~Kanada, M.~Ito, H.~Harashina, M.~Sato, H.~Okumura, K.~Kakurai and H.~Kadowaki, 
\newblock J. Phys. Soc. Jpn. {\bf 71}, 599 (2002) 

\bibitem{luo01}
G.~Luo, S.~T.~Hess and L.~R.~Corruccini
\newblock Physics Letters A {\bf 291}, 306 (2001)

\bibitem{hamaguchi04}
N.~Hamaguchi, T.~Matsushita, N.~Wada, Y.~Yasui, and M.~Sato, 
\newblock Phys. Rev. B {\bf 69}, 132413 (2004)

\bibitem{ofer07}
O.~Ofer, A.~Keren and C.~Baines, 
\newblock J. Phys.: Condens. Matter {\bf 19}, 145270 (2007)

\bibitem{chapuis10}
Y.~Chapuis, A.~Yaouanc, P.~Dalmas de R\'eotier, C.~Marin, 
S.~Vanishri, S.~H.~Curnoe, C.~Vaju and A.~Forget, 
\newblock Phys. Rev. B {\bf 82}, 100402(R), (2010)

\bibitem{ruff07}
J.~P.~C.~Ruff, B.~D.~Gaulin, J~ P.~Castellan, K.~C.~Rule, J.~P.~Clancy, J.~Rodriguez
and H.~A.~Dabkowska,
\newblock Phys. Rev. Lett., 99, 237202 (2007).

\bibitem{ruff-arXiv}
J.~P.~C.~Ruff, Z.~Islam, J.~P.~Clancy, K.~A.~Ross, H.~Nojiri, Y.~H.~Matsuda, H.~A.~Dabkowska, 
A.~D.~Dabkowski and B.~D.~Gaulin
\newblock arXiv 1006.2854v1.

\bibitem{bonville11}
P.~Bonville, I.~Mirebeau, A.~Gukasov, S.~Petit and J.~Robert
\newblock Phys. Rev. B 84, 184409 (2011).

\bibitem{bonville-arXiv}
P.~Bonville, I.~Mirebeau, A.~Gukasov, S.~Petit and J.~Robert
\newblock arXiv:1104.1584v1.

\bibitem{gaulin11}
B.~D.~Gaulin, J.~S.~Gardner, P.~A.~McClarty and M.~J.~P.~Gingras,
\newblock Phys. Rev. B {\bf  84}, 140402(R) (2011).

\bibitem{takatsu12}
H.~Takatsu, H.~Kadowaki, T.~J.~Sato, Y.~Tabata and J.~W.~Lynn,
\newblock  J. Phys.: Condens. Matter {\bf 24}, 052201 (2012)


\bibitem{bloete69}
H.~W.~J.~Bl\"ote, R.~F.~Wielinga, and W.~J.~Huiskamp
\newblock  Physica {\bf 43}, 5 (2011)

\bibitem{bonville04}
P.~Bonville, J.~A.~Hodges, E.~Bertin, J-Ph.~Bouchaud, P.~Dalmas de Reotier, L-P.~Regnault, 
H.~M.~Ronnow, J-P.~Sanchez, S.~Sosin, A.~Yaouanc,
\newblock  Hyperfine Interact. {\bf 156}, 103 (2004).

\bibitem{ross09}
K.~A.~Ross, J.~P.~C.~Ruff, C.~P.~Adams, J.~S. Gardner, H.~A.~Dabkowska,
Y.~Qiu, J.~R.~D. Copley and B.~D.~Gaulin 
\newblock Phys. Rev. Lett. {\bf 103}, 227202 (2009).

\bibitem{ross-arXiv} 
K.~A.~Ross, L.~R.~Yaraskavitch, M.~Laver, J.~S.~Gardner, J.~A.~Quilliam, 
S.~Meng, J.~B.~Kycia, D.~K.~Singh, H.~A.~Dabkowska and B.~D.~Gaulin
\newblock  arXiv:1107.2377v1

\bibitem{yasui03}
Y.~Yasui, M.~Soda, S.~Iikubo, M.~Ito, M.~Sato, N.~Hamaguchi, T.~Matsushita, 
N.~Wada, T.~Takeuchi, N.~Aso and K.~Kakurai, 
\newblock  J. Phys. Soc. Jpn. {\bf 72}, 3014 (2003)

\bibitem{hodges02}
J.~A.~Hodges, P.~Bonville, A.~Forget, A.~Yaouanc, P.~Dalmas de Reotier,
G.~Andre, M.~Rams, K.~Krolas, C.~Ritter, P.~C.~M.~Gubbens, C.~T.~Kaiser,
P.~J.~C. King and C.~Baines
\newblock Phys. Rev. Lett. {\bf 88}, 077204 (2002).

\bibitem{gardner04}
J.~S.~Gardner, G.~Ehlers, N.~Rosov, R.~W.~Erwin, and C.~Petrovic,  
\newblock  Phys. Rev. B {\bf 70}, 180404(R) (2004)


\bibitem{matsuhira04}
K.~Matsuhira, C.~Sekine, C.~Paulsen, Y.~Hinatsu, 
\newblock  J. Magn. Magn. Mater. {\bf 272-276}, 981 (2004)


\bibitem{chakravarty02} 
S.~Chakravarty,
\newblock  Phys. Rev. B {\bf 66}, 224505 (2002).

\bibitem{shannon04} 
N.~Shannon, G.~Misguich, and K.~Penc, 
\newblock  Phys. Rev. B {\bf 69}, 220403(R) (2004). 

\bibitem{syljuasen06} 
O.F.~Syljusen and S.~Chakravarty,
\newblock  Phys. Rev. Lett. {\bf 96}, 147004 (2006). 

\bibitem{pollmann06} 
F.~Pollmann, J.~J.~Betouras, K.~Shtengel and P.~Fulde,
\newblock  Phys. Rev. Lett. {\bf 97}, 170407 (2006).

\bibitem{poilblanc08}
D.~Poilblanc, K.~Penc, and N.~Shannon,
\newblock  Phys. Rev. B {\bf 75}, 220503(R) (2007).

\bibitem{pollmann11} 
F.~Pollmann, J.~J.~Betouras, K.~Shtengel and P.~Fulde,
\newblock  Phys. Rev. B {\bf 83}, 155117 (2011).



\end{thebibliography}



\end{document}